\documentclass[trackchanges]{aastex701}
\usepackage{graphicx}
\usepackage{tcolorbox}
\usepackage{amsmath}
\usepackage{enumitem}
\usepackage{ragged2e}

\newcommand{\astrogen}{\texttt{AstroGenesis}}

\usepackage{amsmath}

\begin{document}

\title{\texttt{AstroGenesis}: A Domain-Specific Multi-Agent AI for Astrophysical Research}

\author[0000-0003-2011-2731]{N. Sahakyan}
\affiliation{ICRANet-Armenia, Marshall Baghramian Avenue 24a, Yerevan 0019, Armenia}
\email{narek.sahakyan@icranet.org}

\author[0009-0007-7798-2072]{M. Khachatryan}
\affiliation{ICRANet-Armenia, Marshall Baghramian Avenue 24a, Yerevan 0019, Armenia}
\email{mherkhachatryan35912@gmail.com}

\author[0000-0003-2242-0244]{A. Mahabal}
\affiliation{California Institute of Technology, Pasadena, CA 91125, USA}
\email{aam@astro.caltech.edu}

\author[0000-0003-1386-7861]{D. Pasham}
\affiliation{Eureka Scientific, Oakland, CA 94602, USA}
\affiliation{Department of Physics, The George Washington University, Washington, DC 20052, USA}
\email{p.dheerajreddy@gmail.com}

\author[0009-0002-9639-283X]{A. Khachatryan}
\affiliation{ICRANet-Armenia, Marshall Baghramian Avenue 24a, Yerevan 0019, Armenia}
\email{artyomkhachatryan117@gmail.com}

\author[0009-0004-8217-7922]{V. Markosyan}
\affiliation{ICRANet-Armenia, Marshall Baghramian Avenue 24a, Yerevan 0019, Armenia}
\email{vikamarkosan@gmail.com}

\author[0009-0007-3500-4414]{N. Unkovskii}
\affiliation{ICRANet-Armenia, Marshall Baghramian Avenue 24a, Yerevan 0019, Armenia}
\email{nii.unkovsky@gmail.com}

\author[0000-0002-3777-7580]{V. Vardanyan}
\affiliation{ICRANet-Armenia, Marshall Baghramian Avenue 24a, Yerevan 0019, Armenia}
\email{vardanyaninbox@gmail.com}

\author[0000-0003-4477-1846]{D. B\'egu\'e}
\affiliation{Department of Physics, Bar-Ilan University, Ramat Gan 5290002, Israel}
\email{cayley38@gmail.com}

\author[0000-0002-2265-5003]{P. Giommi}
\affiliation{Associated to INAF, Osservatorio Astronomico di Brera, Via Brera 28, I-20121 Milano, Italy}
\affiliation{Center for Astrophysics and Space Science (CASS), New York University Abu Dhabi, PO Box 129188, Abu Dhabi, United Arab Emirates}
\email{giommipaolo@gmail.com}

\author[0009-0003-7603-4709]{G. Harutyunyan}
\affiliation{ICRANet-Armenia, Marshall Baghramian Avenue 24a, Yerevan 0019, Armenia}
\email{gevorgharutyunyan97@gmail.com}

\author[0000-0002-5804-6605]{D. Israyelyan}
\affiliation{ICRANet-Armenia, Marshall Baghramian Avenue 24a, Yerevan 0019, Armenia}
\email{davidisrayelyan@gmail.com}

\author[0000-0002-8186-3793]{A. Tramacere}
\affiliation{Department of Astronomy, University of Geneva, Chemin Pegasi 51, 1290 Versoix, Switzerland}
\email{andrea.tramacere@gmail.com}

\author[0009-0007-4522-5501]{A. Casotto}
\affiliation{Siemens Digital Industries Software, 5800 Granite Parkway, Suite 600, Plano, TX 75024, USA}
\email{andrea.casotto@gmail.com}

\author[0000-0001-7959-3387]{Yu Wang}

\affiliation{ICRA, Dipartimento di Fisica, Sapienza Universit\`a di Roma,
Piazzale Aldo Moro 5, I-00185 Roma, Italy}
\affiliation{ICRANet,
Piazza della Repubblica 10, 65122 Pescara, Italy}
\email{yu.wang@icranet.org}
\affiliation{INAF -- Osservatorio Astronomico d'Abruzzo, Via M. Maggini snc, I-64100, Teramo, Italy}

\author[0000-0003-3010-7661]{Di Li}

\affiliation{State Key Laboratory of Radio Astronomy and Technology,
National Astronomical Observatories, Chinese Academy of Sciences,
Beijing 100101, China}

\affiliation{New Cornerstone Science Laboratory,
Department of Astronomy, Tsinghua University,
Beijing 100084, China}

\affiliation{Research Center for Astronomical Computing, Zhejiang Laboratory, Hangzhou 311100, China}

\email{dili@tsinghua.edu.cn}

\author[0000-0001-8939-6862]{H. Hakobyan}
\affiliation{Center for Computational Astrophysics, Flatiron Institute, 162 Fifth Ave., New York, NY 10010, USA}
\email{haykh.astro@gmail.com}

\author[0000-0001-9695-8472]{L. Izzo}
\affiliation{INAF – Osservatorio Astronomico di Capodimonte, Salita
Moiariello 16, 80131 Napoli, Italy}
\email{luca.izzo@inaf.it}



\begin{abstract}
Modern astrophysical research requires the integration of rapidly expanding scientific literature, heterogeneous observational data, and increasingly complex physical models. We introduce \astrogen\ (\url{https://astrogenai.com}), a domain-specific multi-agent AI framework that integrates literature retrieval, multiwavelength data access and analysis, theoretical modeling, and research ideation within a unified research environment. The current implementation focuses on blazar research and is organized around specialized agents coordinated through a Supervisor Agent and a Planner/Replanner architecture. These agents provide capabilities for retrieving and synthesizing literature, accessing and analyzing multiwavelength observations, performing physical modeling, and identifying research directions. A central component is the Theoretical Modeling Agent, which uses pretrained neural-network surrogate models to enable efficient broadband and multimessenger modeling through natural-language interaction. The framework provides natural-language access to science-ready multiwavelength observational data and retrieval-grounded literature through a multi-stage retrieval and ranking pipeline. The literature-retrieval system was evaluated using two benchmarks: a single-paper benchmark, in which each question targets one publication, and a multi-paper benchmark, in which questions may require evidence from several publications. At least one relevant publication was retrieved among the top five results for 76.6\% of the single-paper questions and 79.2\% of the multi-paper questions. Representative workflows demonstrate how literature, observational data, physical modeling, and hypothesis generation can be combined within traceable analyses. The framework is extensible to additional astrophysical domains, with the goal of streamlining research workflows and enabling efficient, connected, and reproducible scientific investigations.

\end{abstract}




\section{Introduction} 

Modern astrophysics is changing rapidly due to the continuous expansion of observational capabilities across the electromagnetic spectrum and beyond \citep{2019NatRP...1..585M}. Large-scale sky surveys such as LSST, Gaia, Euclid, and DESI, together with multi-messenger observatories, are producing increasingly large and heterogeneous data products \citep[e.g.,][]{2019ApJ...873..111I,2023A&A...674A...1G,2025A&A...697A...1E,2025arXiv250314745D}. Instruments operating across the electromagnetic spectrum from the radio, through optical, X-ray, to $\gamma$-ray, as well as in the neutrino and gravitational-wave domains now generate complex datasets that require coordinated analysis and interpretation. At the same time, the number of scientific publications in astrophysics has grown dramatically, making it difficult for researchers to stay informed about recent developments even within specialized subfields. The combination of rapidly expanding data archives and the continuous growth of scientific literature has created a major scalability challenge for modern astrophysical research \citep[for general science publications see ][]{Bornmann2021ScienceGrowth}.

At the same time, astrophysical workflows are becoming more complex \citep{Zhang2026AstroFlow,2025A&A...698A..13N}. Addressing a single scientific problem often requires combining multiple tasks, including literature review, observational data retrieval, preprocessing, statistical analysis, theoretical modeling, and interpretation within the context of previous studies. These tasks are distributed across separate archives, software packages, numerical tools, and analysis pipelines that were not designed to operate within a unified framework \citep{2019AJ....157...98G,2022MNRAS.509.4817G}. As a result, researchers frequently need to manually perform interactions between data repositories, fitting tools, visualization software, and literature databases using homemade scripts challenging the reproducibility of the analysis. Consequently, a substantial research effort is often devoted to workflow management and technical integration rather than scientific interpretation itself.

These challenges are more evident in high-energy astrophysics and in the study of variable, transient, and multimessenger phenomena \citep{2019NatRP...1..585M,2022NatRP...4..697G}. When studying such sources, it is required to combine observational data across radio, optical, X-ray, $\gamma$-ray, neutrino, and gravitational-wave channels obtained with instruments that have different sensitivities, observing cadences, data formats, and analysis procedures \citep{2017ApJ...848L..12A,2018Sci...361.1378I}. Since many of these sources evolve on timescales ranging from seconds to years, the interpretation of the origin of the emission depends on the temporal selection and coordination of the available observations. In some cases, rapid investigation of the variability, flaring activity, or transient emission is necessary both for planning follow-up observations and for studying source emission in different states within theoretical scenarios involving particle acceleration, radiation processes, and relativistic outflows \citep[e.g.,][]{2022MNRAS.509.2102G,2018ApJS..234....3G, 2024ApJS..275....4K}. This is further complicated by the limited accessibility of theoretical modeling tools and by the high computational cost of many physically motivated models, e.g., when parameter-space exploration or statistical fitting is required. As a result, scientific progress depends on the ability to integrate literature retrieval, observational data access, analysis, and theoretical modeling within coherent and efficient workflows.

Recent advances in machine learning \citep{Ball2010DataMining,Fluke2020SurveyingML} and large language models \citep{2024arXiv240714962H, 2023arXiv230318223Z, 2023arXiv230706435N} have introduced new possibilities for supporting scientific research \citep{2026IJMPD..3540009S, 2024arXiv240410019W}. In astrophysics, machine learning methods are already widely used for source classification \citep[e.g.,][]{2023MNRAS.519.3000S, 2017ApJ...837L..28C, 2017ApJS..230...20A, 2022A&A...666A.122M, 2018A&C....25..103G}, anomaly detection \citep{2019MNRAS.484..834G, 2021A&C....3600481L, 2021AJ....162..206S, 2021ApJS..255...24V, 2024ApJ...974..172A}, time-series analysis \citep{2018PhRvD..97d4039G, 2018NatAs...2..151N, 2012cidu.conf...47V, 2018NatAs...2..151N, 2018AJ....156....7H}, parameter inference \citep{2019MNRAS.488.4440A, 2022NatPh..18..112G}, and surrogate modeling of computationally expensive simulations \citep{2022MNRAS.511.1771S, BvL23, 2024ApJ...963...71B, 2024ApJ...971...70S, 2025ApJ...990..222S, 2024A&A...683A.185T}. Large language models have also demonstrated useful capabilities for literature summarization, query interpretation, code generation, and interactive question answering \citep[e.g.,][]{2024arXiv240410019W,Iyer2024Pathfinder}. These developments suggest that AI-based systems can significantly improve the efficiency of complex scientific workflows.

However, general-purpose language models are not designed for rigorous scientific analysis. They typically lack direct integration with astrophysical databases, observational archives, and domain-specific modeling frameworks, and their generated outputs does not remain physically grounded, reproducible, and traceable to supporting evidence. These limitations restrict the direct application of generic AI systems to astrophysical research. Without retrieval grounding or domain-specific validation, large language models may generate unsupported or scientifically incorrect statements. Most existing systems also do not include execution of the structured workflows involving data retrieval, statistical analysis, theoretical modeling, and iterative interpretation. In addition, they generally do not have connection between generated statements and the underlying observational data, literature sources, or modeling assumptions used during the analysis. For scientific applications, AI systems must function not simply as conversational interfaces, but as data-aware, retrieval-grounded, and workflow-oriented environments capable of interacting directly with scientific infrastructure and computational tools.

In this context, multi-agent AI architectures \citep{2025arXiv250106322T} provide a natural framework for addressing these requirements. Rather than relying on a single model to perform all tasks, a multi-agent system splits a scientific request into specialized operations handled by dedicated agents operating within an automated coordination framework. In such systems, different agents for example can focus on literature retrieval, observational data access, theoretical modeling, statistical analysis, or hypothesis generation, while supervisory components coordinate task execution and information exchange across the workflow. This design improves modularity, supports task specialization, and facilitates the integration of heterogeneous tools, databases, and analysis pipelines. It also enables iterative reasoning in which intermediate outputs can be evaluated, refined, and propagated through subsequent stages of the analysis. For astrophysical research, where scientific interpretation often depends on multiple interconnected operations, these properties are particularly important.

Several existing systems already provide important components to create the scientific workflow, including astronomical databases \citep{2000A&AS..143....9W,2000A&AS..143...23O}, virtual observatory tools \citep{2000A&AS..143...33B}, retrieval-augmented generation systems \citep{2020arXiv200511401L}, and specialized modeling frameworks \citep{2023A&A...678A.157D, 2015arXiv150708343V, 2024AJ....168..289S}. However, these tools are typically distributed across separate infrastructures that require substantial manual coordination by the researchers. Some systems focus primarily on literature retrieval (e.g., NASA ADS \footnote{\url{https://ui.adsabs.harvard.edu}}), while others specialize in observational archive access or numerical modeling. There are no platforms that integrates literature reasoning, science-ready data retrieval, theoretical modeling, and research ideation within a unified astrophysical environment. As a result, researchers often still need to bridge disconnected tools and workflows throughout the course of scientific investigations.

In this paper, we introduce \astrogen, a domain-specific multi-agent AI framework designed for astrophysical research. The framework integrates literature retrieval, observational data access, theoretical modeling, and research ideation within a unified environment tailored for multiwavelength and multimessenger astrophysics. \astrogen\ is organized around Domain-Specific Research Modules (DSRMs), each optimized for a particular class of astrophysical sources. A DSRM encapsulates the domain knowledge, data archives, literature collections, modeling tools, and specialized agent workflows required to investigate that source class, so allowing the broader framework to be adapted to different scientific domains through modular components. Rather than functioning as a general-purpose conversational system, the framework supports scientifically grounded analysis through the coordinated execution of specialized agents and the integration of retrieval-augmented generation (RAG) pipelines, vector databases, and science-ready astrophysical archives. In a RAG pipeline, information relevant to a user query is first retrieved from external knowledge sources and then supplied to the language model as contextual evidence for generating its response. This approach reduces dependence on the model internal knowledge alone and helps to produce answer based on the scientific papers. Together, these components enable the retrieval, synthesis, and analysis of relevant literature and observational data within traceable multi-step research workflows. The current implementation focuses on blazar research and supports literature analysis, multiwavelength data retrieval, broadband spectral modeling, and hypothesis generation, while the overall architecture is designed to be extensible to additional source classes and scientific applications.

The paper is organized as follows. Section 2 presents the overall  \astrogen\ architecture and system design. Section 3 describes the structure and functionality of the DSRMs and their associated agents. Section 4 introduces the benchmarking and evaluation framework for the literature retrieval component. Section 5 presents the front-end interface and interaction environment supporting persistent and traceable research sessions. Section 6 demonstrates representative end-to-end scientific workflows using astrophysical use cases. Finally, Section 7 summarizes the main results and discusses future developments of the platform.

\section{AstroGenesis: System Overview and Architecture}

\begin{figure*}
    \centering
    \includegraphics[width=0.98\linewidth]{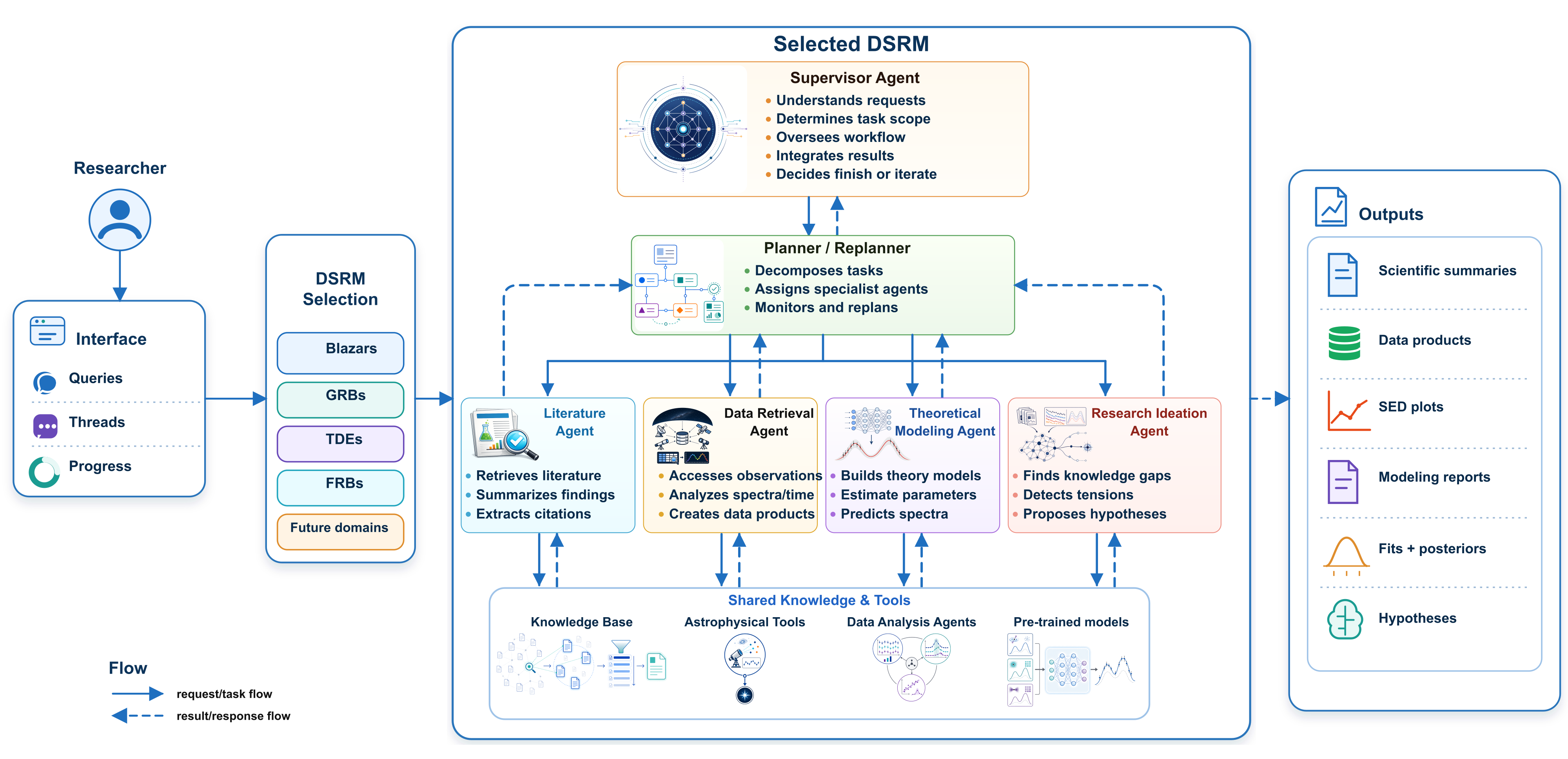}
    \caption{Architecture and workflow of \astrogen. Researchers interact with the system through a chatbot interface and select an appropriate DSRM corresponding to the astrophysical source class under investigation. Within the selected DSRM, a Supervisor Agent coordinates task execution through a Planner/Replanner and a set of specialized agents responsible for literature analysis, observational data retrieval, theoretical modeling, and research ideation. These agents operate using external tools which include, astrophysical tools, data-analysis modules, and pre-trained models to perform modeling of the data, etc.
}
    \label{fig:arch}
\end{figure*}

\astrogen\ is a domain-specific multi-agent AI framework designed to support astrophysical research by integrating data access, analysis, and modeling capabilities within a unified and modular platform. Developed to address the increasing complexity of multiwavelength and multimessenger observations, as well as the rapid growth of the scientific literature, \astrogen\ allows researchers to interact efficiently with large volumes of heterogeneous data and sophisticated theoretical models through natural-language interfaces.

The platform provides a broad range of functionalities necessary for modern astrophysical research. It enables users to retrieve and summarize information from scientific publications, ensuring that responses are grounded in the existing literature, access and analyze observational data across multiple wavebands and messenger types, and use advanced modeling frameworks to investigate astrophysical phenomena. In addition, \astrogen\ supports the exploration of new research directions by combining information from observational data, theoretical models, and scientific publications.

The architecture of \astrogen\ is shown in Figure \ref{fig:arch}. The system is designed to be both flexible and scalable, consisting of a user-facing chatbot interface and a collection of Domain-Specific Research Modules (DSRMs) tailored to different classes of astrophysical sources, such as blazars, gamma-ray bursts (GRBs), tidal disruption events (TDEs), etc. Through the interface, researchers can submit scientific queries, monitor the progress of ongoing tasks, and access previous interactions. Based on the selected scientific domain, the corresponding DSRM is activated and provides access to specialized data resources, modeling frameworks, and knowledge bases relevant to the requested study. Within the selected DSRM, a collection of specialized agents collaboratively retrieve information, analyze data, perform theoretical modeling, and generate scientific outputs, which are subsequently integrated and returned to the researcher through the chatbot interface. The modular organization of \astrogen\ system is intended primarily to support functional extensibility. Each DSRM encapsulates domain-specific resources and workflows, allowing new source classes or scientific capabilities to be added without requiring substantial modification of the existing modules. Similarly, individual agents, data services, and modeling tools can be updated or replaced independently when their interfaces remain compatible with the orchestration layer. This design can facilitate the gradual expansion of the platform; however, computational scalability with respect to the number of users, agents, data products, and concurrent tasks depends on the underlying deployment infrastructure and is not evaluated in the present work.

\subsection{ChatBot Interface}
The ChatBot Interface is the primary user-facing component of \astrogen\ and provides the main entry point through which researchers interact with the system. It supports natural-language requests for literature analysis, observational data retrieval, theoretical modeling, and research ideation. The interface therefore serves as a conversational workspace in which scientific workflows can be initiated, monitored, and refined.

The interface combines persistent request handling with real-time communication. User requests and agent-generated responses are stored through a Representational State Transfer (REST) application programming interface, while progress updates and intermediate results are delivered through WebSocket connections. This hybrid design is particularly useful for scientific tasks that may require extended data processing or model evaluation, as it allows users to follow the execution of a workflow without repeatedly querying the system.

When a request is submitted, it is authenticated and forwarded to the orchestration layer, which assigns the corresponding tasks to the appropriate agents. User messages, intermediate outputs, and final results are stored in a PostgreSQL database, allowing conversations and scientific workflows to be reconstructed after interruptions or reconnections. During task execution, status information and results are exchanged between backend services through Redis \footnote{\url{https://github.com/redis/redis}}, an in-memory data store used for low-latency message distribution, and are streamed to the interface in real time.

This architecture enables the interface to present the sequence of operations performed by the agents, including data retrieval, analysis, and modeling steps. It also preserves the associated inputs and outputs, thereby supporting reproducibility and traceability at the level of the user interaction. The ChatBot Interface consequently functions not only as a communication layer, but also as the workspace through which the scientific products generated by the specialized agents are inspected and integrated.

\section{Domain-Specific Research Modules: Architecture and Structure}
Domain-Specific Research Modules (DSRMs) are the central components of \astrogen, within which the main research tools and specialized agents are organized. Each DSRM is tailored to a specific class of astrophysical sources, such as blazars, GRBs, TDEs, or other well-defined categories. These modules are designed as independent multi-agent systems rather than as a single merged framework. This separation allows each module to operate autonomously and reduces cross-domain interference, allowing the agents to rely on source-specific assumptions, modeling approaches, terminology, and observational constraints. A major motivation for this modular design is that information retrieval that can be improved when source-specific literature corpora, data archives, and modeling pipelines are treated independently. For example, GRB studies involve terminology, physical timescales, and observational strategies that differ substantially from those used in blazar research.Using separate knowledge bases, vector spaces, and analysis pipelines for each DSRM helps keep retrieval focused, semantically relevant, and scientifically coherent. This structure also simplifies the extension of \astrogen\ to additional astrophysical domains, since new DSRMs can be developed and integrated without modifying the logic or performance of existing modules.

It is important to distinguish the functional description of these modules from the computational mechanisms underlying their operation. DSRMs are software orchestration structures rather than autonomous scientific entities: they contain specialized agents, data services, and modeling tools according to predefined workflows. Some components, particularly the Literature Agent, represent text as high-dimensional numerical vectors, referred to as embeddings, and retrieve relevant information by comparing the semantic similarity of these representations. Such operations are statistical and depend on the embedding model, indexed corpus, similarity metric, and ranking procedure; they do not imply that the system possesses an intrinsic physical understanding of the retrieved material. Other components of a DSRM operate through database queries, rule-based validation, conventional statistical methods, pretrained surrogate models, or Bayesian parameter inference. The outputs of the framework should therefore be interpreted as computationally generated results conditioned on the available data, models, retrieval procedures, and workflow configuration, and remain subject to scientific verification by the researcher.

A typical architecture of a DSRM is illustrated in Figure \ref{fig:arch} middle panel. The framework consists of several coordinated agents responsible for different stages of the scientific workflow:

\begin{itemize}
\item \textbf{Supervisor Agent:} coordinates the overall workflow and assigns tasks to the appropriate specialized agents according to the user query.

\item \textbf{Planner and Replanner:} decompose the request into structured steps, manage task execution, and verify that the required stages are completed before the results are returned to the Supervisor Agent.

\item \textbf{Literature Agent:} implements a retrieval-augmented generation pipeline based on vector-space embeddings and reranking methods to identify relevant scientific articles.

\item \textbf{Data Retrieval Agent:} accesses and curates observational datasets from public archives and science-ready databases.

\item \textbf{Theoretical Modeling Agent:} uses pretrained neural networks together with physically motivated modeling frameworks to perform parameter inference and generate model predictions consistent with the available data.

\item \textbf{Research Ideation Agent:} analyzes the outputs produced by the other agents to identify trends, knowledge gaps, and possible directions for further investigation.

\end{itemize}

In this structure, when a user submits a query through the ChatBot Interface and the Subject Detector identifies the relevant topic, the Supervisor Agent then evaluates the query and coordinates the execution of the required tasks by the specialized agents and then the request is passed to the appropriate DSRM. Depending on the nature of the request, the Literature Agent may retrieve and summarize relevant publications from the vector database, the Data Retrieval Agent may collect and preprocess observational data, the Theoretical Modeling Agent may perform parameter inference or generate model predictions, and the Research Ideation Agent may analyze existing trends and identify possible knowledge gaps. The Supervisor Agent when receiving the outputs from the relevant agents formulates into a unified response, which is returned to the user through the ChatBot Interface.

The current implementation of \astrogen, referred to as version 1.0, includes a fully developed DSRM dedicated to blazar research. The following discussion therefore focuses on the architecture, components, and agent structure of this module. Additional DSRMs for other source classes are in preparation and follow the same general design principles. The core structure remains consistent across modules, while domain-specific differences are mainly implemented through the configuration and functionality of individual agents, including data sources, modeling pipelines, and inference strategies. These adaptations reflect the distinct observational and theoretical requirements of different astrophysical phenomena without changing the overall system architecture. Additional DSRMs will be released sequentially as they are finalized, with the goal of extending \astrogen\ into a modular research environment for multiple classes of astrophysical sources.

\subsection{Supervisor Agent}
The Supervisor Agent is the central orchestration component of \astrogen, responsible for managing the flow of information between the user-facing ChatBot Interface and the specialized agents. It interprets user queries, determines which agent or combination of agents should be activated, and integrates their outputs into a unified response. This design allows the system to return coherent and contextually relevant results even when user requests involve multiple stages of analysis.

The Supervisor Agent is implemented using the \texttt{gpt-5.6-luna} \footnote{\url{https://developers.openai.com/api/docs/models/gpt-5.6-luna}} language model, selected for its low-latency performance and ability to handle multi-step workflows. The model operates with a fixed temperature of 0.7, a sampling parameter that controls variability in token selection. Lower values produce more deterministic outputs, whereas higher values increase response diversity; the adopted value permits moderate flexibility when processing ambiguous or incomplete user queries. After receiving a query from the ChatBot Interface, the Supervisor Agent determines whether the request can be answered directly or should be forwarded to one or more specialized agents. Simple interactions, such as clarifications, or short follow-up questions, are handled directly, while more complex tasks are delegated to the appropriate agents. Routing decisions are implemented using a structured-output mechanism based on a Pydantic validation model, \texttt{AgentRouting}. Each routing decision includes the selected target agent, the reasoning behind the decision, and, when needed, a final user-facing response. The workflow can also terminate through the \texttt{END} state once the query has been fully resolved. This routing approach improves transparency in multi-step workflows and helps support debugging and reproducibility.

After the specialized agents return their outputs, the Supervisor Agent combines the results into a single response for the user. It verifies the completeness and consistency of the returned information and preserves relevant references and metadata provided by the sub-agents, including literature citations, dataset identifiers, and other provenance information. The Supervisor Agent also includes error-handling procedures. If a sub-agent fails to return an output, produces incomplete results, or requires additional information, the Supervisor Agent can retry the request with modified parameters, redirect the task to another suitable agent when available, or request clarification from the user through the ChatBot Interface. Once a task is completed, the Supervisor Agent calls the \texttt{END} node and generates the final user-facing response. If a sub-agent returns control without a conclusive response, the Supervisor Agent forces the interaction to the \texttt{END} state with a final message, preventing incomplete or internal agent outputs from being shown to the user.

The Supervisor Agent maintains conversation context across multi-turn interactions, allowing follow-up queries to be interpreted based on previous exchanges. Specialized agents do not communicate directly with each other or with the user; instead, all interactions are coordinated through the Supervisor Agent. This architecture ensures that the final user-facing response integrates the relevant information from each activated component while preserving a modular system design. It also simplifies the integration of new agents, which can be added to the routing framework without disrupting existing workflows.

Finally, the Supervisor Agent includes logging mechanisms for debugging, traceability, and system monitoring. Routing decisions, error events, and sub-agent responses are recorded, providing an execution history that can be inspected during development and evaluation. Such traceability is particularly important for scientific applications, where the reliability, provenance, and accountability of computational workflows must be maintained and assessed.

\subsubsection{Planner and Replanner}

In multi-agent systems which is designed to handle complex scientific tasks, a planning and revision layer is required to translate user queries into coherent sequences of computational actions \citep[e.g., see][]{2025arXiv250309572E,2025arXiv251005592L}. Such architectures enable the system to decompose scientific objectives into structured tasks, coordinate specialized agents, and dynamically adapt the workflow as intermediate results are produced. This capability is particularly important in astrophysical research, where analyses often involve multi-stage procedures including data retrieval, model execution, statistical analysis, and iterative interpretation.

The \textit{Planner} is responsible for the initial decomposition of a user request into executable steps that can be delegated to specialized agents. It analyzes the scientific objective expressed in the query and converts it into an ordered sequence of actions assigned to the appropriate computational nodes (Task Queue). The Planner therefore acts as a reasoning layer that maps high-level scientific questions into executable tasks while preserving the logical ordering of the workflow. The \textit{Replanner} manages adaptive workflow control during execution. After individual agents complete their assigned tasks, it evaluates the returned outputs, updates the execution plan, and determines whether the subsequent steps should proceed as originally designed or whether modifications are required. It also verifies whether all requests issued by the Supervisor Agent have been successfully completed. This mechanism enables the system to assess intermediate results, detect execution failures, and identify missing inputs, which frequently arise in complex scientific queries. By updating task states, modifying subsequent actions, and returning unresolved issues to the Supervisor Agent when clarification is needed, the Replanner maintains the consistency and robustness of the multi-agent workflow. When the tasks are finished, it is returned to the Supervisor Agent, which coordinates task execution and communicates the final results to the user.

\subsection{Literature Agent}
The Literature Agent is a central component of the \astrogen\ architecture, responsible for identifying, ranking, and organizing scientific publications relevant to user queries. Its primary role is to retrieve and summarize information from the scientific literature stored in the vector database. The agent operates within a RAG framework \citep{2023arXiv231210997G}, which combines the capabilities of large language models with external knowledge retrieval from a domain-specific scientific corpus. Within this framework, scientific publications are converted into vector embeddings and indexed in a vector database that supports semantic similarity search. When a user submits a query, the system retrieves the most relevant document fragments through embedding-based retrieval and reranking procedures. The retrieved contexts are then provided to the language model, which generates responses conditioned on the retrieved scientific content rather than relying solely on its internal parameters. This approach improves response quality while reducing hallucinations \citep{2024arXiv240408189B} and enhancing both traceability and domain specificity. It also enables the system to incorporate newly published literature without requiring retraining of the underlying language model.

The Literature Agent is designed specifically for scientific applications, where accurate contextual retrieval and source grounding are essential. By combining semantic retrieval with metadata-aware filtering and reranking, the system can identify publications that are both topically relevant and scientifically informative. In the following subsections, we describe the main components and operational workflow of this agent.

\subsubsection{Document Processing and Vector Database Construction}
To generate domain-specific responses grounded in astrophysical literature, the system incorporates an external scientific knowledge base constructed from refereed publications. The relevant literature is collected through the SAO/NASA Astrophysics Data System (ADS) API\footnote{\url{https://ui.adsabs.harvard.edu/help/api/}}. The literature-collection procedure consists of several sequential steps. First, ADS is queried using a set of domain-specific keywords related to blazar research, including, for example, 'galaxies: active', 'galaxies: jets', 'gamma rays: galaxies', 'BL Lacertae objects: general'. The resulting records are then filtered to retain only refereed publications appearing in astrophysics and physics journals, including The Astrophysical Journal (ApJ), Monthly Notices of the Royal Astronomical Society (MNRAS), Astronomy \& Astrophysics (A\&A), The Astrophysical Journal Letters (ApJL), The Astronomical Journal (AJ), The Astrophysical Journal Supplement Series (ApJS), Nature, Physical Review Letters (PRL), and Physical Review D (PRD). These filtering steps define the final list of publications included in the corpus. For each selected publication, the full text is then retrieved from arXiv\footnote{\url{https://arxiv.org/}} when an arXiv version is available; otherwise, the corresponding published journal version is downloaded. Downloaded files are validated through checking the content to ensure that only valid PDF documents are retained, while corrupted or empty files are removed. As of 24 August 2026, this procedure resulted in a corpus of approximately 21,241 publications relevant to blazar studies. This retrieval process is managed by a dedicated class, \texttt{PaperInfoDownloader}, which provides a structured framework for querying, downloading, and caching publication metadata and full-text documents. For each retrieved paper, the system stores all the metadata including abstracts, bibcodes, DOIs, citation counts from ADS, and reference information. Metadata are cached locally to avoid redundant API requests and to ensure reproducibility of the dataset construction process. Logging is implemented in the pipeline to track download status and record retrieval errors, improving traceability and robustness during corpus assembly.

The downloaded PDF files are subsequently converted into structured text using the OCR framework provided by DeepSeek OCR \citep{wei2025deepseekocr}\footnote{\url{https://www.deepseek-ocr.ai}}. This step is necessary because the PDF format is not directly suitable for semantic indexing and retrieval. DeepSeek OCR employs a vision--language architecture in which document images are first processed by a visual encoder that identifies page structure, including text regions, multi-column layouts, tables, equations, figures, and references. The extracted visual representations are then processed by a language generation module that reconstructs the document content while preserving the order and structure. The tool also can extract the text from the figure and table captions. In practice, the conversion quality was found to be sufficiently reliable for scientific literature processing. Manual inspection of randomly selected documents showed that the sections are well selected, mathematical expressions, tables, figure captions, and complex layouts were generally reconstructed with adequate fidelity for semantic retrieval tasks.

Following text extraction, the documents undergo a cleaning stage designed to standardize the content prior to embedding. This process combines rule-based filtering with LLM-assisted processing to remove non-scientific sections, such as acknowledgments and reference lists, while also correcting formatting inconsistencies introduced during parsing. Scientifically meaningful headings, including 'Introduction', 'Results', and 'Data Collection', are retained to guide the subsequent chunking operations. However, since the 'Introduction' sections of scientific papers often contain broad contextual overviews rather than source-specific technical details, they are excluded from the embedding stage to improve the relevance of retrieved contexts. This multi-stage procedure produces semantically coherent documents optimized for embedding and retrieval.

After preprocessing, the cleaned text files are transformed into vector embeddings to enable semantic search. Each document is divided into overlapping text chunks using a parameterized text splitter. In the current implementation, chunks contain up to 5000 characters with a 200-character overlap to preserve contextual continuity between adjacent segments. A bibcode-based deduplication mechanism is implemented to prevent reprocessing of previously indexed content. Embeddings are generated for each chunk using the OpenAI \texttt{text-embedding-3-small} model \footnote{\url{https://developers.openai.com/api/docs/models/text-embedding-3-small}}, producing 1536-dimensional dense vector representations. Each vector inherits the metadata of its parent document, including bibcode, title, and authorship information, ensuring traceability during retrieval. To improve computational efficiency, embeddings are generated in batches, and only newly added or updated documents are processed during incremental updates.

The resulting embeddings and associated metadata are stored in a vector database. The current implementation uses Weaviate \footnote{\url{https://github.com/weaviate/weaviate}} as the backend due to its support for high-dimensional vector indexing, efficient similarity search, and metadata-aware querying. Each chunk is indexed together with its parent document metadata, enabling both pure semantic retrieval and hybrid searches combining vector similarity with structured metadata constraints. This infrastructure forms the retrieval layer of the RAG framework, allowing the system to identify contextually relevant scientific information that is subsequently provided to the language model during response generation.

In addition to the full-text vector database, a supplementary database is constructed using only the abstracts of the retrieved publications. The same preprocessing, embedding, and metadata indexing pipeline is applied to this abstract-only corpus. Each abstract is treated as an individual document and embedded into the same vector space used for full-text content. The resulting vectors are stored together with metadata such as bibcode, title, authorship, and publication venue. This additional database enables efficient high-level retrieval and facilitates matching between abstract-level and full-text scientific contexts.

\subsubsection{Information Retrieval} \label{info_ret}
When receiving a natural-language query from the Supervisor Agent, the Literature Agent initiates a multi-stage retrieval pipeline designed to maximize contextual relevance and retrieval accuracy. The query is first reformulated by the Literature Agent into a retrieval-oriented form and passed to the search engine. During development of the retrieval pipeline, different query formulations were investigated, including direct use of the reformulated query and query expansion using Hypothetical Document Embeddings (HyDE) \citep{gao2023hyde}. In the HyDE configuration, the reformulated query is expanded into a hypothetical scientific abstract intended to provide a richer semantic representation of the requested information.

HyDE was initially evaluated using an abstract-level retrieval corpus, where the generated hypothetical documents and the indexed abstracts are structurally similar. However, benchmarking showed that retrieval over full-text passages extracted from the bodies of scientific papers provided better performance than abstract-level retrieval, which motivated us the adoption of a chunk-based index. Because the current system performs first-stage retrieval over these full-text chunks, the abstract-like representation generated by HyDE may introduce a mismatch between the expanded query and the indexed textual units. HyDE was therefore also evaluated within the chunk-based retrieval configuration. In this setting, direct use of the reformulated query without HyDE expansion achieved better overall retrieval performance and was consequently adopted in the current implementation. The impact of HyDE-based query expansion and its comparison with the alternative retrieval configurations are examined quantitatively through dedicated retrieval benchmarks in Section~\ref{benchmar}.

The current retrieval process follows a hierarchical strategy that combines passage-level retrieval with paper-level candidate construction. In the first stage, a hybrid dense--sparse search \citep{2022arXiv221011934B} is performed over the chunked full-text collection. The dense component captures semantic similarity between the query and the scientific text, while the sparse component uses BM25 \citep{10.1561/1500000019} to capture lexical similarity and query-specific terminology. The relative contributions of the dense and sparse components were evaluated during development. The adopted configuration uses a hybrid weight of $\alpha=0.8$, where $\alpha$ controls their relative contribution, with $\alpha=1$ corresponding to purely dense retrieval and $\alpha=0$ to purely sparse retrieval.
For each query, the hybrid search retrieves the 1,200 highest-ranked full-text chunks. This relatively large initial set is used to maintain broad candidate coverage before aggregation at the paper level. Since several retrieved chunks may originate from the same publication, the chunk-level results are then grouped by their ADS bibcodes. Each publication is represented by the maximum retrieval score among its matched chunks. This step converts the passage-level search results into a ranked set of paper-level candidates for the subsequent stages of the retrieval pipeline.

The highest-ranked papers from this initial search are then used as pivot documents for citation-graph expansion. In the current implementation, the top 25 candidate papers are selected as pivots. For each pivot paper, the system collects both the references cited by that publication and subsequent publications that cite it. This bidirectional citation expansion broadens the retrieval space beyond documents that exhibit direct semantic or lexical similarity to the query and enables the inclusion of scientifically connected works that may use different terminology or address related aspects of the same problem. The citation-expanded candidates are deduplicated using bibliographic identifiers and merged with the papers obtained from the initial hybrid search. Candidates identified directly through the query retain their retrieval scores, while papers introduced only through citation relationships provide an additional graph-based component of the candidate pool.

Following candidate construction, the system applies a cross-encoder to obtain an additional estimate of query--paper relevance. Although the initial retrieval is performed over full-text chunks, each candidate paper is represented by its abstract at this stage. The chunk-level search therefore determines which papers enter the candidate pool, while the abstracts are used for relevance scoring. This provides a paper-level representation for comparing each candidate with the query. The query and each candidate abstract are jointly processed using the \texttt{cross-encoder/ms-marco-MiniLM-L6-v2} model \citep{2023arXiv230307678W}. Several cross-encoder models, as well as a configuration without cross-encoder scoring, were evaluated during development. The two MiniLM variants showed comparable performance and outperformed the tested BGE models. The selected MiniLM model provides an additional relevance score that complements the dense and BM25 scores obtained during the initial retrieval stage. These three signals are subsequently combined to determine the final ranking of the candidate papers.

Rather than using the cross-encoder score as an independent ranking criterion, the current retrieval architecture combines the relevance signals obtained at different stages of the search. For each candidate paper $d$, the dense similarity score, BM25 score, and cross-encoder score are standardized independently for the current query. For a relevance signal $s$, the standardized score is defined as
\begin{equation}
z_s(d)=
\begin{cases}
\dfrac{s(d)-\mu_s}{\sigma_s}, & d\in O_s, \\[2ex]
0, & d\notin O_s,
\end{cases}
\end{equation}
where $O_s$ is the set of candidates for which signal $s$ is available. The quantities $\mu_s$ and $\sigma_s$ are the mean and standard deviation of that signal, calculated over the candidates in $O_s$. When a signal is unavailable for a candidate, assigning $z_s(d)=0$ places it at the mean of the corresponding standardized distribution rather than treating the missing score as evidence of low relevance.

The standardized scores are then combined into a single relevance score,
\begin{equation}
F(d)=
w_{\rm vec}z_{\rm vec}(d)
+w_{\rm BM25}z_{\rm BM25}(d)
+w_{\rm CE}z_{\rm CE}(d),
\end{equation}
with the adopted weights
\begin{equation}
(w_{\rm vec},w_{\rm BM25},w_{\rm CE})=(1,1,2).
\end{equation}
This weighting gives the cross-encoder relevance score twice the contribution of either the dense or BM25 score, while retaining information from all three retrieval signals. Alternative score-fusion strategies were also evaluated, including reciprocal-rank fusion \citep{10.1145/1571941.1572114} and different relative weights of the three signals. Based on these comparisons, the $z$-normalized weighted fusion was adopted in the current implementation.

To introduce a moderate preference for recent publications, the relevance score can additionally be modulated by a temporal weighting term. In the deployed configuration, the final score is expressed as
\begin{equation}
S(d)=F(d)\left[1+\epsilon\,f_{\rm time}(d)\right]. \label{eq:time_dependent_score}
\end{equation}
where $\epsilon=0.1$ limits the contribution of the temporal weighting and
\begin{equation}
f_{\rm time}(d)=
\frac{1}{
1+\exp\left(
\frac{a_d-\tau}{\tau}
\right)
}.
\end{equation}
where $a_d$ is the age of the publication in years and $\tau=5$ years. This contribution is kept small so that relevance remains the dominant ranking criterion. 

After final ranking, the top candidate papers are selected for full-text resolution, with five papers returned in the current production configuration. The bibliographic identifiers of these papers are used to recover their associated full-text chunks, which provide the scientific context used during LLM-based response synthesis. Thus, although abstracts are used during cross-encoder scoring, the final evidence supplied to the Literature Agent is the full scientific content of the selected publications. The retriever output additionally preserves the corresponding bibliographic metadata, enabling the Literature Agent to provide also the citation to the source from which the answer was generated. The resulting evidence-grounded response is returned to the Supervisor Agent, which integrates it into the broader multi-agent workflow before generating the final response to the user.

Overall, the retrieval architecture combines complementary levels of scientific information: full-text passages are used for high-recall candidate discovery, citation relationships broaden the candidate space beyond direct query matches, abstracts provide compact paper-level representations for context-sensitive reranking, and multi-signal fusion integrates semantic, lexical, and cross-encoder relevance into the final ranking. The adopted configuration was selected after evaluating alternative candidate sources, query-expansion strategies, dense--sparse combinations, cross-encoder models, and score-fusion schemes. This hierarchical approach enables the Literature Agent to preserve broad literature coverage while providing detailed full-text evidence for downstream scientific reasoning and citation-grounded response generation.
\subsection{Data Retrieval Agent}
\label{sec:data_retrieval_agent}

The Data Retrieval Agent is responsible for obtaining science-ready observational data and preparing datasets required for SED construction, visualization, and subsequent analysis within the \astrogen\ framework. The agent resolves source coordinates, retrieves multiwavelength observations from public astronomical databases, and produces science-ready SEDs. It communicates exclusively through the Supervisor Agent and follows predefined protocols for source identification, date handling, and tool invocation. The agent operates through a specialized system prompt tailored for astrophysical data analysis and combines custom Python-based tools with dedicated analysis environments to process heterogeneous observational datasets.

When the Supervisor Agent requests observational data, the Data Retrieval Agent initiates an automated workflow for data retrieval and analysis. The workflow requires the source name and the corresponding temporal interval, which are extracted from the user query by the Supervisor Agent. If no temporal interval is provided, the system returns all data available in the archive. Otherwise, the retrieval is restricted to the specified time interval. Date parsing supports multiple formats, including \texttt{YYYY-MM-DD}, \texttt{DD.MM.YYYY}, \texttt{YYYY.MM.DD}, and \texttt{DD/MM/YYYY}, with automatic conversion to MJD for internal consistency. When ambiguities or invalid date formats are detected, they are logged, and clarification requests are returned to the user when necessary. In order to retrieve the coordinates from the source name, the agent implements a hierarchical fallback strategy. It first queries the Markarian Multiwavelength Data Center (\texttt{MMDC})\footnote{\url{https://mmdc.am}} \citep{2024AJ....168..289S} autocomplete API to retrieve the source right ascension (RA) and declination (Dec). If \texttt{MMDC} does not return valid coordinates, the system falls back to the SIMBAD database through the \texttt{astroquery} interface \citep{2019AJ....157...98G}. If SIMBAD also fails to resolve the source, the agent attempts coordinate retrieval using Astropy \texttt{SkyCoord.from\_name()} \citep{2013A&A...558A..33A,2018AJ....156..123A,2022ApJ...935..167A}. As a final fallback, the coordinates are requested from the NED (NASA/IPAC Extragalactic Database) service. This multi-stage procedure improves the robustness of source identification, particularly in cases involving naming inconsistencies or alternative source designations.

After identification of the source coordinates, the agent queries the \texttt{MMDC} \citep{2024AJ....168..289S}  through a dedicated API endpoint using the resolved coordinates and the requested temporal interval. \texttt{MMDC} serves as the primary observational backend in the current implementation. It contains data from more than 80 instruments and catalogs across the electromagnetic spectrum, with particular emphasis on blazar observations. The available datasets include archival and newly analyzed data from instruments such as Swift-UVOT, Swift-XRT, NuSTAR, and Fermi-LAT. This infrastructure enables the generation of time-resolved SEDs corresponding to different source activity states. As the response to the API request, the agent receives  datasets, spanning multiple instruments and energy bands. Retrieved datasets are indexed and stored through a Redis-backed file management system, enabling efficient access and session consistency across different components of the platform. Before the retrieved data are made available to downstream agents, they pass through a validation and normalization layer. The returned data is checked against a predefined schema that specifies the required fields, accepted data types, units, instrument and filter identifiers, temporal metadata, and uncertainty information. Missing mandatory fields, malformed numerical values, non-finite quantities, inconsistent units, and unrecognized instrument labels are flagged before analysis. The same retrieved dataset is then forwarded to the Streamlit frontend, where it is reconstructed into a tabular representation and passed to the SED plotting routine. The plot shown to the user is therefore rendered by the frontend from the retrieved \texttt{MMDC} data, while the backend remains responsible for data access, indexing, and persistence.

The development of \texttt{MMDC} plays an important role within the \astrogen\ framework by providing an infrastructure for hosting, organizing, and sharing multiwavelength observational data. Such an infrastructure is particularly valuable as it can be used also for the implementation of the other DSRMs, as it enables reusable and standardized access to structured observational datasets across multiple astrophysical source classes. Although the current implementation interfaces primarily with \texttt{MMDC}, the architecture of the Data Retrieval Agent is modular and extensible. The system can be adapted to interact with other observational repositories supporting API-based or Virtual Observatory (VO) interfaces. Planned extensions of this agent include the integration of the access  and reduction of raw observational data, enabling future implementations to support access and analysis of telescope-level data and its theoretical modeling.

Beyond data retrieval, this agent can also perform additional analyses, through sub-agents, required for more in-depth studies of astrophysical sources. This analysis includes to identify flares, or identify trends in the flux photon index connection, or check correlated or anti correlated flux changes in different bands. When such analysis is requested, the system first obtains the relevant light-curve data from \texttt{MMDC} via API request using the resolved source coordinates and any requested temporal or catalog constraints. It should be noted that these light curves are not publicly available through \texttt{MMDC} and can be retrived only via \astrogen. The retrieved observational dataset is stored as an internally accessible CSV file, which can then be reused by specialized downstream analysis sub-agents within the multi-agent planning workflow. This design separates data acquisition from scientific interpretation: the Data Retrieval Agent is responsible for requesting and persisting the light-curve data, while dedicated sub-agents investigate the temporal and spectral behavior. These downstream analyses are particularly useful for multi-band light-curve datasets, where identifying dominant variability bands, detecting flaring episodes, and quantifying spectral evolution provide important constraints for subsequent theoretical modeling. The current implementation includes the following analysis sub-agents:

\begin{itemize}

\item \textbf{Fractional Variability Sub-agent:} The Fractional Variability Sub-agent operates on light-curve datasets prepared by the Data Retrieval Agent and is designed to quantify the intrinsic variability amplitude across different observational bands. Before performing the analysis, the agent accesses the tabular dataset through the shared file-management layer and validates the required schema, including flux measurements, flux uncertainties, catalog labels, and filter-band metadata. The observational bands are then standardized by mapping catalog-specific entries into broader categories such as $\gamma$-ray, X-ray, and optical sub-bands. The fractional variability is subsequently computed independently for each band, if requested by the user, or for all bands.

For each observational band, the intrinsic variability amplitude is estimated using a noise-corrected fractional variability estimator following \citep{2019Galax...7...62S}. The unbiased flux variance $S^{2}$ is corrected by subtracting the mean squared observational uncertainty $\langle\sigma_{\mathrm{err}}^{2}\rangle$, yielding
\begin{equation}
F_{\mathrm{var}} = \frac{\sqrt{S^{2} - \langle\sigma_{\mathrm{err}}^{2}\rangle}}{\langle F\rangle}.
\end{equation}
In the current implementation, the calculation is performed only for bands satisfying a minimum sampling requirement. Cases with non-positive mean flux or non-positive excess variance are treated as non-detections of intrinsic variability. The uncertainty on $F_{\mathrm{var}}$ is estimated through bootstrap resampling of the flux--uncertainty pairs.

\item \textbf{Flaring Event Sub-agent:} The Flaring Event Analyzer performs an analysis of multi-wavelength light curves to identify periods of flux enhancement in different bands. Unlike the Fractional Variability Sub-agent, which characterizes the overall variability amplitude within each band, this agent focuses on localized transient enhancements (flares) and their temporal relationships across energy bands. For each observational band, the agent segments the light curve using the Bayesian Blocks algorithm \citep[e.g.,][]{1998ApJ...504..405S}, producing a piecewise-constant representation of the observed variability. A quiescent flux level is estimated from the distribution of block fluxes, and an activity threshold is defined using the median block flux and a robust scatter estimator. The latter is computed as $1.4826$ times the median absolute deviation (MAD), providing a standard-deviation-like measure that is less sensitive to high-flux outliers. Candidate flares are identified as groups of high-flux blocks containing a well-defined local maximum. In the current implementation, events are retained only when their mean excess above the quiescent level is statistically significant. For each accepted flare, the agent estimates the flare start time, end time, peak time, peak flux, number of contributing blocks, and mean flare flux. The peak time is defined as the center of the brightest block, while the associated timing uncertainty is estimated from the width of that block. After constructing all the information about the flares for the available bands, the agent searches for cross-band flare correlations. Flares are matched using a nearest-peak criterion constrained by both an absolute temporal window and a tolerance proportional to the mean duration of the two candidate flares. For each matched pair, the observed peak-time difference is converted into a source-rest-frame lag,
\begin{equation}
\tau_{\mathrm{rest}} = \frac{t_{\mathrm{peak},2} - t_{\mathrm{peak},1}}{1+z},
\end{equation}
with the corresponding uncertainty propagated from the peak-time uncertainties of the two flares. Positive values of $\tau_{\mathrm{rest}}$ indicate that the second band peaks later than the first. The resulting lag table therefore encodes both the temporal ordering of emission across bands and the statistical reliability of each matched flare pair.

\item \textbf{Spectral Trend Sub-agent:} The Spectral Trend Sub-agent investigates the evolution of the spectral index with flux during localized flaring activities. The analysis is performed in both the X-ray and $\gamma$-ray bands, which have measured spectral index, from the internally stored light-curve dataset. After accessing these dataset, the agent validates the time, flux, spectral index, catalog labels, and filter information when available. The temporal intervals used in the analysis are defined by the Bayesian-Block algorithm. For each identified X-ray or $\gamma$-ray flare, the corresponding interval is expanded by a fixed temporal window to include the immediate pre-flare and post-flare evolution. The agent extracts all flux and spectral-index measurements within the expanded interval and fits the spectral index as a linear function of $\log_{10}$ flux. For each independent flare, it records the slope, intercept, number of Bayesian blocks, corresponding to the number of data point, Pearson correlation coefficient, Spearman rank correlation coefficient, and the associated statistical significance values. The sign and statistical significance of the fitted relation are used to classify the spectral behavior of each flare. A negative slope corresponds to a harder-when-brighter trend, in which the photon index decreases as the flux increases, while a positive slope indicates softer-when-brighter behavior. Flares with weak correlations, large p-values, or insufficient sampling are classified as flat or undetermined rather than assigned a physical interpretation.

\end{itemize}

The numerical results produced by the three sub-agents are subsequently passed to their interpretive components, which generate structured scientific summaries before returning the results to the Supervisor Agent. For the Fractional Variability Sub-agent, the interpretation includes the variability strength and statistical significance in each band, comparisons of variability amplitudes across energy bands, and identification of unreliable measurements caused by sparse sampling, large uncertainties, non-positive mean fluxes, or non-positive excess variance. For the Flaring Event Sub-agent, the interpretation summarizes flare and lag properties for each band pair, evaluates the significance and consistency of the measured lags, distinguishes between simultaneous, leading, lagging, and statistically inconclusive behavior, and checks whether individual flares are repeatedly used in multiple associations. For the Spectral Trend Sub-agent, the interpretation summarizes the spectral-index--flux relations in the X-ray and $\gamma$-ray bands, identifies statistically significant harder-when-brighter or softer-when-brighter trends, and highlights cases affected by weak correlations or limited sampling. In all three cases, the physical interpretation is kept conservative and is restricted to trends supported by the derived measurements. General connections with particle acceleration, radiative cooling, or multiple emission regions may be discussed when justified by the results, while detailed emission mechanisms or source parameters are not inferred from these analyses alone. The resulting summaries are returned to the Supervisor Agent for integration into the final user-facing response.

The statistical analysis tools described above operate as independent sub-agents and provide characterizations of the source variability. Their outputs are subsequently interpreted and summarized before being incorporated into the broader scientific workflow. These analyses provide information on the amplitude of variability across energy bands, the occurrence and temporal correspondence of flaring events, possible inter-band time delays, and the evolution of the spectral index with flux. This combined information provides a basis for identifying periods of correlated multi-band activity and for examining whether the observed variability patterns are qualitatively consistent with processes such as particle acceleration, radiative cooling, or changes in the emitting regions. Such interpretations are restricted to trends supported by the statistical measurements and are not used to infer detailed physical properties without additional modeling. The current set of analysis sub-agents is not final, and additional specialized analysis tools will be incorporated as the system evolves.

The performance of the Data Retrieval Agent is controlled through a validation layer designed to identify incomplete or underspecified requests. If any required information is missing, the agent requests clarification through the Supervisor Agent before proceeding. This reduces the likelihood of incomplete or ambiguous outputs and improves the robustness of complex scientific workflows. All stages of the workflow, including coordinate resolution, data retrieval, time parsing, file management, and plotting, are logged through a centralized logging system (\texttt{default\_logger}) to ensure traceability, reproducibility, and debugging capability. By combining modular tool interfaces with domain-specific analysis procedures, the Data Retrieval Agent enables efficient construction and interpretation of observational datasets for astrophysical research. 

\subsection{Theoretical Modeling Agent}
\label{sec:theoretical_modeling}

The Theoretical Modeling Agent is responsible for converting science-ready observational SED data into physically interpretable model products. Operating under the direction of the Supervisor Agent, it supports two complementary workflows: (i) generation of theoretical spectra for user-specified physical parameters and (ii) fitting observational SEDs with physical emission models. In both cases, the agent interfaces with \texttt{MMDC} to perform the computationally intensive modeling tasks, while managing input validation, request construction, result retrieval, visualization, and user interaction. 

Currently, the modeling backend of \texttt{MMDC} relies on pretrained convolutional neural network (CNN) surrogate models for blazar broadband emission. These surrogates are trained to reproduce the outputs of computationally expensive radiative simulations over broad ranges of physical parameters. \texttt{MMDC} supports Synchrotron Self-Compton \citep[SSC;][]{2024ApJ...963...71B}, External Inverse Compton \citep[EIC;][]{2024ApJ...971...70S}, hadronic \citep{2025ApJ...990..222S}, and hybrid lepto-hadronic models. For parameter inference, the CNN surrogates are coupled with Bayesian sampling through MultiNest \citep{2009MNRAS.398.1601F}, allowing efficient exploration of the parameter space and estimation of posterior distributions, best-fit parameters, and associated uncertainties. The surrogate models retain the physical information contained in the underlying radiative calculations, which include particle injection, self-consistent cooling, and the relevant leptonic or hadronic emission processes. They therefore allow the agent to evaluate physically detailed emission models without repeatedly executing the original numerical simulations during parameter inference. This substantially reduces the computational cost of model fitting and makes it possible to compare theoretical predictions with observational data while exploring a broad parameter space within an interactive workflow. This capability is particularly important for hadronic and multimessenger modeling, for which direct numerical calculations are computationally demanding. By using the pretrained surrogate models, \astrogen\ can perform rapid joint modeling of broadband spectral energy distributions and neutrino emission within an interactive agent-driven workflow. To our knowledge, the combined capability for real-time hadronic SED fitting, neutrino modeling, and Bayesian parameter inference is currently unique to \astrogen and these modeling tasks can be performed through natural-language interaction.

When the Supervisor Agent receives a request requiring theoretical modeling, the task is delegated to the Theoretical Modeling Agent. The agent supports two main modes of operation: generation of a theoretical spectrum for a specified set of model parameters, and fitting of a theoretical model to an observed SED. For fixed-parameter spectrum generation, the agent verifies that the required model parameters, source redshift, EBL configuration, and model type are available. Once these inputs are validated, the request is submitted to the synchronous modeling endpoint of \texttt{MMDC}, which evaluates the selected emission model for the specified parameter values. The resulting theoretical spectrum is then retrieved and returned to the workflow. For observational-data fitting, the agent first verifies that an SED dataset is available, confirms or resolves the source redshift, identifies the requested model class, validates the EBL configuration, and processes any user-defined fixed parameters. If the redshift is unavailable but the source name is provided, the agent attempts to resolve it through catalog queries before proceeding. The observational SED is then submitted to the appropriate inference backend of \texttt{MMDC}, where the remaining free model parameters are constrained by the data. For SSC and EIC modeling, the agent uses the leptonic batch-inference workflow. For hadronic modeling, it uses the hadronic inference backend and requires an additional neutrino constraint. This constraint may be specified through natural-language instructions and can be represented either as a Poisson likelihood, using the expected number of neutrino events and exposure time, or as a band-flux likelihood, using a neutrino energy interval and flux constraint. The hadronic backend predicts both electromagnetic and neutrino emission, allowing the agent to perform joint multimessenger fitting of blazar sources.

After submission, \texttt{MMDC} executes the selected CNN-based surrogate model and, for fitting requests, performs the parameter-inference procedure. The Theoretical Modeling Agent continuously monitors the job status until the batch result becomes available. The returned products include best-fit parameters, parameter uncertainties, posterior samples, MultiNest statistics, a PDF summary report, a CSV file containing the best-fit parameters, and a CSV file containing the best-fit model spectrum. The agent then generates visualizations combining the model prediction and observational data before returning the results to the Supervisor Agent in a user-facing form. The agent also manages intermediate artifacts  generated during the workflow. Uploaded or \texttt{MMDC}-retrieved SED files and generated plot images are stored in Redis-backed short-term storage, while model outputs are registered in the internal modeling cache for subsequent reuse. Before returning the results, the agent verifies the consistency of units, labels, model type, EBL configuration, and downloadable artifacts. Users may subsequently request additional visualization modifications, such as changes in plotting ranges or display styles. 

If any required information is missing or inconsistent, the Theoretical Modeling Agent does not submit an unreliable modeling request. Instead, it initiates a clarification loop through the Supervisor Agent to obtain the missing source name, redshift, model type, observational file, fixed-parameter specification, or hadronic likelihood configuration. If \texttt{MMDC} is unreachable, validation fails, or the backend reports an execution error, the agent returns a concise diagnostic message together with corrective guidance when possible. 

A key feature of the Theoretical Modeling Agent is its ability to combine electromagnetic SED modeling and multimessenger interpretation within an interactive natural-language workflow. Rather than requiring users to manually construct backend requests or execute dedicated modeling scripts, \astrogen\ allows researchers to interact directly with the modeling system through scientific language, including selecting SSC, EIC, or hadronic models; fixing or releasing parameters; enabling or disabling EBL absorption; and requesting revised visualizations or follow-up fits. The architecture is also designed to be extensible. \texttt{MMDC} provides the infrastructure for hosting additional pretrained surrogate models, enabling future versions of the Theoretical Modeling Agent to incorporate new physical scenarios, improved emission models, and source classes beyond blazars. Consequently, the agent functions not as a fixed collection of hard-coded tools, but as a natural-language interface to an evolving library of physically motivated pretrained theoretical models. 

\subsection{Research Ideation Agent}
The Research Ideation Agent addresses a fundamental challenge in modern astrophysical research: identifying scientifically promising questions within a rapidly expanding body of literature and increasingly complex multiwavelength and multimessenger datasets. Relevant evidence is often distributed across numerous publications and heterogeneous observational archives, making it difficult to systematically connect existing knowledge, observed source behavior, and unexplored research opportunities. To address this challenge, the Research Ideation Agent transforms broad scientific questions into a structured hypothesis-generation workflow that integrates information from both published literature and observational data with the aim to identify potential gaps, unresolved tensions, and underexplored scientific directions. The agent employs a coordinated multi-agent workflow that combines literature retrieval, automated data analysis, and LLM-based scientific reasoning to generate candidate research hypotheses. The workflow is implemented as a LangGraph-based execution graph, in which agents exchange structured state information and intermediate outputs, including topic descriptions, retrieved publications, source selections, variability diagnostics, and candidate hypotheses. 

When a user submits a research-ideation query, the Supervisor Agent delegates the request to the Research Ideation Agent. The workflow begins with the request reformulated into a scientifically structured topic description that defines the research scope, relevant physical processes, motivation, and constraints. This description is subsequently used as the query for the literature retrieval system described in Section~\ref{info_ret}. The retrieved publications establish the scientific context for the subsequent stages and are passed, together with the topic description, to the observational analysis workflow. The next stage is performed by the Source Data Analysis Agent, which operates in two phases. First, the agent identifies the astrophysical source most relevant to the scientific topic using the topic description and retrieved literature as contextual input. If no suitable source can be identified, the workflow proceeds using only the literature-derived context. When a source is selected, the system retrieves the corresponding light-curve data from \texttt{MMDC} and invokes the Fractional Variability Sub-agent, Flaring Event Sub-agent, and Spectral Trend Sub-agent to characterize its temporal and spectral behavior. In the second phase, the resulting diagnostics is synthesized into a structured summary for downstream hypothesis generation. The summary emphasizes statistically significant variability patterns, flare characteristics, spectral trends, data-quality limitations, and potential physical implications while remaining closely tied to the measured observational properties. If the selected source lacks sufficient observational data, the source-selection procedure is repeated once; if no suitable dataset can be identified, the workflow continues using only the literature-based context.

To ensure reliable execution of the multi-agent workflow, intermediate LLM outputs are required to satisfy to predefined structured formats. For example, the source-selection stage must return fields containing the selection rationale, source identifier, and optional synthesis summary. The system validates both the format and content of these outputs and requests regeneration when required fields are missing or incorrectly formatted, up to a predefined retry limit. This validation procedure prevents malformed outputs from propagating through subsequent stages and improves workflow robustness and interpretability.

The synthesized literature review and observational summary are then passed to the seed-idea generation stage. Here, the system generates an exploratory set of 30 candidate research ideas. Each candidate consists of four components: the scientific motivation, the proposed methodological approach, the relevant observational or experimental setting, and the anticipated advantage relative to existing approaches. The purpose of this stage is to explore a broad range of potential research directions, including variability studies, SED modeling, multimessenger investigations, population analyses, simulation-based studies, and data-driven methodologies. The generated ideas are subsequently evaluated rather than being returned directly to the user. The first evaluation stage is performed by the Feasibility Agent, which assesses each candidate according to technical feasibility, practical implementability, scalability beyond a limited proof-of-concept study, and the absence of prohibitive resource or infrastructure requirements. Ideas that do not satisfy these criteria are discarded. For the remaining candidates, the agent preserves the original scientific motivation, methodology, observational setting, and expected advantages while providing a concise justification of their feasibility. The filtered ideas are then passed to the Novelty Agent which behaves as a judge  ranks them according to originality, scientific significance, and potential impact. Up to five top-ranked candidates are retained and forwarded to the Supervisor Agent, which prepares the final response to the user.

By combining literature retrieval with automated analysis of observational data, the Research Ideation Agent generates hypotheses informed by both published scientific literature and measurable properties of astrophysical sources. This enables the identification of research directions motivated by unresolved discrepancies in the literature, previously unexplored observational behaviour, and methodological opportunities suggested by the available data. Consequently, the Research Ideation Agent provides a structured framework for exploring and prioritizing potential research directions based on existing scientific knowledge and observational evidence. The generated hypotheses are intended as exploratory research suggestions rather than established scientific conclusions. The role of the Research Ideation Agent is to assist researchers by synthesizing existing knowledge and observational evidence to generate candidate research hypotheses, while the scientific assessment, validation, and interpretation of these hypotheses remain the responsibility of the researcher.

\section{Benchmarking and Evaluation}\label{benchmar}

Because the scientific information returned by \astrogen\ depends on several retrieval, ranking, and selection stages, it is important to evaluate how reliably the system identifies and recovers the relevant scientific literature. We therefore assess the literature-retrieval component using dedicated benchmarks designed to test its ability to return the publications needed to support a scientific response. To evaluate the literature-retrieval capabilities of \astrogen, we construct two complementary benchmarks that target distinct retrieval behaviors in controlled settings. The first benchmark tests whether the system can identify a single primary source, while the second evaluates whether it can recover a coherent evidence set when synthesis across multiple papers is required. Together, these benchmarks provide a clearer and more scientifically relevant assessment of the retrieval system.

\subsection{Single-paper Retrieval Benchmark}
To generate evaluation queries, an automated pipeline is used to construct the single-paper retrieval benchmark. The benchmark is built from the same set of scientific papers, in Markdown format, used to fill in the vector database. For each randomly selected paper, the abstract is removed from the extracted text, and the system scans the paper to identify evidence-rich sentences containing concrete results, observational findings, modeling choices, data or instrument usage, or physical interpretations. These statements, together with the paper metadata and relevant body text, are then provided to the question-generation model. A prompt is sent via the API to a configured GPT model (\texttt{gpt-5.1-codex-mini}), which generates candidate benchmark questions in structured JSON format. Each candidate contains a question, a short reference answer, a question-type label, supporting evidence notes, a difficulty label, and revision notes. The prompt instructs the model to generate realistic, researcher-style literature-search questions rather than quiz-like or summary-based questions. It also requires the questions to exclude explicit title or author information and references to figures, tables, equations, appendices, or section numbers, while ensuring that each question is centered on a single primary paper. In addition, the model is instructed to avoid broad review-style questions, rare phrases copied directly from the source, and questions that could be answered using only the abstract.

Each candidate question is then passed through a multistage quality-control process. First, the system measures possible overlap with the paper title, abstract, author names, and distinctive source \(n\)-grams to detect source-revealing wording. It checks content-word overlap with the title and abstract, exact phrase reuse, author-name mentions, distinctive four- to six-word source phrases, and overall similarity to the paper title. If moderate overlap is identified, a separate rewriting prompt is used to preserve the scientific intent while reducing overlap and removing title- or author-specific references. The rewritten question is then checked again for leakage. Candidates with high-severity leakage are not automatically rewritten and are normally rejected.

Next, each candidate question is evaluated through an automated LLM-as-a-judge step implemented using a self-review prompt and the \texttt{gpt-5.1-codex-mini} model. The LLM judge assesses whether the question is grounded in a single paper, scientifically meaningful, realistically phrased, and supported by the extracted evidence. It also checks whether the question is suitable for retrieval evaluation rather than being exam-style, broad, trivial, ambiguous, or dependent on unavailable figures or tables. In addition, the judge considers whether the question could be answered through simple keyword matching with the source. For each candidate, it returns a structured decision—\texttt{approve}, \texttt{revise}, or \texttt{reject}-together with a score, comments, specific failure reasons, and, when applicable, a suggestion to revise it. Deterministic post-processing can further downgrade or reject candidates because of high lexical leakage, prohibited document-summary phrasing, insufficient specificity, a low review score, or near-duplication with another candidate generated from the same paper. Finally, the surviving candidates are ranked, and at most one question is selected for each paper and exported with its bibcode and reference answer.

Using this procedure, 500 candidate questions were generated, of which 200 that satisfied the approval criteria were retained for the initial benchmark. To expand the benchmark and increase the diversity of the questions, an additional set of papers was randomly sampled and manually reviewed for inclusion. Questions for these papers were generated using four different LLMs: \texttt{gpt-4.1}, \texttt{grok-3-beta}, \texttt{gemini-2.5-pro-exp-03-25}, and \texttt{gemini-2.5-flash-preview-04-17}. The generated questions were manually reviewed and added to the initial set, resulting in a single-paper retrieval benchmark containing 256 questions.

Further verification identified nine questions that contained references to internal document elements, and these questions were removed. The final single-paper benchmark therefore contained 247 questions, all of which were submitted to \astrogen\ for evaluation. Of these, 209 were subsequently sent by the planner to the literature-retrieval agent, whereas the remaining 38 were directed to other processing routes or resulted in clarification requests. The planner requested clarification for 27 questions containing contextual references such as ``this study'' or ``the paper,'' classified eight questions as answerable from general knowledge, routed two to the observational-data retriever, and sent one calculation-based question to the Data Analyzer agent.  These cases show that the planner determines the appropriate processing route based on the information required by each question rather than automatically invoking literature retrieval. The 27 clarification cases represent a limitation of the routing stage because the corresponding gold papers were available in the indexed corpus but the questions were not passed to the literature-retrieval agent. We therefore evaluate routing and retrieval separately. Routing recall measures the fraction of the 247 benchmark questions routed to the literature-retrieval agent, while retrieval performance is evaluated on the 209 questions that reached this agent. For these 209 questions, we first evaluate whether the gold paper is recovered in the initial candidate pool. This is measured by pool recall, defined as the fraction of gold papers present in this pool. Since only papers included in the candidate pool can be considered by the subsequent ranking stage, pool recall sets the maximum retrieval recall that the ranking stage can achieve. We then evaluate the final ranked results using $\mathrm{Hit}@k$, defined as the fraction of questions for which the gold paper appears among the top $k$ retrieved results:
\begin{equation}
\mathrm{Hit@}k =
\frac{1}{N}
\sum_{i=1}^{N}
\mathbf{1}\!\left(r_i \leq k\right).
\end{equation}
where $N$ is the number of evaluated questions and $r_i$ is the rank of the gold paper for question $i$. Mean reciprocal rank (MRR) additionally measures how highly the gold paper is ranked:
\begin{equation}
\mathrm{MRR} =
\frac{1}{N}
\sum_{i=1}^{N}
\frac{1}{r_i}.
\end{equation}
A higher MRR indicates that the gold paper tends to appear closer to the top of the retrieved results. Ranking quality was also evaluated using normalized discounted cumulative gain (nDCG), which discounts relevant results according to their rank in the retrieved list. In the single-paper benchmark, where each question has exactly one gold paper, nDCG therefore reflects the rank of that paper with a logarithmic discount, assigning progressively lower scores when the gold paper appears deeper in the ranking.

During this benchmark evaluation, the temporal weighting term used in the deployed retrieval system was disabled, see Equation \eqref{eq:time_dependent_score}. This was necessary because the gold paper is defined as the specific source paper from which each benchmark question was constructed, irrespective of its publication date. Applying a recency preference could systematically disadvantage older gold papers when newer publications address similar topics, thereby conflating retrieval accuracy with the intended preference of the system for recent literature. The benchmark therefore evaluates whether the retrieval system can recover and correctly rank the target gold paper based on relevance alone, while temporal weighting remains part of the deployed system.

\begin{table}[t]
\centering
\caption{Retrieval results for the single-paper and multi-paper benchmarks.}
\label{tab:retrieval_results}
\begin{tabular}{lcc}
\hline
\textbf{Metric} & \textbf{Single-paper} & \textbf{Multi-paper} \\
\hline
Questions constructed/evaluated & 247/209 & 200/154 \\
Mean gold papers per question   & 1.00    & 6.10 \\
Routing recall                  & 0.846   & 0.945 \\
Pool recall                     & 0.962   & 0.920$^{a}$ \\
Hit@1                           & 0.560   & 0.422 \\
Hit@5                           & 0.766   & 0.792 \\
Hit@10                          & 0.809   & 0.864 \\
Recall@10                       & 0.809$^{b}$ & 0.497 \\
MRR@10                          & 0.647   & 0.563 \\
nDCG@10                         & 0.686   & 0.441 \\
Evidence Coverage@10            & --      & 0.354 \\
\hline
\end{tabular}

\vspace{0.5em}
\begin{minipage}{0.95\linewidth}
\footnotesize
$^{a}$ Multi-paper pool recall should be interpreted cautiously because the gold evidence was constructed from the same captured candidate pool.

$^{b}$ Recall@10 and Hit@10 are identical for the single-paper benchmark because each question has exactly one gold paper.
\end{minipage}
\end{table}

The benchmarking results for a single paper retrieval are summarized in Table \ref{tab:retrieval_results}. The gold paper was ranked first for 56.0\% of the evaluated questions and appeared among the top three results for 72.3\%. This fraction increased to 76.6\% within the top five results and 80.9\% within the top ten. Because the deployed system returns five documents for each retrieval request, $\mathrm{Hit}@5=0.766$ provides the most direct measure of retrieval performance under the production configuration. $\mathrm{Hit}@10$ is additionally reported to characterize ranking performance beyond the retrieval depth used by the deployed system. The $\mathrm{MRR}@10$ of 0.647 further indicates that the gold paper generally appeared near the top of the ranked results, while the $\mathrm{nDCG}@10$ of 0.686 provides a complementary rank-discounted measure of the position of the target paper. Pool recall reached 96.2\%, indicating that the candidate-generation stage successfully included the gold paper for the large majority of evaluated questions. Only 3.8\% of the gold papers did not enter the candidate pool and were therefore unavailable to the ranking stage. The difference between pool recall and $\mathrm{Hit}@10$ is 15.3 percentage points, corresponding to cases in which the gold paper was present in the candidate pool but was not ranked among the top ten results. These results show that candidate generation provides high coverage of the target literature and that the current ranking performance is already sufficient to support the literature-retrieval workflow. Further improvements to the ranking stage could increase the fraction of relevant papers appearing among the highest-ranked results and further improve the overall retrieval performance of the system. However, the reported retrieval metrics should be interpreted within the specific benchmark and retrieval setup used in this work. A direct quantitative comparison with results reported for other systems \citep[for example][]{Iyer2024Pathfinder, 2026arXiv260803860A, 2025arXiv250808742C} is not straightforward, since retrieval performance depends strongly on the composition and size of the indexed literature corpus, the construction of the benchmark questions and gold references, the candidate-generation procedure, and the number of documents returned for evaluation.

\subsection{Multi-paper Retrieval Benchmark}
A separate multi-paper benchmark was constructed to evaluate whether the literature-retrieval system can recover multiple publications that provide the evidence needed to address a scientific question. This is a more demanding retrieval setting than the single-paper benchmark, where the objective is to identify one specific publication. In the multi-paper case, the necessary information is distributed across several studies, requiring the retrieval system not only to identify an appropriate publication but also to recover a sufficiently complete set of supporting papers.

The benchmark questions were constructed from passages in the same scientific literature corpus used by the retrieval system, following the same general procedure as for the single-paper benchmark but with different selection criteria. Source papers were segmented into scientific sections, and passages containing at least two distinct citations were identified. Candidate passages were required to express a scientific claim, comparison, interpretation, or other relation between the cited studies rather than simply listing references. The selection therefore favored passages containing synthesis-related language or claim-bearing statements, while historical citation lists, routine methodological descriptions, and passages dominated by weakly connected references were excluded. For each candidate passage, the cited references were linked to the corresponding bibliography entries and matched to papers in the local corpus using bibliographic identifiers and metadata, including DOI, arXiv identifier, title, author and year, and journal information. Only passages containing at least two well-matched references were retained, while those dominated by weak citation matches or bibliographic lists were excluded. Each retained candidate was then assigned to one of six question categories: observational, theoretical, comparative, methodological, source-centered, or event-centered.

For each approved candidate, the source passage, additional context from the corresponding section of the source paper, and selected content from the associated evidence papers were collected and provided to the question-generation model. Context from the evidence papers was preferentially selected from scientifically informative sections, such as the Discussion, Conclusion, and Results, with Methods or other sections used when relevant. Paper titles, bibcodes, and source paths were removed from the context presented to the model, and the associated publications were represented as unidentified evidence sources. The source passage itself was retained to preserve the scientific relation from which the question was derived. Based on this information, the model was instructed to generate a focused question whose answer requires the synthesis of information from multiple supporting publications. Each question was required to contain sufficient scientific specificity for literature retrieval without revealing the publications from which it was generated. Questions containing citation metadata, paper titles or identifiers, cited-author names, references to figures or tables, yes/no formulations, generic review-style requests, or wording closely reproducing the source passage were excluded. The model also generated a concise reference answer for each question. Benchmark questions were generated using \texttt{gpt-5.4-mini}.

Generated questions were further subjected to automatic quality control before inclusion in the candidate benchmark. In addition to basic requirements on question length and form, the validation required synthesis-oriented wording and sufficient overlap with scientifically informative terms from the source passage and associated evidence papers. Questions were rejected when they were too similar to the original source statement, contained citation years or identifying metadata, exposed paper titles or cited authors, referred to internal figures or tables. The corresponding reference answers were also checked for identifying information and internal references. These filters were designed to reduce source leakage and remove questions not suitable for retrieval evaluation while keeping the scientific information needed to identify relevant literature.

After applying the quality filters described above, 200 questions with their corresponding data were selected. At this stage, however, the papers associated with each generated question are preliminary evidence sets rather than independently validated retrieval targets. They originated from references linked to the source passage and their subsequent matches to papers in the corpus; their association with the generated question therefore did not establish that each paper provided relevant evidence for answering that question. Moreover, the generation-stage validation evaluated question form, leakage, and lexical specifications but did not independently evaluate the scientific relevance of each question--paper pair. The preliminary evidence sets were therefore subjected to further independent relevance evolution before being used as retrieval targets. For each of the 200 generated questions, the papers associated with the source passage were evaluated together with the source paper from which the passage had been extracted. Each question--paper pair was independently assessed by two LLM judges: \texttt{gpt-5.6-luna} \footnote{\url{https://developers.openai.com/api/docs/models/gpt-5.6-luna}} as the primary judge and \texttt{deepseek-v4-flash} \footnote{\url{https://huggingface.co/deepseek-ai/DeepSeek-V4-Flash}} as an independent cross-family judge. The judges assigned a relevance grade from 0 to 3 to each candidate. A grade of 3 indicates that the paper directly addresses the question and can serve as a primary scientific source, while grade 2 indicates that it provides substantial supporting evidence. Grade 1 corresponds to related background information, and grade 0 indicates that the paper is irrelevant. This initial assessment covered 1,649 question--paper pairs. Based on the evaluation, 19 questions were removed because none of their evaluated papers, including the corresponding source paper, received a relevance grade of at least 2 from both judges. This left 181 questions after the initial relevance-validation stage.

The assessment also showed that the citation-derived evidence sets could not be used directly as the gold standard for retrieval evaluation. Only 28.3\% of the evidence papers originally associated with the source passages received a relevance grade of at least 2, whereas the source paper, which had been excluded from the evidence set during question generation, was judged relevant in 69.5\% of cases. This difference reflects the way the benchmark questions were constructed: the citations in a source passage identify papers that contribute to the scientific discussion in that passage, but these papers are not necessarily the most relevant sources for answering the generated question. Treating all citation-derived evidence papers as gold papers would therefore evaluate the system ability to retrieve the references cited by the source paper rather than its ability to retrieve the evidence needed to answer the question.

Before the final retrieval evaluation, the 181 validated questions were submitted to \astrogen. Of these, the planner routed 171 to the literature-retrieval agent, corresponding to a routing recall of 0.945. The remaining 10 questions were directed to the observational-data retriever and therefore did not produce a literature-retrieval query. As in the single-paper benchmark, these cases were treated as routing outcomes rather than retrieval failures and were excluded from the subsequent retrieval evaluation. For the 171 questions that reached the literature-retrieval agent, the relevance assessment was then extended to a broader set of candidate papers. For each question, this set combined the highest-ranked papers returned by the different retrieval configurations described in Subsection~\ref{info_ret}, the original citation-derived evidence papers, and papers randomly sampled from the larger candidate pool. This resulted in 9,761 question--paper pairs, with a median of 57 candidate papers per question. The random sample was included to reduce the dependence of the relevance set on any individual retrieval configuration and to provide an estimate of relevance among papers outside the highest-ranked candidates.

Every question--paper pair in this expanded set was independently evaluated by the same two LLM judges, \texttt{gpt-5.6-luna} and \texttt{deepseek-v4-flash}, using the 0--3 relevance scale. Agreement between the judges was substantial, with 75.9\% exact agreement, 99.5\% agreement within one grade, and a linear-weighted Cohen $\kappa$ of 0.748. The mean difference between the two judges was 0.02 grade. A conservative criterion was adopted for selecting  the final retrieval targets: a paper was included in the gold set only when both judges assigned grade 3. The source paper was not included automatically and was evaluated according to the same criterion as all other candidate papers. Applying this criterion removed a further 17 questions for which no paper satisfied the final gold requirement. The resulting retrieval evaluation set therefore contained 154 questions and 940 validated gold papers, corresponding to an average of 6.1 gold publications per question. Only 99 of these gold assignments originated from the initial citation-derived evidence sets, showing that most relevant retrieval targets were identified only through the subsequent relevance-validation procedure.

Retrieval performance in this setting requires metrics that account for both the presence and the completeness of the returned evidence. Hit@$k$ measures whether at least one validated gold paper occurs among the top $k$ retrieved publications,
\begin{equation}
\mathrm{Hit}@k =
\frac{1}{N}
\sum_{i=1}^{N}
\mathbb{1}
\left(
|G_i \cap R_i^{(k)}|>0
\right),
\end{equation}
where $\mathbb{1}(\cdot)$ denotes the indicator function, which equals 1 when the condition is true and 0 otherwise, $G_i$ is the validated gold set for question $i$ and $R_i^{(k)}$ denotes the top-$k$ retrieved papers. In contrast to the single-paper benchmark, however, a successful Hit@$k$ does not imply that the required evidence set has been recovered. Recall@$k$ therefore measures the fraction of validated gold papers retrieved,

\begin{equation}
\mathrm{Recall}@k =
\frac{1}{N}
\sum_{i=1}^{N}
\frac{|G_i \cap R_i^{(k)}|}{|G_i|}.
\end{equation}

This distinction is central to the multi-paper evaluation: retrieval of one relevant publication constitutes a successful hit, but answering a synthesis question may require several publications distributed throughout the ranking. Ranking quality was additionally quantified using MRR and nDCG, as in the case of single paper benchmarking. In this case,  MRR measures the position of the first relevant publication, whereas nDCG accounts for the positions of multiple relevant papers and therefore provides a more informative measure when several publications contribute to the evidence set. Evidence Coverage@$k$ was also used to quantify how much of the graded scientific evidence represented in the relevance assessment is recovered within the retrieval window, giving greater weight to primary than to supporting evidence. Together, these metrics separate the ability to retrieve at least one relevant publication from the more demanding requirement of recovering a sufficiently complete and appropriately ranked evidence set.

It should be noted that although different variant of retrieval were investigated to asses the performance of the system, the results reported here correspond to the retrieval configuration consistent with that used for the single-paper evaluation. Candidate papers were obtained from body-text chunks using hybrid dense--sparse retrieval with $\alpha=0.8$, collapsed to the paper level, and expanded through the citation graph. Candidate ranking combined standardized dense, BM25, and cross-encoder scores with weights $(1,1,2)$, with the cross-encoder applied to paper abstracts. For the offline benchmark, deeper captured candidate pools were retained to evaluate performance beyond the production retrieval depth. The temporal weighting, as in the single paper retrieval, used in deployment was disabled during offline scoring so that differences reflected retrieval and ranking rather than publication recency.

The multi-paper retrial benchmarking results are listed in Table \ref{tab:retrieval_results} which shows a clear difference between identifying some relevant literature and recovering the broader evidence set. At rank 1, at least one gold publication was retrieved for 42.2\% of questions. This increased to 79.2\% at the deployed depth of five papers and to 86.4\% within the top 10. The corresponding MRR@10 was 0.563. Thus, for most questions the system places at least one primary relevant publication within the small set of documents returned to the user. Recovery of the complete evidence set is more demanding. Recall@5 was 0.347 and Recall@10 reached 0.497, increasing to 0.634 at rank 20 and 0.787 at rank 50. At $k=10$, nDCG was 0.441 and weighted Evidence Coverage was 0.354. The difference between Hit@10 of 0.864 and Recall@10 of 0.497 is particularly informative: although at least one primary source is found for most questions, approximately half of the validated gold publications are recovered within the first ten results. At the production depth of five documents, this distinction is stronger, with Hit@5 of 0.792 but Recall@5 of 0.347. The lower recall cannot be explained primarily by the number of gold papers per question. Although the benchmark contains an average of 6.1 gold papers, only 18.2\% of questions contain more than ten gold papers. A perfect ranking would therefore attain a Recall@10 of 0.944 given the observed gold-set sizes, compared with the measured value of 0.497. Increasing the retrieval depth produces a substantial improvement: Recall rises from 0.497 at rank 10 to 0.787 at rank 50 and 0.868 at rank 100. This indicates that a substantial fraction of the evidence missed near the top of the ranking is present deeper within the candidate set rather than being intrinsically inaccessible at the chosen retrieval depth.

This interpretation must nevertheless be qualified by the construction of the relevance set. The reported candidate-pool recall is 0.920, but the expanded relevance frame itself was partly constructed from papers in the captured retrieval pool. Pool recall is therefore not an independent estimate of corpus-level evidence coverage and should not be interpreted in the same manner as in the single-paper benchmark. The retrieval results more directly characterize the ability of the system  to rank relevant evidence within the captured candidate space. Establishing corpus-level recall would require relevance assessment extending independently beyond that pool.

Overall, the multi-paper benchmark indicates that the retrieval system is substantially more successful at identifying at least one relevant primary publication than at recovering a complete set of evidence for scientific synthesis. At the production depth of five papers, a relevant primary source is present for approximately four fifths of the evaluated questions, whereas only about one third of the complete gold evidence set is recovered. The continued increase in recall at larger retrieval depths indicates that improving the ordering and selection of candidate papers, and potentially adapting the number of returned publications to the evidence requirements of individual questions, could improve the completeness of multi-paper retrieval. A more detailed investigation of this aspect will be performed in future work, including adaptive retrieval strategies in which the number of returned papers is determined dynamically according to the complexity and evidence requirements of the query.

\subsubsection{Query Expansion with HyDE}

The effect of query expansion using HyDE \citep{gao2023hyde} was evaluated using both retrieval benchmarks. In HyDE, the original retrieval query is expanded into a hypothetical document representing a plausible answer, and this generated text is used for retrieval. The expanded representation can provide additional semantic context, but its effectiveness depends on how well it corresponds to the documents or passages stored in the retrieval index. This is particularly relevant for the present system because candidate retrieval is performed over chunks extracted from the bodies of scientific papers rather than over abstracts. HyDE generally produces a compact, abstract-like synthesis, but the indexed chunks contain more specific descriptions of observations, methods, measurements, and intermediate scientific arguments. The expanded query may therefore provide broader topical context while being less well matched to the individual passages used during first-stage retrieval.

To quantify this effect, retrieval with the HyDE-expanded query was compared directly with retrieval using the unexpanded query, while keeping the remaining retrieval and ranking configuration unchanged. For the single-paper benchmark, HyDE reduced Recall@10 from 0.803 to 0.744, nDCG@10 from 0.677 to 0.586, and MRR@10 from 0.637 to 0.534. The differences in these ranking metrics were statistically significant under paired bootstrap resampling. Pool recall also decreased from 0.961 to 0.941, although this difference was not statistically significant.

A different behavior was observed for the multi-paper benchmark. HyDE increased pool recall from 0.920 to 0.943, with the difference being statistically significant, indicating that the expanded query can recover a broader set of potentially relevant papers when the required evidence is distributed across multiple publications. However, this improvement in candidate-pool coverage did not lead to improved ranking. Recall@10 was 0.487 with HyDE and 0.497 with the unexpanded query, while the differences in nDCG@10 (0.447 versus 0.441) and MRR@10 (0.569 versus 0.563) were small and not statistically significant.

These results show that the effect of HyDE depends on the retrieval stage and benchmark setting. For single-paper retrieval, query expansion provides no improvement in candidate coverage and significantly degrades the ranking of the target paper. For multi-paper retrieval, it improves candidate-pool coverage but provides no measurable improvement in the subsequent ranking. One possible explanation is the difference between the abstract-like representation generated by HyDE and the body-text chunks used for candidate retrieval. In addition, expansion may dilute discriminative terms from the original query or introduce plausible details that match individual passages without improving identification of the most relevant papers.

Based on these results, the deployed retrieval path uses the unexpanded query for chunk-level retrieval. The improvement in pool recall observed for the multi-paper benchmark nevertheless suggests that query expansion may be useful specifically for candidate discovery. A hybrid strategy will therefore be investigated in future work, in which the HyDE-expanded query is used to construct a broader candidate pool, while the original query is retained for subsequent scoring and ranking.

\section{Front-End Interface}

\astrogen\ is accessible through a web-based interface designed to support end-to-end astrophysical research workflows, available at \url{https://astrogenai.com}. The front-end layer provides access to the platform and its scientific capabilities, integrating literature analysis, observational data retrieval, theoretical modeling, and hypothesis generation within a unified environment. Through this interface, user requests can be translated into coordinated multi-agent workflows, while the overall design promotes transparency, traceability, and iterative scientific exploration.

The front-end is organized into several pages that provide information about the platform and its scientific objectives:

\begin{itemize}
\item \textit{Home.} Provides a high-level overview of \astrogen\ and its core capabilities, including literature reasoning, observational data retrieval, theoretical modeling, and research ideation.

\item \textit{About.} Describes the scientific motivation behind \astrogen, outlining the challenges posed by fragmented data resources, distributed literature, and disconnected analysis tools, and presents the platform as an integrated research framework.

\item \textit{Research.} Introduces DSRMs which organize the platform around different astrophysical source classes, such as blazars, GRBs, and TDEs, each associated with dedicated datasets, models, and analysis workflows.

\item \textit{Team.} Presents the multidisciplinary team involved in the development of the platform, including expertise in astrophysics, data science, and software engineering.

\item \textit{Partnership.} Describes opportunities for collaboration with research institutions and other stakeholders interested in contributing to or utilizing the platform.

\item \textit{Pricing.} Provides information on platform access options and subscription plans.

\item \textit{Contact.} Provides channels for scientific inquiries, technical support, and collaboration requests.
\end{itemize}

While these pages provide contextual information and access to platform resources, the primary scientific functionality is delivered through the interactive chat interface, where users can initiate and monitor multi-agent research workflows.

\subsubsection{Research Chat Interface.}
The central component of the \astrogen\ front-end is the research chat interface, which serves as the primary interaction point between the user and the multi-agent system. Unlike conventional conversational interfaces, this environment is designed as a structured research workspace in which user queries are interpreted as scientific tasks and executed through coordinated agent workflows.

The interface consists of a sidebar, a central message stream, and a query composer. The sidebar organizes research activities into threads and folders, allowing users to manage multiple projects and analyses. It supports search, tagging, and management operations such as renaming, archiving, and grouping threads. This organization facilitates long-term management of scientific workflows rather than isolated query execution.

The central message stream serves as the main research workspace where all interactions between the user and \astrogen\ are displayed and recorded. It shows the complete progression of a scientific task, including the original user request, intermediate execution states, generated artifacts, and final results. User requests are represented as query units, while system responses are returned as structured outputs that may contain visualizations, reports, retrieved datasets, model prediction, and feedback controls. Because all interactions are preserved, users can inspect and revisit the full execution history of a task, from the initial query to the final scientific output. The interface supports both free-text queries and structured interaction modes. In addition to natural-language input, users can explicitly select a DSRM (e.g., blazar, GRB, or TDE) before submitting a request. This selection guides the system toward the appropriate data sources, theoretical models, and analysis pipelines associated with the chosen source class. If no DSRM is specified, the Planner Agent automatically identifies the most appropriate module based on the content and scientific context of the query.

When a query is submitted, the system initiates a multi-stage execution workflow. Progress is exposed through a sequence of intermediate states corresponding to operations such as literature retrieval, archive queries, data normalization, SED construction, model selection, parameter estimation, and result synthesis. These states provide transparency into the execution process and allow users to monitor the workflow in real time. The final outputs may include SED visualizations, modeling reports, textual summaries, and derived data products. Each output is linked to the originating query, the retrieved data, and the models used during the analysis, providing traceability and supporting reproducibility. The interface also includes feedback mechanisms within the workflow. Users can evaluate responses, provide feedback, and request regeneration of results, enabling iterative refinement of both the query and the generated output. These feedback reports are periodically reviewed by the development team to improve the performance and reliability of \astrogen.

All interactions are stored within persistent research sessions represented as structured threads. This allows users to revisit, refine, and extend previous analyses, transforming the interaction history into a reproducible research record rather than a transient conversation. Such persistent research sessions support iterative scientific investigation by enabling users to build upon previous results without repeating earlier stages of the analysis.

\subsection{Back-end}

The back-end of \astrogen\ is organized around an asynchronous, stateful execution architecture that connects user requests to the multi-agent research framework. Each submitted request is first associated with a persistent conversation thread and registered as an execution job, after which an orchestration layer manages its processing independently of the client request. The job is passed to the LangGraph-based multi-agent workflow \footnote{\url{https://www.langchain.com/blog/langgraph-multi-agent-workflows}}, where planning and supervisory components coordinate the specialized agents required for the task and their access to literature, observational data, physical models, and computational tools. During execution, intermediate agent and tool outputs are propagated through the system, while the evolving workflow state is checkpointed to preserve the context and results of the research process. Once execution is completed, the final response and associated metadata or generated products are stored and returned to the user, while the retained conversational and agent state provides the context for subsequent requests within the same research session. This architecture therefore separates request handling, workflow orchestration, scientific execution, and state management, allowing complex and potentially long-running research tasks to be performed while maintaining continuity across multi-turn interactions.

\subsection{Backend Infrastructure}

The backend of \astrogen\ is implemented as a service-oriented architecture designed to support authenticated user access, persistent scientific conversations, asynchronous agent execution, real-time communication, file management, and scalable deployment. The backend consists of multiple specialized services that communicate through well-defined interfaces, allowing individual components to be developed, maintained, and scaled independently. The architecture combines a research-oriented processing stack with a dedicated authentication and subscription-management service.

The core research backend is implemented using FastAPI \footnote{\url{https://fastapi.tiangolo.com}} and serves as the central entry point for scientific workflows. It is responsible for managing conversation threads, user messages, file metadata, job execution, and interactions with the various agents that comprise the \astrogen\ framework. To support computationally intensive tasks, user requests are processed asynchronously. Rather than waiting for the completion of an entire workflow, the system immediately registers a job and executes the associated agent pipeline in the background. This design enables efficient handling of long-running tasks such as literature retrieval, document analysis, observational data processing, and theoretical modeling.

Real-time communication between the backend and frontend is implemented through WebSocket connections. While conventional REST endpoints are used for thread management and message submission, WebSocket channels stream intermediate execution updates, progress information, and final results to the user interface. This allows users to monitor the status of complex multi-step workflows as they execute. Redis serves as the underlying messaging and publish-subscribe layer, decoupling job execution from client communication and enabling scalable event-driven interactions.

Persistent storage is provided through PostgreSQL \footnote{\url{https://www.postgresql.org}}, which stores user accounts, conversation threads, messages, job metadata, uploaded-file records, and other application data. Redis is additionally used for caching, event coordination, and temporary communication between services. The backend also incorporates MongoDB \footnote{\url{https://www.mongodb.com}} for auxiliary metadata storage and Weaviate as a vector database supporting semantic retrieval and retrieval-augmented generation workflows. This separation between relational data, vector representations, and transient messaging allows each storage technology to be optimized for its specific role within the platform.

File management is implemented through an object-storage workflow that supports large scientific datasets and user-uploaded documents. Uploaded files are stored independently from the application services and are referenced through database records, enabling efficient handling of observational datasets, scientific articles, and intermediate analysis products. The architecture supports validation, tracking, and reuse of uploaded resources throughout agent-driven research workflows.

User authentication, authorization, and subscription management are implemented through a separate Django REST Framework service \footnote{\url{https://www.django-rest-framework.org}}. This service manages user registration, login, password recovery, email verification, third-party authentication, token issuance, subscription plans, and payment processing. Security is enforced through JSON Web Tokens (JWTs), while service-to-service authentication enables secure communication between the research backend and authentication components. The separation of identity management from scientific processing improves modularity and isolates sensitive account and billing information from research-related data.

The entire backend infrastructure is deployed using containerized services orchestrated through Docker. Nginx acts as the external gateway, providing HTTPS termination, request routing, WebSocket support, and integration with the individual backend services. Health monitoring endpoints and observability tools are incorporated to facilitate deployment management, fault detection, and operational monitoring. This architecture provides a scalable and maintainable foundation capable of supporting persistent research sessions, concurrent agent execution, real-time interaction, and future expansion of the \astrogen\ platform.

\subsection{The server}
\astrogen\ is currently deployed on a dedicated server equipped with an AMD EPYC 9534 processor providing 128 computational threads, 251 GB of RAM, and approximately 3.5 TB of local storage. The processor incorporates 256 MB of L3 cache and supports advanced vector instruction sets, enabling efficient execution of parallel scientific workloads and large-scale data processing tasks. The available memory allows efficient handling of large document collections, vector databases, and observational datasets, while the high degree of parallelism supports the concurrent execution of multiple agent workflows. This infrastructure enables simultaneous execution of literature retrieval, observational data analysis, theoretical modeling, and agent coordination tasks while maintaining responsive user interactions. The current deployment serves as the primary computational node of the platform. Additional servers are planned for future integration to increase computational and storage capacity and to support a growing user base through a distributed computing infrastructure.

\section{AstroGenesis in Action: End-to-End Workflows}
To illustrate the practical operation of \astrogen, we present several representative workflows initiated by a user query. They are presented from the least to the most demanding. Rather than operating as a collection of independent components, the system executes a coordinated multi-stage process in which tasks are dynamically planned, delegated to specialized agents, and iteratively refined based on intermediate results. This example demonstrates how the architectural components described above interact to support realistic scientific research workflows.

\subsection{Scientific Information Retrieval}
To illustrate the literature-retrieval capabilities of \astrogen, we consider the representative query:

\textbf{Query.}
\textit{``Explain the association between the blazar TXS 0506+056 and high-energy neutrino emission?''}

\begin{figure}
    \centering
    \includegraphics[width=1\textwidth]{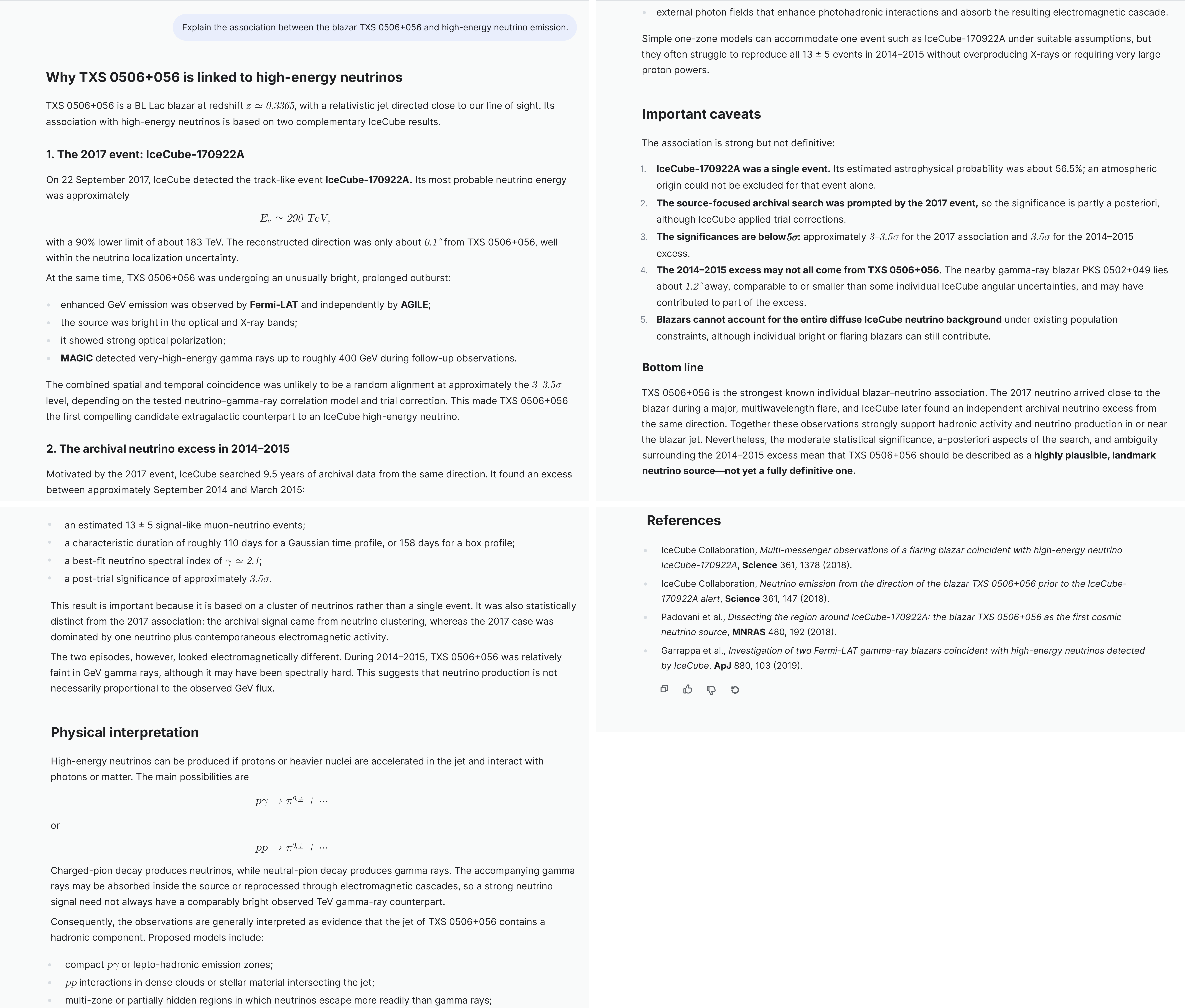}
    \caption{Example of a literature-retrieval response generated by \astrogen\ for the query on the association between TXS~0506+056 and high-energy neutrino emission described in Section \ref{sec:data_retrieval_example}. The chat interface presents a structured synthesis of the retrieved studies, highlighting the main evidences, key limitations, and leading physical scenarios discussed in the literature.}
    \label{fig:literatura_retrival}
\end{figure}

Answering this question requires identifying relevant publications, integrating evidence from multiple studies, and synthesizing the current scientific understanding of the source. When the query is received, the Supervisor Agent classifies it as a literature-centered task and forwards it to the Planner Agent, which decomposes the request into retrieval and synthesis stages. The Literature Agent then executes the retrieval pipeline described above, including vector-based search and reranking to identify the most relevant publications. In this example, the retrieved context includes key observational and theoretical studies of TXS~0506+056, including investigations of the IceCube-170922A event, in this case \citep{2018Sci...361.1378I, 2018Sci...361..147I, 2018MNRAS.480..192P, 2019ApJ...880..103G} and subsequent analyses of its potential association with high-energy neutrino production. For the detailed answer see Figure \ref{fig:literatura_retrival}.

After selecting the relevant documents, the Literature Agent extracts information from the full texts and constructs a structured synthesis that integrates observational evidence, statistical considerations, and proposed physical interpretations. The resulting summary is returned to the Supervisor Agent, while the Replanner verifies that the planned retrieval and synthesis steps have been completed before the final response is generated. The output, shown in Figure~\ref{fig:literatura_retrival}, is the response received through the chat interface and provides a traceable summary of the current literature, highlighting the principal observational findings, outstanding uncertainties, and leading physical scenarios discussed in the retrieved studies as well as the references which were used to generate the answer.

This example illustrates how \astrogen\ treats scientific information retrieval as a coordinated multi-stage workflow rather than a single search operation. By combining query expansion, retrieval, reranking, and synthesis within a multi-agent framework, the system produces literature based responses in a structured and interpretable form.

\subsection{Observational Data Retrieval and Analysis}
\label{sec:data_retrieval_example}

To illustrate the observational data retrieval capabilities of \astrogen, we consider a representative request targeting broadband spectral data for a well-studied blazar:

\textbf{Query.}
\textit{``Retrieve the SED of 3C 279.''}

This task requires identification of the astrophysical source, retrieval of multiwavelength observational data from external archives, and construction of a structured dataset necessary for further analysis. When receiving the query, the Supervisor Agent classifies it as a data-oriented task and forwards it to the Planner Agent. The Planner decomposes the request into three stages: \textit{(i)} source identification, \textit{(ii)} data retrieval, and \textit{(iii)} dataset construction.

\begin{figure}
    \centering
    \includegraphics[width=1\textwidth]{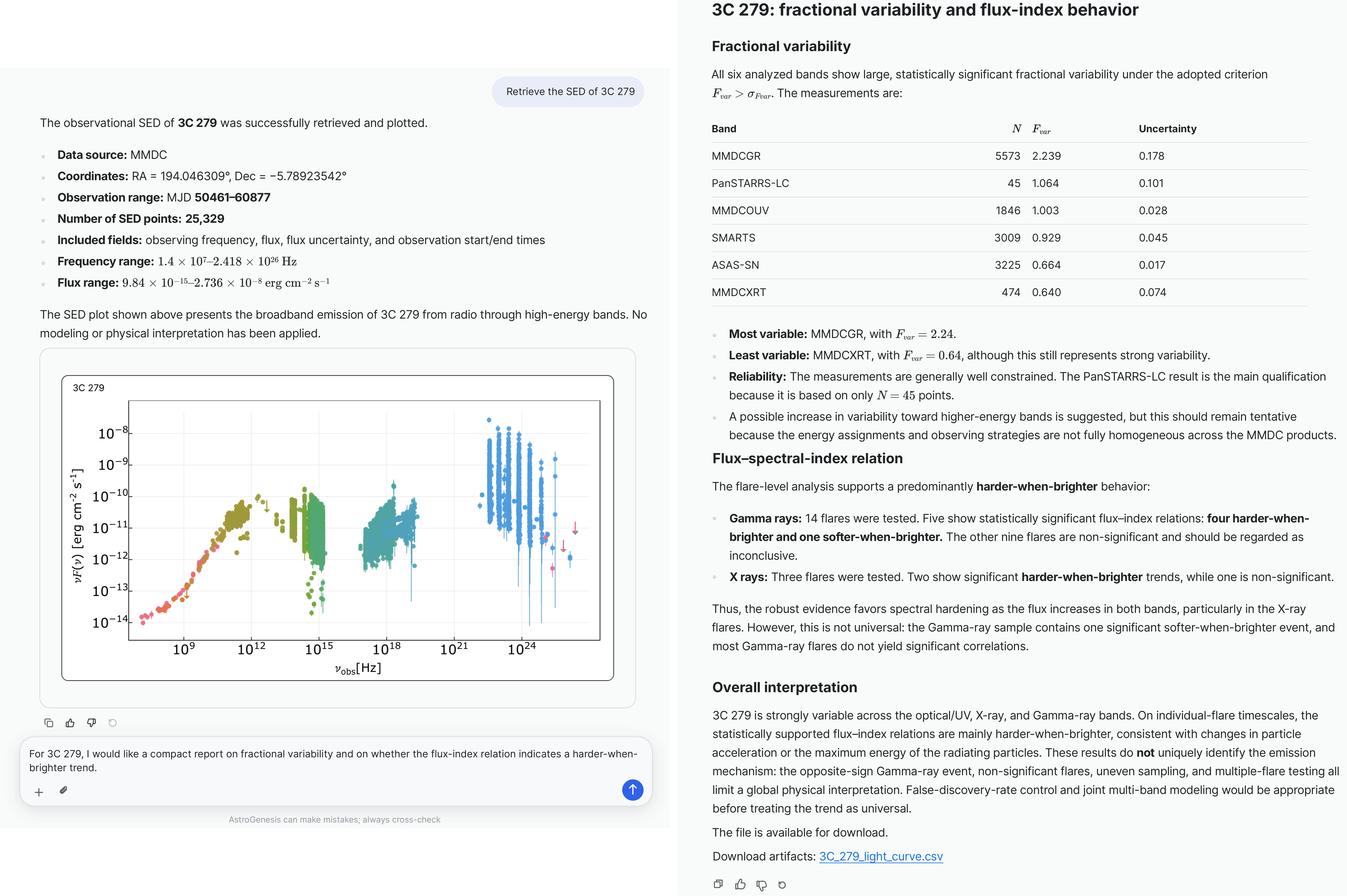}
    \caption{ Example of a data-retrieval response generated by \astrogen\ for the blazar 3C~279. The left panel shows the retrieval and visualization of the broadband observational SED, including basic source metadata and the retrieved measurements. The right panel shows a subsequent analysis of the retrieved data, in which \astrogen\ summarizes the fractional variability across the available bands and evaluates the flux--spectral-index relation during individual flaring episodes. The example illustrates how the Data Retrieval Agent can combine observational data access, visualization, quantitative analysis, and a concise scientific interpretation within a single conversational workflow.}
    \label{request_data}
\end{figure}

The workflow begins with the source resolution. The system identifies the target object (3C~279) and determines its sky coordinates (RA $\simeq 194.0463^\circ$, Dec $\simeq -5.7892^\circ$), which are used to query the underlying observational database. After the relevant measurements have been retrieved, the system automatically constructs a broadband SED from the collected dataset. The resulting plot is generated directly from the observational data and delivered to the front-end interface for visualization. An example of the generated output is shown in Figure~\ref{request_data}. The visualization is interactive, allowing users to zoom into specific energy ranges and inspect individual measurements in greater detail.

Once the broadband SED has been constructed, the retrieved dataset can serve as the basis for additional investigations tailored to the scientific objectives of the user. In this sense, the initial SED request represents the starting point of a broader analysis workflow rather than its final product. For example, the user can request an SED corresponding to a restricted time interval in order to isolate a particular activity state, flare episode, or observing period and thereby construct a time-resolved spectral dataset.

Beyond modifying the temporal selection, if requested additional analyses through specialized agents coordinated by the Supervisor Agent can be performed. One example is shown Figure ~\ref{request_data} where \textit{``For 3C 279, I would like a compact report on fractional variability and on whether the flux-index relation indicates a harder-when-brighter trend.''} query was sent to \astrogen. Initially the sub-agent evaluates the fractional variability amplitude, $F_{\rm var}$, across different wavebands while accounting for measurement uncertainties. Then, the sub-agent examines whether higher flux states are associated with systematically harder or softer spectra in the X-ray and $\gamma$-ray bands. For more details on these two agents, see Section \ref{sec:data_retrieval_agent}. The outputs of these analyses are aggregated into a structured response that combines quantitative results with scientific interpretation. The results are presented through the chat interface in a structured format, including tables, statistical summaries, and concise interpretations of the observed source behavior (see the right-hand panel of Figure~\ref{request_data}).

This example demonstrates that \astrogen\ extends beyond observational data retrieval by supporting iterative analysis workflows. Users can refine temporal selections, request additional diagnostics, and obtain both textual and visual outputs within a unified multi-agent environment.

\subsection{Theoretical Modeling}
\label{sec:theoretical_modeling_example}

To illustrate the theoretical modeling capabilities of \astrogen, we consider a representative request in which the user asks the system to interpret a time-resolved broadband SED within a physical emission model:

\textbf{Query.}
\textit{``Obtain the SED of PKS 2155$-$304 during 10--14 May 2010 and model it within the synchrotron self-Compton (SSC) scenario.''}

This task requires several steps, including retrieval of the observational SED for the requested epoch, identification of the relevant source properties, construction of a radiative model, and estimation of the physical parameters describing the emitting region. When the request is received,  the Supervisor Agent classifies the request as a theoretical-modeling task and forwards it to the Planner Agent. The Planner decomposes the workflow into \textit{(i)} time-resolved SED retrieval, \textit{(ii)} source metadata extraction, \textit{(iii)} model setup, \textit{(iv)} parameter estimation, and (v) response generation.

The workflow starts by retrieving the observational SED of PKS 2155$-$304 over the interval MJD 55326--55330, corresponding to 10--14 May 2010. The resulting dataset contains broadband flux measurements and associated uncertainties spanning approximately $10^{13}$--$10^{25}\mathrm{Hz}$. The system then automatically retrieves the source redshift, $z=0.116$, and combines this information with the observational dataset to construct the input required for sending theoretical modeling. The assembled data are subsequently passed to the \texttt{MMDC} \citep{2024AJ....168..289S}, which performs parameter estimation within a single-zone SSC framework.

\begin{figure}
    \centering
    \includegraphics[width=1\textwidth]{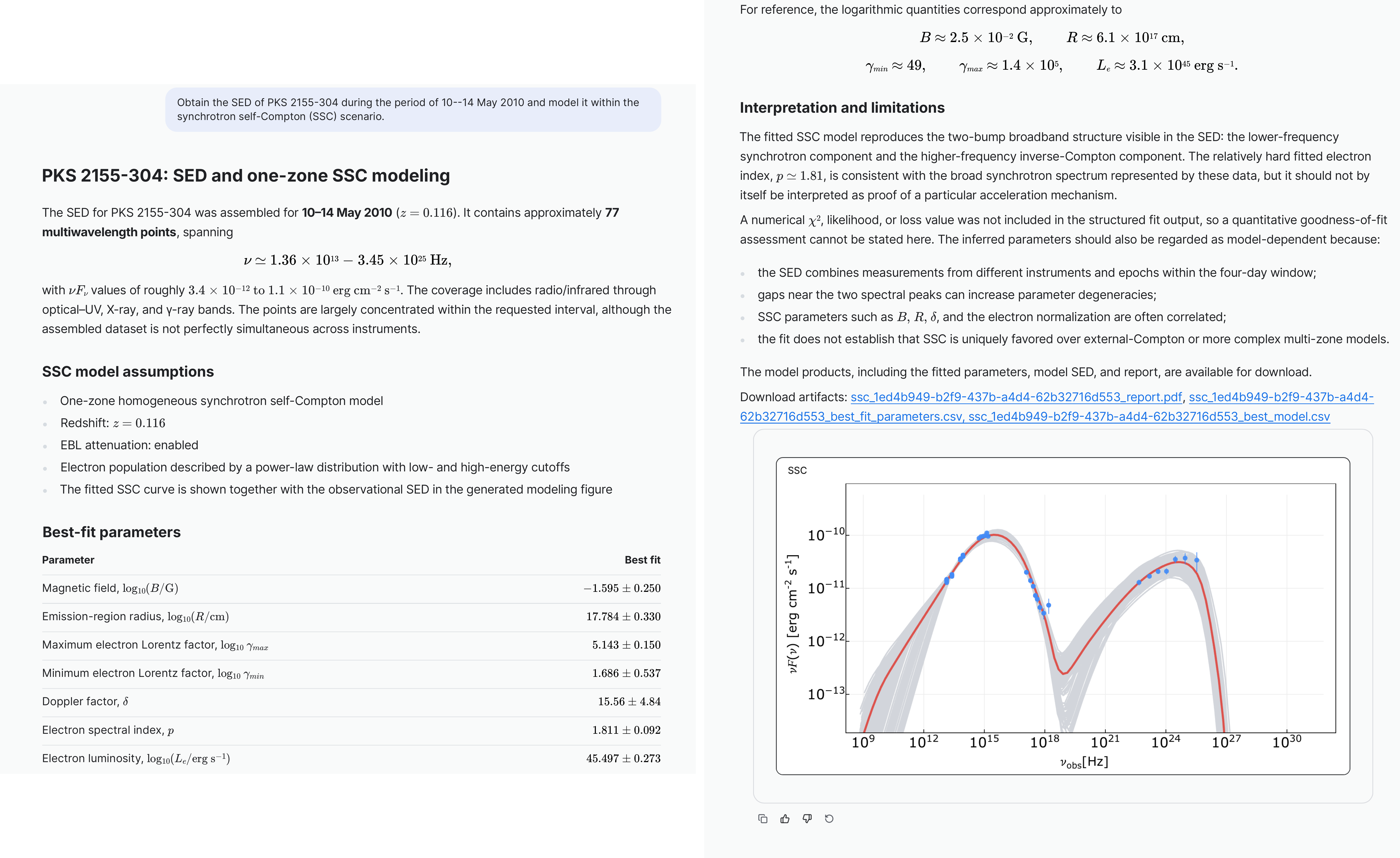}
    \caption{Example of a theoretical-modeling response generated by \astrogen\ for PKS\,2155$-$304 for the data accumulated during 10--14 May 2010. The interface shows the retrieved time-resolved SED together with the best-fit single-zone synchrotron self-Compton model, the inferred physical parameters, and downloadable fitting products.}
    \label{theoretical_modeling}
\end{figure}

An example of the user query and the corresponding response is shown in Figure~\ref{theoretical_modeling}. In addition to the numerical parameter summary, the response includes a visualization of the best-fit model overlaid on the retrieved observational data. This figure is displayed within the chat interface alongside the textual summary, while downloadable artifacts provide the fitted spectrum and parameter tables for further analysis. Before the final response is returned, the Replanner verifies that all retrieval, modeling, and response-generation steps have been successfully completed.

In this workflow, the modeling stage serves as an interpretive layer that translates the observed emission state into a physically motivated set of source parameters. By combining automated data retrieval with radiative-model fitting, \astrogen\ enables users to move from science-ready observations, either provided by \texttt{MMDC} or supplied by themselves,  to quantitative physical interpretation through natural-language interaction alone. The entire workflow, from data retrieval and model selection to parameter estimation and result presentation, is initiated and controlled through conversational requests, avoiding the need for direct interaction with specialized modeling software or analysis pipelines.

\subsection{Research Ideation}
To illustrate the research-ideation capabilities of \astrogen, we consider a representative question in which the user asks the system to identify potentially unexplained trends in a well-studied blazar and suggest promising directions for future investigation:

\textbf{Query.}
\textit{``Investigate the multiwavelength variability of 3C 279 and propose research ideas that could help distinguish between competing explanations for its extreme gamma-ray flares. Provide a focused and conceive summary.''}

This task requires more than literature retrieval or observational data analysis alone. The system must identify potentially unresolved observational patterns, compare them with the context of existing scientific knowledge, generate physically motivated hypotheses, and evaluate whether the resulting ideas are both novel and scientifically testable. When receiving the query, the Supervisor Agent classifies the request as an ideation task and forwards it to the Planner Agent. The Planner splits the workflow into \textit{(i)} observational data retrieval, \textit{(ii)} literature-context retrieval, \textit{(iii)} observational-diagnostic analysis, \textit{(iv)} hypothesis generation, \textit{(v)} novelty assessment, \textit{(vi)} feasibility evaluation, and \textit{(vii)} response construction. The implementation of these steps requires requesting several specialized agents, including the Data Retrieval Agent, Flare Segmentation Sub-agent, Fractional Variability Sub-agent, Novelty Agent, Feasibility Agent, Replanner, and Supervisor Agent.

\begin{figure}
    \centering
    \includegraphics[width=1\linewidth]{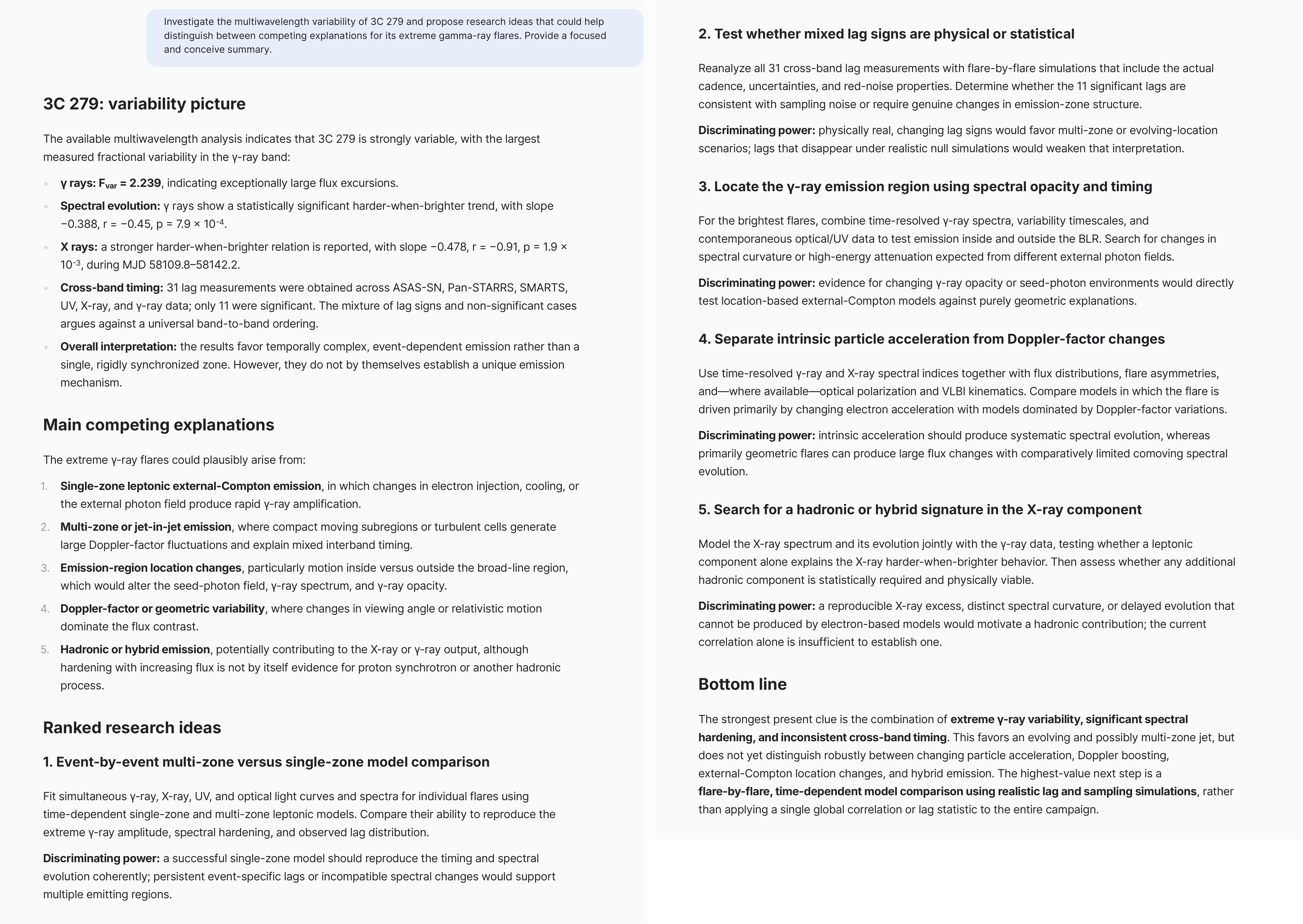}
    \caption{Example of a research-ideation response generated by \astrogen\ for 3C 279. Using retrieved multiwavelength observations and literature context, the system identifies unresolved observational trends and proposes scientifically motivated research directions. Each suggestion is accompanied by its motivation, methodological approach, and potential scientific impact.}
    \label{research_ideation}
\end{figure}

The generated ideation output is shown in Figure~\ref{research_ideation}. The process starts by retrieving the available observational datasets for 3C 279, including the broadband SED and multiwavelength light curves. In parallel, the system retrieves literature relevant to 3C 279. The retrieved observational data and literature are then jointly evaluated to identify potential research directions. The observational diagnostics derived from the specialized analysis agents are then considered. In particular, the fractional-variability analysis reveals strong variability across the available bands, with the largest amplitude measured in the $\gamma$-ray band. The cross-band timing analysis yields 31 lag measurements, of which only 11 are statistically significant. Moreover, the mixture of positive and negative lags, together with the large number of non-significant measurements, does not indicate a consistent band-to-band temporal ordering. These results point toward a complex, event-dependent variability pattern rather than a simple, universally synchronized emission scenario. Using the combined literature and observational context, the Novelty Agent identifies the combination of extreme $\gamma$-ray variability, significant spectral hardening, and heterogeneous cross-band timing in 3C 279 as a key unresolved problem and generates candidate research directions together with their scientific motivation and observational predictions. The agent proposes several testable directions aimed at distinguishing between competing interpretations, including single-zone versus multi-zone emission, changes in the location of the $\gamma$-ray emitting region, intrinsic particle acceleration versus Doppler-factor variations, and possible hadronic or hybrid contributions. It then formulates concrete observational and modeling strategies, with particular emphasis on flare-by-flare, time-dependent analyses combining multiwavelength spectral and timing information to discriminate among these scenarios.

This example demonstrates that \astrogen\ treats research ideation as a structured scientific workflow rather than an unconstrained brainstorming exercise. Through natural-language interaction, the system combines literature retrieval, observational data analysis, novelty assessment, and feasibility evaluation within a coordinated multi-agent framework. The generated ideas are grounded in both observational evidence and existing scientific knowledge and are for starting points for further investigation rather than established physical conclusions.

\subsection{Future Developments and Planned Functionality}

The current version of \astrogen\ is the first stage of a broader development plan for extending the framework to additional astrophysical source classes, observational data levels, and modeling capabilities. The planned development is in several phases, with each phase extending the existing architecture while preserving the same modular structure of DSRMs, specialized agents, data services, and modeling tools.

The first phase is the current blazar-focused implementation described in this work. It combines literature retrieval and analysis, science-ready observational data access, neural network-based emission modeling, and coordinated multi-agent workflows within a single DSRM. This phase establishes the main infrastructure of \astrogen\ and provides the basic framework required for the integration of additional data products, modeling methods, and scientific analysis tools. The blazar DSRM serves both as the first scientific application of the system and as the reference implementation for the development of future modules.

An intermediate extension of this phase will be the integration of raw observational data pipelines. In the current implementation, the Data Retrieval Agent is designed to operate on science-ready data products available through \texttt{MMDC} and other public databases. The planned extension will allow the system to access instrument-level observational data directly from mission and observatory archives and route them to specialized data-analysis agents. These agents will apply instrument-specific reduction and calibration procedures, including standard preprocessing, event or image filtering, background estimation, calibration, and quality-control steps, as appropriate for each instrument. The resulting calibrated, science-ready products will then be passed to the existing analysis and modeling components of \astrogen. This extension would therefore allow the framework to support a more complete workflow, starting from archival observations and progressing through data reduction and calibration to scientific analysis and physical interpretation.

The subsequent phases will extend \astrogen\ beyond blazars to three major classes of transient sources: GRBs, TDEs, and FRBs. Dedicated DSRMs will be developed for each source class following the general multi-agent architecture defined for blazars, while integrating source-specific literature collections, observational databases and archives, analysis procedures, and pretrained theoretical models. The individual agents and their interactions will be adapted to the different observational characteristics, relevant physical processes, and characteristic timescales of each transient class. For GRBs, the main focus will be on the rapid and highly variable evolution of the emission, requiring time-dependent data retrieval, temporal and spectral analysis, and the combination of heterogeneous observations from multiple instruments covering the prompt and afterglow phases. For TDEs, the framework will instead address evolution over substantially longer timescales, supporting the construction and analysis of long-term multiwavelength light curves, spectral evolution, and the combination of observations obtained across different epochs and wavelength bands. The FRB DSRM will focus on millisecond-duration radio transients and will implement analysis of burst temporal and spectral properties, polarization, dispersion and rotation measures, repeating activity, and associations with host galaxies or counterparts at other wavelengths. The corresponding theoretical modeling components will be specialized for each source class, allowing the observed temporal and spectral properties to be connected with the relevant physical scenarios and emission mechanisms. Extending \astrogen\ across GRBs, TDEs, and FRBs will therefore provide an important test of the generality and flexibility of the proposed multi-agent architecture, spanning transient phenomena with characteristic timescales ranging from milliseconds to months or years and with substantially different observational requirements and underlying physical processes.

The longer-term objective is to develop \astrogen\ into a general astrophysical research infrastructure in which literature analysis, observational data retrieval, raw-data processing, statistical analysis, and theoretical modeling can be performed within a common environment. Different DSRMs will provide source-specific knowledge, data resources, analysis tools, and modeling capabilities while sharing the same underlying orchestration and interaction framework. In such a configuration, researchers will be able to move from the formulation of a scientific question to literature retrieval, data identification and processing, statistical characterization, model execution, and physical interpretation without requiring independent workflows for each stage.

An important component of this development will be the introduction of specialized agents designed to interact with current and planned large-scale sky surveys and time-domain facilities, for example Nancy Grace Roman Space Telescope (Roman)\footnote{\url{https://science.nasa.gov/mission/roman-space-telescope/}}, the Zwicky
Transient Facility \citep[][]{2019PASP..131a8002B},  the Vera C. Rubin Observatory \citep{2019ApJ...873..111I} and facilities being developed within the Eric and Wendy Schmidt Observatory System \footnote{\url{https://www.schmidtsciences.org/schmidt-observatory-system/}}, such as the Argus Array, Deep Synoptic Array (DSA), and Lazuli Space Observatory \citep{2019ApJ...873..111I,2022SPIE12182E..4HL, 2019BAAS...51g.255H, 2026arXiv260102556R}. Modern surveys are producing increasingly large and rapidly evolving data in which potentially interesting astrophysical events must be identified and characterized on timescales much shorter than is practical through manual analysis. Dedicated \astrogen\ agents will therefore be developed to ingest survey products and alerts, perform rapid initial analysis and classification, identify potentially interesting or anomalous sources, and place new detections in the context of existing archival observations and scientific literature. These agents could also coordinate follow-up analysis across different facilities and wavelength bands, allowing newly detected events to be connected rapidly with available multiwavelength and, where relevant, multimessenger information. The integration of pretrained classification and foundation models will further allow specialized tasks, such as transient classification, light-curve characterization, source identification, and anomaly detection, to be incorporated directly into these workflows.

This architecture would allow \astrogen\ to evolve from a system primarily responding to researcher-defined queries into a framework that can also interact with continuously changing observational data streams. By combining survey-driven event identification with literature knowledge, archival data, automated analysis, and physical modeling, the system could support rapid scientific assessment of newly detected sources and help prioritize those requiring more detailed analysis or follow-up observations. In this way, the same multi-agent infrastructure could support both targeted investigations of individual astrophysical sources and data-intensive discovery workflows associated with the next generation of large astronomical surveys.

\subsubsection{Planned Functionality}
In addition to the development of new DSRMs, the functionalities of \astrogen\ will be gradually expanded by introducing additional agents and tools that support common research tasks. These capabilities will be added independently to the existing architecture and made available to the Supervisor Agent when required by a user request. The purpose is to extend the framework beyond the current literature, data-analysis, and modeling workflows while keeping the individual functions modular.

One planned capability is direct analysis of user-provided scientific references. When a DOI, arXiv identifier, ADS reference, or link to a scientific publication is provided, a dedicated agent will resolve the reference, retrieve the corresponding paper when accessible, and answer questions using the content of that specific publication. This will allow researchers to inspect individual papers, request explanations of particular results or methods, and obtain responses directly grounded in the selected source rather than in a broader literature search while maintaining context-aware knowledge of the field for critical analysis.

Another planned functionality is automated monitoring of newly released scientific literature. For example, an agent will be able to inspect the arXiv submissions for a selected day or scientific category, identify papers relevant to a research topic, and provide a concise summary of the most relevant new results. Such functionality can also be adapted to follow selected source classes, methods, instruments, or scientific topics, providing researchers with a simple way to follow new developments without manually reviewing large numbers of publications.

The framework will also support the integration of additional pretrained and foundation models developed for specific astronomical tasks. These may include models such as SELDON for transient and supernova light-curve analysis \citep{2026arXiv260304392W}, AstroCLIP for multimodal image and spectral analysis \citep{2024MNRAS.531.4990P}, and other models for galaxy classification, source characterization, time-series analysis, and related applications. Rather than reproducing these capabilities internally, \astrogen can provide a common natural-language interface through which suitable external models are selected and executed as part of a scientific workflow. New models can therefore be incorporated as they become available, provided that their inputs, outputs, and execution procedures can be connected to the agent framework.

A further planned extension of the theoretical-modeling capabilities is the integration of JetSeT \footnote{\url{https://github.com/andreatramacere/jetset}}, \citep{2020ascl.soft09001T, 2011ApJ...739...66T, 2009A&A...501..879T} which will provide access to a more general framework for modeling and fitting emission from astrophysical sources. While the current implementation focuses on the pre-trained neural networks, JetSeT will allow \astrogen\ to construct and fit more flexible models, including combinations of predefined components and user-defined functions. Through a dedicated interface, researchers will be able to specify the model or functional form to be fitted, define parameters and constraints, provide observational data, and request the corresponding optimization or parameter-estimation procedure. This will extend the modeling functionality beyond a fixed set of emission scenarios and provide a more general connection between natural-language scientific requests and configurable fitting workflows. The integration will also allow additional models and functions to be incorporated without requiring modifications to the overall agent architecture, providing a flexible basis for progressively expanding the range of theoretical and phenomenological analyses supported by \astrogen.

A further planned direction is the integration of numerical simulation frameworks. Modern astrophysical research makes extensive use of N-body, hydrodynamic and magnetohydrodynamic, stellar-evolution, numerical-relativity, Monte Carlo, and particle-in-cell simulations \citep[see, e.g.,][]{1998ARA&A..36..599B,2011ApJS..192....3P,2014ARA&A..52..661L,2019FrASS...6...51T,2019LRCA....5....1N,2006PhPl...13e5904L}. However, the use of these frameworks often requires substantial effort for code selection, configuration, compilation, deployment, and adaptation to a specific scientific problem. Future versions of \astrogen\ will include specialized agents designed to assist researchers throughout this process, from selecting an appropriate simulation framework and translating scientific objectives into numerical problem definitions to preparing input configurations and executing simulations on available computational resources. These agents will combine domain knowledge of the underlying astrophysical problem with detailed information about the supported simulation codes, including their architecture, APIs, configuration systems, build procedures, documentation, tutorials, and established workflows. The framework will also incorporate knowledge of common high-performance computing environments, including compiler toolchains, MPI implementations, accelerator backends such as Kokkos, AMReX, and RAJA, job schedulers, and facility-specific execution requirements. This will provide a direct connection between natural-language scientific objectives and reproducible numerical simulation workflows, allowing \astrogen\ to generate code- and infrastructure-aware configurations rather than generic programming recommendations. By combining astrophysical, software-specific, and computational knowledge within the same agent framework, \astrogen\ is intended to simplify the transition from the formulation of a numerical experiment to its practical implementation and execution on modern computing infrastructure.

More generally, the set of agents available within \astrogen\ is intended to evolve together with the needs of astrophysical research and the development of new computational tools. Additional agents can be introduced for specialized data analysis, access to new archives, statistical methods, external models, simulation codes, or other research operations without changing the overall architecture of the system. In this way, \astrogen\ is intended to develop into a research assistant through which researchers can access a growing collection of scientific capabilities within a common environment. By reducing the need to move manually between literature services, observational archives, analysis software, pretrained models, and numerical tools, the framework aims to make complex astrophysical research workflows more efficient, connected, and reproducible.

\section{Summary}
In this paper, we have presented \astrogen, a domain-specific multi-agent AI framework designed to integrate scientific literature, observational data, data analysis, theoretical modeling, and research ideation within a unified environment for astrophysical research. The framework is organized around DSRMs, which encapsulate the knowledge bases, data resources, modeling tools, and specialized agents required for particular classes of astrophysical sources. This modular architecture allows individual scientific capabilities and source-specific modules to be developed and extended independently while sharing a common orchestration framework.

The current implementation, \astrogen\ version 1.0, focuses on blazar research and integrates several specialized capabilities within a common multi-agent workflow. A central and particularly advanced component is the Theoretical Modeling Agent, which connects natural-language scientific requests directly to pretrained neural-network surrogate models of physically motivated blazar emission scenarios. The current modeling infrastructure supports SSC, EIC, hadronic, and hybrid lepto-hadronic models and combines these surrogates with Bayesian parameter inference, allowing observational SEDs to be fitted and physical source parameters to be estimated efficiently. In particular, the use of pretrained surrogates makes computationally demanding hadronic and multimessenger modeling accessible within an interactive workflow, including the joint treatment of electromagnetic and neutrino constraints. This modeling capability is complemented by a literature-retrieval system based on hybrid dense--sparse search, citation expansion, and cross-encoder reranking; automated access to science-ready multiwavelength observations; dedicated tools for variability, flare, and spectral-trend analysis; and a Research Ideation Agent that combines literature and observational diagnostics to identify candidate research directions. These components are coordinated through the Supervisor Agent and Planner/Replanner architecture, enabling multi-stage research workflows to be initiated and refined through natural-language interaction.

The literature-retrieval component was evaluated using complementary single-paper and multi-paper benchmarks, demonstrating strong performance across both targeted and synthesis-oriented scientific queries. In the single-paper benchmark, the system achieved a routing recall of 0.846 and a pool recall of 0.962, with the target publication retrieved within the top five results for 76.6\% of the evaluated questions and within the top ten for 80.9\%. In the more demanding multi-paper benchmark, routing recall reached 0.945, while at least one primary relevant publication was recovered within the top five results for 79.2\% of questions and within the top ten for 86.4\%. These results demonstrate that the retrieval architecture can reliably identify highly relevant scientific literature within a small set of returned publications, including for questions requiring evidence from multiple studies. The multi-paper benchmark further yielded a Recall\@10 of 0.497 and Evidence Coverage\@10 of 0.354, reflecting the greater challenge of recovering complete evidence sets for scientific synthesis. The substantial increase in recall at larger retrieval depths shows that much of this additional evidence is already present deeper in the ranked candidate set, providing a clear path for further improvement through enhanced ranking and adaptive retrieval strategies.

The end-to-end examples demonstrate how the same framework can connect literature retrieval, observational data access, statistical analysis, physical modeling, and hypothesis generation within persistent and traceable research workflows. The present implementation should therefore be regarded as a first step toward a broader astrophysical research environment rather than a replacement for scientific judgment. Future development will extend the framework to additional source classes, including GRBs, TDEs, and FRBs, and incorporate raw-data processing, additional modeling and fitting frameworks, pretrained scientific models, survey-driven workflows, and numerical simulation tools. The overall objective is to provide a modular interface through which increasingly complex astrophysical research workflows can be executed in a more connected, reproducible, and efficient manner.

\begin{acknowledgments}
N.S. acknowledges support from the Higher Education and Science Committee of MESCS RA (Research Project No. 23LCG-1C004).

This research has made use of NASA's Astrophysics Data System
Bibliographic Services.
\end{acknowledgments}

\bibliography{sample701}{}

\begin{thebibliography}{}
\expandafter\ifx\csname natexlab\endcsname\relax\def\natexlab#1{#1}\fi
\providecommand{\url}[1]{\href{#1}{#1}}
\providecommand{\dodoi}[1]{doi:~\href{http://doi.org/#1}{\nolinkurl{#1}}}
\providecommand{\doeprint}[1]{\href{http://ascl.net/#1}{\nolinkurl{http://ascl.net/#1}}}
\providecommand{\doarXiv}[1]{\href{https://arxiv.org/abs/#1}{\nolinkurl{https://arxiv.org/abs/#1}}}

\bibitem[{B.~P. {Abbott} {et~al.}(2017){Abbott}, {Abbott}, {Abbott},
  {Acernese}, {Ackley}, {Adams}, {Adams}, {Addesso}, {Adhikari}, {Adya},
  {et~al.}}]{2017ApJ...848L..12A}
{Abbott}, B.~P., {Abbott}, R., {Abbott}, T.~D., {et~al.} 2017,
  \bibinfo{title}{{Multi-messenger Observations of a Binary Neutron Star
  Merger},} \apjl, 848, L12, \dodoi{10.3847/2041-8213/aa91c9}

\bibitem[{P.~D. {Aleo} {et~al.}(2024){Aleo}, {Engel}, {Narayan}, {Angus},
  {Malanchev}, {Auchettl}, {Baldassare}, {Berres}, {de Boer}, {Boyd},
  {Chambers}, {Davis}, {Esquivel}, {Farias}, {Foley}, {Gagliano}, {Gall},
  {Gao}, {Gomez}, {Grayling}, {Jones}, {Lin}, {Magnier}, {Mandel}, {Matheson},
  {Raimundo}, {Shah}, {Soraisam}, {de Soto}, {Vicencio}, {Villar}, \&
  {Wainscoat}}]{2024ApJ...974..172A}
{Aleo}, P.~D., {Engel}, A.~W., {Narayan}, G., {et~al.} 2024,
  \bibinfo{title}{{Anomaly Detection and Approximate Similarity Searches of
  Transients in Real-time Data Streams},} \apj, 974, 172,
  \dodoi{10.3847/1538-4357/ad6869}

\bibitem[{J. {Alsing} {et~al.}(2019){Alsing}, {Charnock}, {Feeney}, \&
  {Wandelt}}]{2019MNRAS.488.4440A}
{Alsing}, J., {Charnock}, T., {Feeney}, S., \& {Wandelt}, B. 2019,
  \bibinfo{title}{{Fast likelihood-free cosmology with neural density
  estimators and active learning},} \mnras, 488, 4440,
  \dodoi{10.1093/mnras/stz1960}

\bibitem[{K. {Anas Apurba} {et~al.}(2026){Anas Apurba}, {Hasibul Hasan}, {Alam
  Shehab}, \& {Azad}}]{2026arXiv260803860A}
{Anas Apurba}, K., {Hasibul Hasan}, M., {Alam Shehab}, R., \& {Azad}, A. 2026,
  \bibinfo{title}{{SciRet: A Compute-Aware Empirical Study of Retrieval and
  Reranking for Scientific RAG},} arXiv e-prints, arXiv:2608.03860,
  \dodoi{10.48550/arXiv.2608.03860}

\bibitem[{A.~K. {Aniyan} \& K. {Thorat}(2017){Aniyan} \&
  {Thorat}}]{2017ApJS..230...20A}
{Aniyan}, A.~K., \& {Thorat}, K. 2017, \bibinfo{title}{{Classifying Radio
  Galaxies with the Convolutional Neural Network},} \apjs, 230, 20,
  \dodoi{10.3847/1538-4365/aa7333}

\bibitem[{ {Astropy Collaboration} {et~al.}(2013){Astropy Collaboration},
  {Robitaille}, {Tollerud}, {Greenfield}, {Droettboom}, {Bray}, {Aldcroft},
  {Davis}, {Ginsburg}, {Price-Whelan}, {Kerzendorf}, {Conley}, {Crighton},
  {Barbary}, {Muna}, {Ferguson}, {Grollier}, {Parikh}, {Nair}, {Unther},
  {Deil}, {Woillez}, {Conseil}, {Kramer}, {Turner}, {Singer}, {Fox}, {Weaver},
  {Zabalza}, {Edwards}, {Azalee Bostroem}, {Burke}, {Casey}, {Crawford},
  {Dencheva}, {Ely}, {Jenness}, {Labrie}, {Lim}, {Pierfederici}, {Pontzen},
  {Ptak}, {Refsdal}, {Servillat}, \& {Streicher}}]{2013A&A...558A..33A}
{Astropy Collaboration}, {Robitaille}, T.~P., {Tollerud}, E.~J., {et~al.} 2013,
  \bibinfo{title}{{Astropy: A community Python package for astronomy},} \aap,
  558, A33, \dodoi{10.1051/0004-6361/201322068}

\bibitem[{ {Astropy Collaboration} {et~al.}(2018){Astropy Collaboration},
  {Price-Whelan}, {Sip{\H{o}}cz}, {G{\"u}nther}, {Lim}, {Crawford}, {Conseil},
  {Shupe}, {Craig}, {Dencheva}, {Ginsburg}, {VanderPlas}, {Bradley},
  {P{\'e}rez-Su{\'a}rez}, {de Val-Borro}, {Aldcroft}, {Cruz}, {Robitaille},
  {Tollerud}, {Ardelean}, {Babej}, {Bach}, {Bachetti}, {Bakanov}, {Bamford},
  {Barentsen}, {Barmby}, {Baumbach}, {Berry}, {Biscani}, {Boquien}, {Bostroem},
  {Bouma}, {Brammer}, {Bray}, {Breytenbach}, {Buddelmeijer}, {Burke},
  {Calderone}, {Cano Rodr{\'\i}guez}, {Cara}, {Cardoso}, {Cheedella}, {Copin},
  {Corrales}, {Crichton}, {D'Avella}, {Deil}, {Depagne}, {Dietrich}, {Donath},
  {Droettboom}, {Earl}, {Erben}, {Fabbro}, {Ferreira}, {Finethy}, {Fox},
  {Garrison}, {Gibbons}, {Goldstein}, {Gommers}, {Greco}, {Greenfield},
  {Groener}, {Grollier}, {Hagen}, {Hirst}, {Homeier}, {Horton}, {Hosseinzadeh},
  {Hu}, {Hunkeler}, {Ivezi{\'c}}, {Jain}, {Jenness}, {Kanarek}, {Kendrew},
  {Kern}, {Kerzendorf}, {Khvalko}, {King}, {Kirkby}, {Kulkarni}, {Kumar},
  {Lee}, {Lenz}, {Littlefair}, {Ma}, {Macleod}, {Mastropietro}, {McCully},
  {Montagnac}, {Morris}, {Mueller}, {Mumford}, {Muna}, {Murphy}, {Nelson},
  {Nguyen}, {Ninan}, {N{\"o}the}, {Ogaz}, {Oh}, {Parejko}, {Parley}, {Pascual},
  {Patil}, {Patil}, {Plunkett}, {Prochaska}, {Rastogi}, {Reddy Janga},
  {Sabater}, {Sakurikar}, {Seifert}, {Sherbert}, {Sherwood-Taylor}, {Shih},
  {Sick}, {Silbiger}, {Singanamalla}, {Singer}, {Sladen}, {Sooley},
  {Sornarajah}, {Streicher}, {Teuben}, {Thomas}, {Tremblay}, {Turner},
  {Terr{\'o}n}, {van Kerkwijk}, {de la Vega}, {Watkins}, {Weaver}, {Whitmore},
  {Woillez}, {Zabalza}, \& {Astropy Contributors}}]{2018AJ....156..123A}
{Astropy Collaboration}, {Price-Whelan}, A.~M., {Sip{\H{o}}cz}, B.~M., {et~al.}
  2018, \bibinfo{title}{{The Astropy Project: Building an Open-science Project
  and Status of the v2.0 Core Package},} \aj, 156, 123,
  \dodoi{10.3847/1538-3881/aabc4f}

\bibitem[{ {Astropy Collaboration} {et~al.}(2022){Astropy Collaboration},
  {Price-Whelan}, {Lim}, {Earl}, {Starkman}, {Bradley}, {Shupe}, {Patil},
  {Corrales}, {Brasseur}, {N{\"o}the}, {Donath}, {Tollerud}, {Morris},
  {Ginsburg}, {Vaher}, {Weaver}, {Tocknell}, {Jamieson}, {van Kerkwijk},
  {Robitaille}, {Merry}, {Bachetti}, {G{\"u}nther}, {Aldcroft},
  {Alvarado-Montes}, {Archibald}, {B{\'o}di}, {Bapat}, {Barentsen},
  {Baz{\'a}n}, {Biswas}, {Boquien}, {Burke}, {Cara}, {Cara}, {Conroy},
  {Conseil}, {Craig}, {Cross}, {Cruz}, {D'Eugenio}, {Dencheva}, {Devillepoix},
  {Dietrich}, {Eigenbrot}, {Erben}, {Ferreira}, {Foreman-Mackey}, {Fox},
  {Freij}, {Garg}, {Geda}, {Glattly}, {Gondhalekar}, {Gordon}, {Grant},
  {Greenfield}, {Groener}, {Guest}, {Gurovich}, {Handberg}, {Hart},
  {Hatfield-Dodds}, {Homeier}, {Hosseinzadeh}, {Jenness}, {Jones}, {Joseph},
  {Kalmbach}, {Karamehmetoglu}, {Ka{\l}uszy{\'n}ski}, {Kelley}, {Kern},
  {Kerzendorf}, {Koch}, {Kulumani}, {Lee}, {Ly}, {Ma}, {MacBride}, {Maljaars},
  {Muna}, {Murphy}, {Norman}, {O'Steen}, {Oman}, {Pacifici}, {Pascual},
  {Pascual-Granado}, {Patil}, {Perren}, {Pickering}, {Rastogi}, {Roulston},
  {Ryan}, {Rykoff}, {Sabater}, {Sakurikar}, {Salgado}, {Sanghi}, {Saunders},
  {Savchenko}, {Schwardt}, {Seifert-Eckert}, {Shih}, {Jain}, {Shukla}, {Sick},
  {Simpson}, {Singanamalla}, {Singer}, {Singhal}, {Sinha}, {Sip{\H{o}}cz},
  {Spitler}, {Stansby}, {Streicher}, {{\v{S}}umak}, {Swinbank}, {Taranu},
  {Tewary}, {Tremblay}, {de Val-Borro}, {Van Kooten}, {Vasovi{\'c}}, {Verma},
  {de Miranda Cardoso}, {Williams}, {Wilson}, {Winkel}, {Wood-Vasey}, {Xue},
  {Yoachim}, {Zhang}, {Zonca}, \& {Astropy Project
  Contributors}}]{2022ApJ...935..167A}
{Astropy Collaboration}, {Price-Whelan}, A.~M., {Lim}, P.~L., {et~al.} 2022,
  \bibinfo{title}{{The Astropy Project: Sustaining and Growing a
  Community-oriented Open-source Project and the Latest Major Release (v5.0) of
  the Core Package},} \apj, 935, 167, \dodoi{10.3847/1538-4357/ac7c74}

\bibitem[{N.~M. {Ball} \& R.~J. {Brunner}(2010){Ball} \&
  {Brunner}}]{Ball2010DataMining}
{Ball}, N.~M., \& {Brunner}, R.~J. 2010, \bibinfo{title}{{Data Mining and
  Machine Learning in Astronomy},} International Journal of Modern Physics D,
  19, 1049, \dodoi{10.1142/S0218271810017160}

\bibitem[{P. {B{\'e}chard} \& O. {Marquez Ayala}(2024){B{\'e}chard} \& {Marquez
  Ayala}}]{2024arXiv240408189B}
{B{\'e}chard}, P., \& {Marquez Ayala}, O. 2024, \bibinfo{title}{{Reducing
  hallucination in structured outputs via Retrieval-Augmented Generation},}
  arXiv e-prints, arXiv:2404.08189, \dodoi{10.48550/arXiv.2404.08189}

\bibitem[{D. {B{\'e}gu{\'e}} {et~al.}(2024){B{\'e}gu{\'e}}, {Sahakyan},
  {Dereli-B{\'e}gu{\'e}}, {Giommi}, {Gasparyan}, {Khachatryan}, {Casotto}, \&
  {Pe'er}}]{2024ApJ...963...71B}
{B{\'e}gu{\'e}}, D., {Sahakyan}, N., {Dereli-B{\'e}gu{\'e}}, H., {et~al.} 2024,
  \bibinfo{title}{{Modeling Blazar Broadband Emission with a Convolutional
  Neural Network. I. Synchrotron Self-Compton Model},} \apj, 963, 71,
  \dodoi{10.3847/1538-4357/ad19cf}

\bibitem[{E.~C. {Bellm} {et~al.}(2019){Bellm}, {Kulkarni}, {Graham}, {Dekany},
  {Smith}, {Riddle}, {Masci}, {Helou}, {Prince}, {Adams}, {Barbarino},
  {Barlow}, {Bauer}, {Beck}, {Belicki}, {Biswas}, {Blagorodnova}, {Bodewits},
  {Bolin}, {Brinnel}, {Brooke}, {Bue}, {Bulla}, {Burruss}, {Cenko}, {Chang},
  {Connolly}, {Coughlin}, {Cromer}, {Cunningham}, {De}, {Delacroix}, {Desai},
  {Duev}, {Eadie}, {Farnham}, {Feeney}, {Feindt}, {Flynn}, {Franckowiak},
  {Frederick}, {Fremling}, {Gal-Yam}, {Gezari}, {Giomi}, {Goldstein},
  {Golkhou}, {Goobar}, {Groom}, {Hacopians}, {Hale}, {Henning}, {Ho}, {Hover},
  {Howell}, {Hung}, {Huppenkothen}, {Imel}, {Ip}, {Ivezi{\'c}}, {Jackson},
  {Jones}, {Juric}, {Kasliwal}, {Kaspi}, {Kaye}, {Kelley}, {Kowalski},
  {Kramer}, {Kupfer}, {Landry}, {Laher}, {Lee}, {Lin}, {Lin}, {Lunnan},
  {Giomi}, {Mahabal}, {Mao}, {Miller}, {Monkewitz}, {Murphy}, {Ngeow},
  {Nordin}, {Nugent}, {Ofek}, {Patterson}, {Penprase}, {Porter}, {Rauch},
  {Rebbapragada}, {Reiley}, {Rigault}, {Rodriguez}, {van Roestel}, {Rusholme},
  {van Santen}, {Schulze}, {Shupe}, {Singer}, {Soumagnac}, {Stein}, {Surace},
  {Sollerman}, {Szkody}, {Taddia}, {Terek}, {Van Sistine}, {van Velzen},
  {Vestrand}, {Walters}, {Ward}, {Ye}, {Yu}, {Yan}, \&
  {Zolkower}}]{2019PASP..131a8002B}
{Bellm}, E.~C., {Kulkarni}, S.~R., {Graham}, M.~J., {et~al.} 2019,
  \bibinfo{title}{{The Zwicky Transient Facility: System Overview, Performance,
  and First Results},} \pasp, 131, 018002, \dodoi{10.1088/1538-3873/aaecbe}

\bibitem[{E. {Bertschinger}(1998){Bertschinger}}]{1998ARA&A..36..599B}
{Bertschinger}, E. 1998, \bibinfo{title}{{Simulations of Structure Formation in
  the Universe},} \araa, 36, 599, \dodoi{10.1146/annurev.astro.36.1.599}

\bibitem[{O.~M. {Boersma} \& J. {van Leeuwen}(2023){Boersma} \& {van
  Leeuwen}}]{BvL23}
{Boersma}, O.~M., \& {van Leeuwen}, J. 2023, \bibinfo{title}{{DeepGlow: An
  efficient neural network emulator of physical afterglow models for gamma-ray
  bursts and gravitational-wave events},} \pasa, 40, e030,
  \dodoi{10.1017/pasa.2023.32}

\bibitem[{F. {Bonnarel} {et~al.}(2000){Bonnarel}, {Fernique}, {Bienaym{\'e}},
  {Egret}, {Genova}, {Louys}, {Ochsenbein}, {Wenger}, \&
  {Bartlett}}]{2000A&AS..143...33B}
{Bonnarel}, F., {Fernique}, P., {Bienaym{\'e}}, O., {et~al.} 2000,
  \bibinfo{title}{{The ALADIN interactive sky atlas. A reference tool for
  identification of astronomical sources},} \aaps, 143, 33,
  \dodoi{10.1051/aas:2000331}

\bibitem[{L. {Bornmann} {et~al.}(2021){Bornmann}, {Haunschild}, \&
  {Mutz}}]{Bornmann2021ScienceGrowth}
{Bornmann}, L., {Haunschild}, R., \& {Mutz}, R. 2021, \bibinfo{title}{{Growth
  rates of modern science: a latent piecewise growth curve approach to model
  publication numbers from established and new literature databases},}
  Humanities and Social Sciences Communications, 8, 224,
  \dodoi{10.1057/s41599-021-00903-w}

\bibitem[{S. {Bruch} {et~al.}(2022){Bruch}, {Gai}, \&
  {Ingber}}]{2022arXiv221011934B}
{Bruch}, S., {Gai}, S., \& {Ingber}, A. 2022, \bibinfo{title}{{An Analysis of
  Fusion Functions for Hybrid Retrieval},} arXiv e-prints, arXiv:2210.11934,
  \dodoi{10.48550/arXiv.2210.11934}

\bibitem[{T. {Charnock} \& A. {Moss}(2017){Charnock} \&
  {Moss}}]{2017ApJ...837L..28C}
{Charnock}, T., \& {Moss}, A. 2017, \bibinfo{title}{{Deep Recurrent Neural
  Networks for Supernovae Classification},} \apjl, 837, L28,
  \dodoi{10.3847/2041-8213/aa603d}

\bibitem[{H. {Chen} {et~al.}(2025){Chen}, {Long}, {Xiao}, {Luo}, {Ju}, {Wang},
  {Wang}, {Zhou}, \& {Zhu}}]{2025arXiv250808742C}
{Chen}, H., {Long}, Q., {Xiao}, M., {et~al.} 2025,
  \bibinfo{title}{{SciRerankBench: Benchmarking Rerankers Towards Scientific
  Retrieval-Augmented Generated LLMs},} arXiv e-prints, arXiv:2508.08742,
  \dodoi{10.48550/arXiv.2508.08742}

\bibitem[{G.~V. Cormack {et~al.}(2009)Cormack, Clarke, \&
  Buettcher}]{10.1145/1571941.1572114}
Cormack, G.~V., Clarke, C. L.~A., \& Buettcher, S. 2009,
  \bibinfo{title}{Reciprocal rank fusion outperforms condorcet and individual
  rank learning methods,} in Proceedings of the 32nd International ACM SIGIR
  Conference on Research and Development in Information Retrieval, SIGIR '09
  (New York, NY, USA: Association for Computing Machinery), 758–759,
  \dodoi{10.1145/1571941.1572114}

\bibitem[{ {DESI Collaboration} {et~al.}(2025){DESI Collaboration}, {Karim},
  {Adame}, {Aguilar}, {Ahlen}, {Alam}, {Alobaid}, {Alves}, {Anand},
  {Armengaud}, {et~al.}}]{2025arXiv250314745D}
{DESI Collaboration}, {Karim}, M.~A., {Adame}, A.~G., {et~al.} 2025,
  \bibinfo{title}{{DESI Data Release 1},} arXiv e-prints, arXiv:2503.14745,
  \dodoi{10.48550/arXiv.2503.14745}

\bibitem[{A. {Donath} {et~al.}(2023){Donath}, {Terrier}, {Remy}, {Sinha},
  {Nigro}, {Pintore}, {Kh{\'e}lifi}, {Olivera-Nieto}, {Ruiz}, {Br{\"u}gge},
  {Linhoff}, {Contreras}, {Acero}, {Aguasca-Cabot}, {Berge}, {Bhattacharjee},
  {Buchner}, {Boisson}, {Carreto Fidalgo}, {Chen}, {de Bony de Lavergne}, {de
  Miranda Cardoso}, {Deil}, {F{\"u}{\ss}ling}, {Funk}, {Giunti}, {Hinton},
  {Jouvin}, {King}, {Lefaucheur}, {Lemoine-Goumard}, {Lenain},
  {L{\'o}pez-Coto}, {Mohrmann}, {Morcuende}, {Panny}, {Regeard}, {Saha},
  {Siejkowski}, {Siemiginowska}, {Sip{\H{o}}cz}, {Unbehaun}, {van Eldik},
  {Vuillaume}, \& {Zanin}}]{2023A&A...678A.157D}
{Donath}, A., {Terrier}, R., {Remy}, Q., {et~al.} 2023,
  \bibinfo{title}{{Gammapy: A Python package for gamma-ray astronomy},} \aap,
  678, A157, \dodoi{10.1051/0004-6361/202346488}

\bibitem[{L. {Eren Erdogan} {et~al.}(2025){Eren Erdogan}, {Lee}, {Kim}, {Moon},
  {Furuta}, {Anumanchipalli}, {Keutzer}, \& {Gholami}}]{2025arXiv250309572E}
{Eren Erdogan}, L., {Lee}, N., {Kim}, S., {et~al.} 2025,
  \bibinfo{title}{{Plan-and-Act: Improving Planning of Agents for Long-Horizon
  Tasks},} arXiv e-prints, arXiv:2503.09572, \dodoi{10.48550/arXiv.2503.09572}

\bibitem[{ {Euclid Collaboration} {et~al.}(2025){Euclid Collaboration},
  {Mellier}, {Abdurro'uf}, {Acevedo Barroso}, {Ach{\'u}carro}, {Adamek},
  {Adam}, {Addison}, {Aghanim}, {Ahumada}, {et~al.}}]{2025A&A...697A...1E}
{Euclid Collaboration}, {Mellier}, Y., {Abdurro'uf}, {et~al.} 2025,
  \bibinfo{title}{{Euclid. I. Overview of the Euclid mission},} \aap, 697, A1,
  \dodoi{10.1051/0004-6361/202450810}

\bibitem[{F. {Feroz} {et~al.}(2009){Feroz}, {Hobson}, \&
  {Bridges}}]{2009MNRAS.398.1601F}
{Feroz}, F., {Hobson}, M.~P., \& {Bridges}, M. 2009,
  \bibinfo{title}{{MULTINEST: an efficient and robust Bayesian inference tool
  for cosmology and particle physics},} \mnras, 398, 1601,
  \dodoi{10.1111/j.1365-2966.2009.14548.x}

\bibitem[{C.~J. {Fluke} \& C. {Jacobs}(2020){Fluke} \&
  {Jacobs}}]{Fluke2020SurveyingML}
{Fluke}, C.~J., \& {Jacobs}, C. 2020, \bibinfo{title}{{Surveying the reach and
  maturity of machine learning and artificial intelligence in astronomy},}
  WIREs Data Mining and Knowledge Discovery, 10, e1349,
  \dodoi{10.1002/widm.1349}

\bibitem[{H. {Gabbard} {et~al.}(2022){Gabbard}, {Messenger}, {Heng},
  {Tonolini}, \& {Murray-Smith}}]{2022NatPh..18..112G}
{Gabbard}, H., {Messenger}, C., {Heng}, I.~S., {Tonolini}, F., \&
  {Murray-Smith}, R. 2022, \bibinfo{title}{{Bayesian parameter estimation using
  conditional variational autoencoders for gravitational-wave astronomy},}
  Nature Physics, 18, 112, \dodoi{10.1038/s41567-021-01425-7}

\bibitem[{ {Gaia Collaboration} {et~al.}(2023){Gaia Collaboration},
  {Vallenari}, {Brown}, {Prusti}, {de Bruijne}, {Arenou}, {Babusiaux},
  {Bailer-Jones}, {Biermann}, {Creevey}, {et~al.}}]{2023A&A...674A...1G}
{Gaia Collaboration}, {Vallenari}, A., {Brown}, A.~G.~A., {et~al.} 2023,
  \bibinfo{title}{{Gaia Data Release 3. Summary of the content and survey
  properties},} \aap, 674, A1, \dodoi{10.1051/0004-6361/202243940}

\bibitem[{L. {Gao} {et~al.}(2023){Gao}, {Ma}, {Lin}, \& {Callan}}]{gao2023hyde}
{Gao}, L., {Ma}, X., {Lin}, J., \& {Callan}, J. 2023, \bibinfo{title}{{Precise
  Zero-Shot Dense Retrieval without Relevance Labels},} in Proceedings of the
  61st Annual Meeting of the Association for Computational Linguistics (Volume
  1: Long Papers), 1762--1777, \dodoi{10.18653/v1/2023.acl-long.99}

\bibitem[{Y. {Gao} {et~al.}(2023){Gao}, {Xiong}, {Gao}, {Jia}, {Pan}, {Bi},
  {Dai}, {Sun}, {Wang}, \& {Wang}}]{2023arXiv231210997G}
{Gao}, Y., {Xiong}, Y., {Gao}, X., {et~al.} 2023,
  \bibinfo{title}{{Retrieval-Augmented Generation for Large Language Models: A
  Survey},} arXiv e-prints, arXiv:2312.10997, \dodoi{10.48550/arXiv.2312.10997}

\bibitem[{L.~J. {Garcia} {et~al.}(2022){Garcia}, {Timmermans}, {Pozuelos},
  {Ducrot}, {Gillon}, {Delrez}, {Wells}, \& {Jehin}}]{2022MNRAS.509.4817G}
{Garcia}, L.~J., {Timmermans}, M., {Pozuelos}, F.~J., {et~al.} 2022,
  \bibinfo{title}{{prose: A Python framework for modular astronomical images
  processing},} \mnras, 509, 4817, \dodoi{10.1093/mnras/stab3113}

\bibitem[{S. {Garrappa} {et~al.}(2019){Garrappa}, {Buson}, {Franckowiak},
  {Fermi-LAT Collaboration}, {Shappee}, {Beacom}, {Dong}, {Holoien},
  {Kochanek}, {Prieto}, {Stanek}, {Thompson}, {ASAS-SN Collaboration},
  {Aartsen}, {Ackermann}, {Adams}, {Aguilar}, {Ahlers}, {Ahrens}, {Alispach},
  {Andeen}, {Anderson}, {Ansseau}, {Anton}, {Arg{\"u}elles}, {Auffenberg},
  {Axani}, {Backes}, {Bagherpour}, {Bai}, {Barbano}, {Barwick}, {Baum}, {Bay},
  {Beatty}, {Becker}, {Becker Tjus}, {BenZvi}, {Berley}, {Bernardini},
  {Besson}, {Binder}, {Bindig}, {Blaufuss}, {Blot}, {Bohm}, {B{\"o}rner},
  {B{\"o}ser}, {Botner}, {Bourbeau}, {Bourbeau}, {Bradascio}, {Braun}, {Bretz},
  {Bron}, {Brostean-Kaiser}, {Burgman}, {Busse}, {Carver}, {Chen}, {Cheung},
  {Chirkin}, {Clark}, {Classen}, {Collin}, {Conrad}, {Coppin}, {Correa},
  {Cowen}, {Cross}, {Dave}, {de Andr{\'e}}, {De Clercq}, {DeLaunay},
  {Dembinski}, {Deoskar}, {De Ridder}, {Desiati}, {de Vries}, {de Wasseige},
  {de With}, {DeYoung}, {Diaz}, {D{\'\i}az-V{\'e}lez}, {Dujmovic}, {Dunkman},
  {Dvorak}, {Eberhardt}, {Ehrhardt}, {Eller}, {Evenson}, {Fahey}, {Fazely},
  {Felde}, {Filimonov}, {Finley}, {Franckowiak}, {Friedman}, {Fritz},
  {Gaisser}, {Gallagher}, {Ganster}, {Garrappa}, {Gerhardt}, {Ghorbani},
  {Glauch}, {Gl{\"u}senkamp}, {Goldschmidt}, {Gonzalez}, {Grant}, {Griffith},
  {G{\"u}nder}, {G{\"u}nd{\"u}z}, {Haack}, {Hallgren}, {Halve}, {Halzen},
  {Hanson}, {Hebecker}, {Heereman}, {Helbing}, {Hellauer}, {Henningsen},
  {Hickford}, {Hignight}, {Hill}, {Hoffman}, {Hoffmann}, {Hoinka},
  {Hokanson-Fasig}, {Hoshina}, {Huang}, {Huber}, {Hultqvist}, {H{\"u}nnefeld},
  {Hussain}, {In}, {Iovine}, {Ishihara}, {Jacobi}, {Japaridze}, {Jeong},
  {Jero}, {Jones}, {Kang}, {Kappes}, {Kappesser}, {Karg}, {Karl}, {Karle},
  {Katz}, {Kauer}, {Keivani}, {Kelley}, {Kheirandish}, {Kim}, {Kintscher},
  {Kiryluk}, {Kittler}, {Klein}, {Koirala}, {Kolanoski}, {K{\"o}pke}, {Kopper},
  {Kopper}, {Koskinen}, {Kowalski}, {Krings}, {Kr{\"u}ckl}, {Kulacz}, {Kunwar},
  {Kurahashi}, {Kyriacou}, {Labare}, {Lanfranchi}, {Larson}, {Lauber}, {Lazar},
  {Leonard}, {Leuermann}, {Liu}, {Lohfink}, {Lozano Mariscal}, {Lu},
  {Lucarelli}, {L{\"u}nemann}, {Luszczak}, {Madsen}, {Maggi}, {Mahn}, {Makino},
  {Mallot}, {Mancina}, {Mari{\textcommabelow s}}, {Maruyama}, {Mase}, {Maunu},
  {Meagher}, {Medici}, \& {Medina}}]{2019ApJ...880..103G}
{Garrappa}, S., {Buson}, S., {Franckowiak}, A., {et~al.} 2019,
  \bibinfo{title}{{Investigation of Two Fermi-LAT Gamma-Ray Blazars Coincident
  with High-energy Neutrinos Detected by IceCube},} \apj, 880, 103,
  \dodoi{10.3847/1538-4357/ab2ada}

\bibitem[{S. {Gasparyan} {et~al.}(2022){Gasparyan}, {B{\'e}gu{\'e}}, \&
  {Sahakyan}}]{2022MNRAS.509.2102G}
{Gasparyan}, S., {B{\'e}gu{\'e}}, D., \& {Sahakyan}, N. 2022,
  \bibinfo{title}{{Time-dependent lepto-hadronic modelling of the emission from
  blazar jets with SOPRANO: the case of TXS 0506+056, 3HSP J095507.9+355101,
  and 3C 279},} \mnras, 509, 2102, \dodoi{10.1093/mnras/stab2688}

\bibitem[{J.-J. {Geng} {et~al.}(2018){Geng}, {Huang}, {Wu}, {Zhang}, \&
  {Zong}}]{2018ApJS..234....3G}
{Geng}, J.-J., {Huang}, Y.-F., {Wu}, X.-F., {Zhang}, B., \& {Zong}, H.-S. 2018,
  \bibinfo{title}{{Low-energy Spectra of Gamma-Ray Bursts from Cooling
  Electrons},} \apjs, 234, 3, \dodoi{10.3847/1538-4365/aa9e84}

\bibitem[{D. {George} \& E.~A. {Huerta}(2018){George} \&
  {Huerta}}]{2018PhRvD..97d4039G}
{George}, D., \& {Huerta}, E.~A. 2018, \bibinfo{title}{{Deep neural networks to
  enable real-time multimessenger astrophysics},} \prd, 97, 044039,
  \dodoi{10.1103/PhysRevD.97.044039}

\bibitem[{D. {Giles} \& L. {Walkowicz}(2019){Giles} \&
  {Walkowicz}}]{2019MNRAS.484..834G}
{Giles}, D., \& {Walkowicz}, L. 2019, \bibinfo{title}{{Systematic serendipity:
  a test of unsupervised machine learning as a method for anomaly detection},}
  \mnras, 484, 834, \dodoi{10.1093/mnras/sty3461}

\bibitem[{A. {Ginsburg} {et~al.}(2019){Ginsburg}, {Sip{\H{o}}cz}, {Brasseur},
  {Cowperthwaite}, {Craig}, {Deil}, {Guillochon}, {Guzman}, {Liedtke}, {Lian
  Lim}, {Lockhart}, {Mommert}, {Morris}, {Norman}, {Parikh}, {Persson},
  {Robitaille}, {Segovia}, {Singer}, {Tollerud}, {de Val-Borro}, {Valtchanov},
  {Woillez}, \& {Astroquery Collaboration}}]{2019AJ....157...98G}
{Ginsburg}, A., {Sip{\H{o}}cz}, B.~M., {Brasseur}, C.~E., {et~al.} 2019,
  \bibinfo{title}{{{astroquery}: An Astronomical Web-querying Package in
  Python},} \aj, 157, 98, \dodoi{10.3847/1538-3881/aafc33}

\bibitem[{R.~E. {Gonz{\'a}lez} {et~al.}(2018){Gonz{\'a}lez}, {Mu{\~n}oz}, \&
  {Hern{\'a}ndez}}]{2018A&C....25..103G}
{Gonz{\'a}lez}, R.~E., {Mu{\~n}oz}, R.~P., \& {Hern{\'a}ndez}, C.~A. 2018,
  \bibinfo{title}{{Galaxy detection and identification using deep learning and
  data augmentation},} Astronomy and Computing, 25, 103,
  \dodoi{10.1016/j.ascom.2018.09.004}

\bibitem[{C. {Gu{\'e}pin} {et~al.}(2022){Gu{\'e}pin}, {Kotera}, \&
  {Oikonomou}}]{2022NatRP...4..697G}
{Gu{\'e}pin}, C., {Kotera}, K., \& {Oikonomou}, F. 2022,
  \bibinfo{title}{{High-energy neutrino transients and the future of
  multi-messenger astronomy},} Nature Reviews Physics, 4, 697,
  \dodoi{10.1038/s42254-022-00504-9}

\bibitem[{D. {Haileselassie Hagos} {et~al.}(2024){Haileselassie Hagos},
  {Battle}, \& {Rawat}}]{2024arXiv240714962H}
{Haileselassie Hagos}, D., {Battle}, R., \& {Rawat}, D.~B. 2024,
  \bibinfo{title}{{Recent Advances in Generative AI and Large Language Models:
  Current Status, Challenges, and Perspectives},} arXiv e-prints,
  arXiv:2407.14962, \dodoi{10.48550/arXiv.2407.14962}

\bibitem[{G. {Hallinan} {et~al.}(2019){Hallinan}, {Ravi}, {Weinreb}, {Kocz},
  {Huang}, {Woody}, {Lamb}, {D'Addario}, {Catha}, {Law}, {Kulkarni}, {Phinney},
  {Eastwood}, {Bouman}, {McLaughlin}, {Ransom}, {Siemens}, {Cordes}, {Lynch},
  {Kaplan}, {Brazier}, {Bhatnagar}, {Myers}, {Walter}, \&
  {Gaensler}}]{2019BAAS...51g.255H}
{Hallinan}, G., {Ravi}, V., {Weinreb}, S., {et~al.} 2019, \bibinfo{title}{{The
  DSA-2000 {\textemdash} A Radio Survey Camera},} in Bulletin of the American
  Astronomical Society, Vol.~51 (AIP), 255, \dodoi{10.48550/arXiv.1907.07648}

\bibitem[{T.~A. {Hinners} {et~al.}(2018){Hinners}, {Tat}, \&
  {Thorp}}]{2018AJ....156....7H}
{Hinners}, T.~A., {Tat}, K., \& {Thorp}, R. 2018, \bibinfo{title}{{Machine
  Learning Techniques for Stellar Light Curve Classification},} \aj, 156, 7,
  \dodoi{10.3847/1538-3881/aac16d}

\bibitem[{ {IceCube Collaboration} {et~al.}(2018{\natexlab{a}}){IceCube
  Collaboration}, {Aartsen}, {Ackermann}, {Adams}, {Aguilar}, {Ahlers},
  {Ahrens}, {Al Samarai}, {Altmann}, {Andeen}, {Anderson}, {Ansseau}, {Anton},
  {Arg{\"u}elles}, {Auffenberg}, {Axani}, {Bagherpour}, {Bai}, {Barron},
  {Barwick}, {Baum}, {Bay}, {Beatty}, {Becker Tjus}, {Becker}, {BenZvi},
  {Berley}, {Bernardini}, {Besson}, {Binder}, {Bindig}, {Blaufuss}, {Blot},
  {Bohm}, {B{\"o}rner}, {Bos}, {B{\"o}ser}, {Botner}, {Bourbeau}, {Bourbeau},
  {Bradascio}, {Braun}, {Brenzke}, {Bretz}, {Bron}, {Brostean-Kaiser},
  {Burgman}, {Busse}, {Carver}, {Cheung}, {Chirkin}, {Christov}, {Clark},
  {Classen}, {Coenders}, {Collin}, {Conrad}, {Coppin}, {Correa}, {Cowen},
  {Cross}, {Dave}, {Day}, {de Andr{\'e}}, {De Clercq}, {DeLaunay}, {Dembinski},
  {De Ridder}, {Desiati}, {de Vries}, {de Wasseige}, {de With}, {DeYoung},
  {D{\'\i}az-V{\'e}lez}, {di Lorenzo}, {Dujmovic}, {Dumm}, {Dunkman}, {Dvorak},
  {Eberhardt}, {Ehrhardt}, {Eichmann}, {Eller}, {Evenson}, {Fahey}, {Fazely},
  {Felde}, {Filimonov}, {Finley}, {Flis}, {Franckowiak}, {Friedman}, {Fritz},
  {Gaisser}, {Gallagher}, {Gerhardt}, {Ghorbani}, {Glauch}, {Gl{\"u}senkamp},
  {Goldschmidt}, {Gonzalez}, {Grant}, {Griffith}, {Haack}, {Hallgren},
  {Halzen}, {Hanson}, {Hebecker}, {Heereman}, {Helbing}, {Hellauer},
  {Hickford}, {Hignight}, {Hill}, {Hoffman}, {Hoffmann}, {Hoinka},
  {Hokanson-Fasig}, {Hoshina}, {Huang}, {Huber}, {Hultqvist}, {H{\"u}nnefeld},
  {Hussain}, {In}, {Iovine}, {Ishihara}, {Jacobi}, {Japaridze}, {Jeong},
  {Jero}, {Jones}, {Kalaczynski}, {Kang}, {Kappes}, {Kappesser}, {Karg},
  {Karle}, {Katz}, {Kauer}, {Keivani}, {Kelley}, {Kheirandish}, {Kim}, {Kim},
  {Kintscher}, {Kiryluk}, {Kittler}, {Klein}, {Koirala}, {Kolanoski},
  {K{\"o}pke}, {Kopper}, {Kopper}, {Koschinsky}, {Koskinen}, {Kowalski},
  {Krings}, {Kroll}, {Kr{\"u}ckl}, {Kunwar}, {Kurahashi}, {Kuwabara},
  {Kyriacou}, {Labare}, {Lanfranchi}, {Larson}, {Lauber}, {Leonard},
  {Lesiak-Bzdak}, {Leuermann}, {Liu}, {Lozano Mariscal}, {Lu}, {L{\"u}nemann},
  {Luszczak}, {Madsen}, {Maggi}, {Mahn}, {Mancina}, {Maruyama}, {Mase},
  {Maunu}, {Meagher}, {Medici}, {Meier}, {Menne}, {Merino}, {Meures},
  {Miarecki}, {Micallef}, {Moment{\'e}}, {Montaruli}, {Moore}, {Morse},
  {Moulai}, {Nahnhauer}, {Nakarmi}, {Naumann}, \& {Neer}}]{2018Sci...361.1378I}
{IceCube Collaboration}, {Aartsen}, M.~G., {Ackermann}, M., {et~al.}
  2018{\natexlab{a}}, \bibinfo{title}{{Multimessenger observations of a flaring
  blazar coincident with high-energy neutrino IceCube-170922A},} Science, 361,
  eaat1378, \dodoi{10.1126/science.aat1378}

\bibitem[{ {IceCube Collaboration} {et~al.}(2018{\natexlab{b}}){IceCube
  Collaboration}, {Aartsen}, {Ackermann}, {Adams}, {Aguilar}, {Ahlers},
  {Ahrens}, {Samarai}, {Altmann}, {Andeen}, {Anderson}, {Ansseau}, {Anton},
  {Arg{\"u}elles}, {Arsioli}, {Auffenberg}, {Axani}, {Bagherpour}, {Bai},
  {Barron}, {Barwick}, {Baum}, {Bay}, {Beatty}, {Becker Tjus}, {Becker},
  {BenZvi}, {Berley}, {Bernardini}, {Besson}, {Binder}, {Bindig}, {Blaufuss},
  {Blot}, {Bohm}, {B{\"o}rner}, {Bos}, {B{\"o}ser}, {Botner}, {Bourbeau},
  {Bourbeau}, {Bradascio}, {Braun}, {Brenzke}, {Bretz}, {Bron},
  {Brostean-Kaiser}, {Burgman}, {Busse}, {Carver}, {Cheung}, {Chirkin},
  {Christov}, {Clark}, {Classen}, {Coenders}, {Collin}, {Conrad}, {Coppin},
  {Correa}, {Cowen}, {Cross}, {Dave}, {Day}, {de Andr{\'e}}, {De Clercq},
  {DeLaunay}, {Dembinski}, {DeRidder}, {Desiati}, {de Vries}, {de Wasseige},
  {de With}, {DeYoung}, {D{\'\i}az-V{\'e}lez}, {di Lorenzo}, {Dujmovic},
  {Dumm}, {Dunkman}, {Dvorak}, {Eberhardt}, {Ehrhardt}, {Eichmann}, {Eller},
  {Evenson}, {Fahey}, {Fazely}, {Felde}, {Filimonov}, {Finley}, {Flis},
  {Franckowiak}, {Friedman}, {Fritz}, {Gaisser}, {Gallagher}, {Gerhardt},
  {Ghorbani}, {Giommi}, {Glauch}, {Gl{\"u}senkamp}, {Goldschmidt}, {Gonzalez},
  {Grant}, {Griffith}, {Haack}, {Hallgren}, {Halzen}, {Hanson}, {Hebecker},
  {Heereman}, {Helbing}, {Hellauer}, {Hickford}, {Hignight}, {Hill}, {Hoffman},
  {Hoffmann}, {Hoinka}, {Hokanson-Fasig}, {Hoshina}, {Huang}, {Huber},
  {Hultqvist}, {H{\"u}nnefeld}, {Hussain}, {In}, {Iovine}, {Ishihara},
  {Jacobi}, {Japaridze}, {Jeong}, {Jero}, {Jones}, {Kalaczynski}, {Kang},
  {Kappes}, {Kappesser}, {Karg}, {Karle}, {Katz}, {Kauer}, {Keivani}, {Kelley},
  {Kheirandish}, {Kim}, {Kim}, {Kintscher}, {Kiryluk}, {Kittler}, {Klein},
  {Koirala}, {Kolanoski}, {K{\"o}pke}, {Kopper}, {Kopper}, {Koschinsky},
  {Koskinen}, {Kowalski}, {Krammer}, {Krings}, {Kroll}, {Kr{\"u}ckl}, {Kunwar},
  {Kurahashi}, {Kuwabara}, {Kyriacou}, {Labare}, {Lanfranchi}, {Larson},
  {Lauber}, {Leonard}, {Lesiak-Bzdak}, {Leuermann}, {Liu}, {Lozano Mariscal},
  {Lu}, {L{\"u}nemann}, {Luszczak}, {Madsen}, {Maggi}, {Mahn}, {Mancina},
  {Maruyama}, {Mase}, {Maunu}, {Meagher}, {Medici}, {Meier}, {Menne}, {Merino},
  {Meures}, {Miarecki}, {Micallef}, {Moment{\'e}}, {Montaruli}, {Moore},
  {Morse}, {Moulai}, \& {Nahnhauer}}]{2018Sci...361..147I}
{IceCube Collaboration}, {Aartsen}, M.~G., {Ackermann}, M., {et~al.}
  2018{\natexlab{b}}, \bibinfo{title}{{Neutrino emission from the direction of
  the blazar TXS 0506+056 prior to the IceCube-170922A alert},} Science, 361,
  147, \dodoi{10.1126/science.aat2890}

\bibitem[{{\v{Z}}. {Ivezi{\'c}} {et~al.}(2019){Ivezi{\'c}}, {Kahn}, {Tyson},
  {Abel}, {Acosta}, {Allsman}, {Alonso}, {AlSayyad}, {Anderson}, {Andrew},
  {et~al.}}]{2019ApJ...873..111I}
{Ivezi{\'c}}, {\v{Z}}., {Kahn}, S.~M., {Tyson}, J.~A., {et~al.} 2019,
  \bibinfo{title}{{LSST: From Science Drivers to Reference Design and
  Anticipated Data Products},} \apj, 873, 111, \dodoi{10.3847/1538-4357/ab042c}

\bibitem[{K.~G. {Iyer} {et~al.}(2024){Iyer}, {Yunus}, {O'Neill}, {Ye}, {Hyk},
  {McCormick}, {Ciuc{\u{a}}}, {Wu}, {Accomazzi}, {Astarita},
  {et~al.}}]{Iyer2024Pathfinder}
{Iyer}, K.~G., {Yunus}, M., {O'Neill}, C., {et~al.} 2024,
  \bibinfo{title}{{Pathfinder: A Semantic Framework for Literature Review and
  Knowledge Discovery in Astronomy},} \apjs, 275, 38,
  \dodoi{10.3847/1538-4365/ad7c43}

\bibitem[{M. {Klinger} {et~al.}(2024){Klinger}, {Rudolph}, {Rodrigues}, {Yuan},
  {Fichet de Clairfontaine}, {Fedynitch}, {Winter}, {Pohl}, \&
  {Gao}}]{2024ApJS..275....4K}
{Klinger}, M., {Rudolph}, A., {Rodrigues}, X., {et~al.} 2024,
  \bibinfo{title}{{AM$^{3}$: An Open-source Tool for Time-dependent
  Lepto-hadronic Modeling of Astrophysical Sources},} \apjs, 275, 4,
  \dodoi{10.3847/1538-4365/ad725c}

\bibitem[{G. {Lapenta} {et~al.}(2006){Lapenta}, {Brackbill}, \&
  {Ricci}}]{2006PhPl...13e5904L}
{Lapenta}, G., {Brackbill}, J.~U., \& {Ricci}, P. 2006,
  \bibinfo{title}{{Kinetic approach to microscopic-macroscopic coupling in
  space and laboratory plasmasa)},} Physics of Plasmas, 13, 055904,
  \dodoi{10.1063/1.2173623}

\bibitem[{N. {Law} {et~al.}(2022){Law}, {Vasquez Soto}, {Corbett}, {Galliher},
  {Gonzalez}, {Machia}, \& {Walters}}]{2022SPIE12182E..4HL}
{Law}, N., {Vasquez Soto}, A., {Corbett}, H., {et~al.} 2022,
  \bibinfo{title}{{The inside-out, upside-down telescope: the Argus Array's new
  pseudofocal design},} in Society of Photo-Optical Instrumentation Engineers
  (SPIE) Conference Series, Vol. 12182, Ground-based and Airborne Telescopes
  IX, ed. H.~K. {Marshall}, J.~{Spyromilio}, \& T.~{Usuda}, 121824H,
  \dodoi{10.1117/12.2630037}

\bibitem[{L. {Lehner} \& F. {Pretorius}(2014){Lehner} \&
  {Pretorius}}]{2014ARA&A..52..661L}
{Lehner}, L., \& {Pretorius}, F. 2014, \bibinfo{title}{{Numerical Relativity
  and Astrophysics},} \araa, 52, 661,
  \dodoi{10.1146/annurev-astro-081913-040031}

\bibitem[{P. {Lewis} {et~al.}(2020){Lewis}, {Perez}, {Piktus}, {Petroni},
  {Karpukhin}, {Goyal}, {K{\"u}ttler}, {Lewis}, {Yih}, {Rockt{\"a}schel},
  {Riedel}, \& {Kiela}}]{2020arXiv200511401L}
{Lewis}, P., {Perez}, E., {Piktus}, A., {et~al.} 2020,
  \bibinfo{title}{{Retrieval-Augmented Generation for Knowledge-Intensive NLP
  Tasks},} arXiv e-prints, arXiv:2005.11401, \dodoi{10.48550/arXiv.2005.11401}

\bibitem[{Z. {Li} {et~al.}(2025){Li}, {Zhang}, {Han}, {Liu}, {Xie}, {Zhang},
  {Choi}, {Zou}, \& {Lu}}]{2025arXiv251005592L}
{Li}, Z., {Zhang}, H., {Han}, S., {et~al.} 2025, \bibinfo{title}{{In-the-Flow
  Agentic System Optimization for Effective Planning and Tool Use},} arXiv
  e-prints, arXiv:2510.05592, \dodoi{10.48550/arXiv.2510.05592}

\bibitem[{M. {Lochner} \& B.~A. {Bassett}(2021){Lochner} \&
  {Bassett}}]{2021A&C....3600481L}
{Lochner}, M., \& {Bassett}, B.~A. 2021, \bibinfo{title}{{ASTRONOMALY:
  Personalised active anomaly detection in astronomical data},} Astronomy and
  Computing, 36, 100481, \dodoi{10.1016/j.ascom.2021.100481}

\bibitem[{G. {Maravelias} {et~al.}(2022){Maravelias}, {Bonanos}, {Tramper}, {de
  Wit}, {Yang}, \& {Bonfini}}]{2022A&A...666A.122M}
{Maravelias}, G., {Bonanos}, A.~Z., {Tramper}, F., {et~al.} 2022,
  \bibinfo{title}{{A machine-learning photometric classifier for massive stars
  in nearby galaxies. I. The method},} \aap, 666, A122,
  \dodoi{10.1051/0004-6361/202141397}

\bibitem[{P. {M{\'e}sz{\'a}ros} {et~al.}(2019){M{\'e}sz{\'a}ros}, {Fox},
  {Hanna}, \& {Murase}}]{2019NatRP...1..585M}
{M{\'e}sz{\'a}ros}, P., {Fox}, D.~B., {Hanna}, C., \& {Murase}, K. 2019,
  \bibinfo{title}{{Multi-messenger astrophysics},} Nature Reviews Physics, 1,
  585, \dodoi{10.1038/s42254-019-0101-z}

\bibitem[{B. {Naul} {et~al.}(2018){Naul}, {Bloom}, {P{\'e}rez}, \& {van der
  Walt}}]{2018NatAs...2..151N}
{Naul}, B., {Bloom}, J.~S., {P{\'e}rez}, F., \& {van der Walt}, S. 2018,
  \bibinfo{title}{{A recurrent neural network for classification of unevenly
  sampled variable stars},} Nature Astronomy, 2, 151,
  \dodoi{10.1038/s41550-017-0321-z}

\bibitem[{H. {Naveed} {et~al.}(2023){Naveed}, {Ullah Khan}, {Qiu}, {Saqib},
  {Anwar}, {Usman}, {Akhtar}, {Barnes}, \& {Mian}}]{2023arXiv230706435N}
{Naveed}, H., {Ullah Khan}, A., {Qiu}, S., {et~al.} 2023, \bibinfo{title}{{A
  Comprehensive Overview of Large Language Models},} arXiv e-prints,
  arXiv:2307.06435, \dodoi{10.48550/arXiv.2307.06435}

\bibitem[{U.~M. {Noebauer} \& S.~A. {Sim}(2019){Noebauer} \&
  {Sim}}]{2019LRCA....5....1N}
{Noebauer}, U.~M., \& {Sim}, S.~A. 2019, \bibinfo{title}{{Monte Carlo radiative
  transfer},} Living Reviews in Computational Astrophysics, 5, 1,
  \dodoi{10.1007/s41115-019-0004-9}

\bibitem[{J. {Nordin} {et~al.}(2025){Nordin}, {Brinnel}, {van Santen},
  {Reusch}, \& {Kowalski}}]{2025A&A...698A..13N}
{Nordin}, J., {Brinnel}, V., {van Santen}, J., {Reusch}, S., \& {Kowalski}, M.
  2025, \bibinfo{title}{{AMPEL workflows for LSST: Modular and reproducible
  real-time photometric classification},} \aap, 698, A13,
  \dodoi{10.1051/0004-6361/202452481}

\bibitem[{F. {Ochsenbein} {et~al.}(2000){Ochsenbein}, {Bauer}, \&
  {Marcout}}]{2000A&AS..143...23O}
{Ochsenbein}, F., {Bauer}, P., \& {Marcout}, J. 2000, \bibinfo{title}{{The
  VizieR database of astronomical catalogues},} \aaps, 143, 23,
  \dodoi{10.1051/aas:2000169}

\bibitem[{P. {Padovani} {et~al.}(2018){Padovani}, {Giommi}, {Resconi},
  {Glauch}, {Arsioli}, {Sahakyan}, \& {Huber}}]{2018MNRAS.480..192P}
{Padovani}, P., {Giommi}, P., {Resconi}, E., {et~al.} 2018,
  \bibinfo{title}{{Dissecting the region around IceCube-170922A: the blazar TXS
  0506+056 as the first cosmic neutrino source},} \mnras, 480, 192,
  \dodoi{10.1093/mnras/sty1852}

\bibitem[{L. {Parker} {et~al.}(2024){Parker}, {Lanusse}, {Golkar}, {Sarra},
  {Cranmer}, {Bietti}, {Eickenberg}, {Krawezik}, {McCabe}, {Morel}, {Ohana},
  {Pettee}, {R{\'e}galdo-Saint Blancard}, {Cho}, {Ho}, \& {Polymathic AI
  Collaboration}}]{2024MNRAS.531.4990P}
{Parker}, L., {Lanusse}, F., {Golkar}, S., {et~al.} 2024,
  \bibinfo{title}{{AstroCLIP: a cross-modal foundation model for galaxies},}
  \mnras, 531, 4990, \dodoi{10.1093/mnras/stae1450}

\bibitem[{B. {Paxton} {et~al.}(2011){Paxton}, {Bildsten}, {Dotter}, {Herwig},
  {Lesaffre}, \& {Timmes}}]{2011ApJS..192....3P}
{Paxton}, B., {Bildsten}, L., {Dotter}, A., {et~al.} 2011,
  \bibinfo{title}{{Modules for Experiments in Stellar Astrophysics (MESA)},}
  \apjs, 192, 3, \dodoi{10.1088/0067-0049/192/1/3}

\bibitem[{S. Robertson \& H. Zaragoza(2009)Robertson \&
  Zaragoza}]{10.1561/1500000019}
Robertson, S., \& Zaragoza, H. 2009, \bibinfo{title}{The Probabilistic
  Relevance Framework: BM25 and Beyond,} Found. Trends Inf. Retr., 3,
  333–389, \dodoi{10.1561/1500000019}

\bibitem[{A. {Roy} {et~al.}(2026){Roy}, {Feldman}, {Klupar}, {DiPalma},
  {Perlmutter}, {Douglas}, {Aldering}, {Furesz}, {Ingraham}, {Stefansson},
  {Kelly}, {Yang}, {Wevers}, {Arulanantham}, {Lasker}, {Rigault}, {Schlawin},
  {Zandbergen}, {Worden}, {Anche}, {Choi}, {Crossfield}, {Derby}, {Edelstein},
  {Eiklenborg}, {Gezari}, {Giuliano}, {Hom}, {Hoyt}, {Kang}, {Kim},
  {Kunnumkai}, {Lacroix}, {Males}, {Maccarone}, {Milani}, {Miller}, {Miller},
  {Nicolas}, {Palmese}, {Pero}, {Pueyo}, {Rinaldi}, {Sand}, {Schneider},
  {Sabhlok}, {Smith}, {Stefan}, {Kalyani Subramanian}, {Van Gorkom}, {Wong},
  {Yoo}, {Abdullah Al Zaman}, \& {the Lazuli Science
  Team}}]{2026arXiv260102556R}
{Roy}, A., {Feldman}, S., {Klupar}, P., {et~al.} 2026, \bibinfo{title}{{The
  Lazuli Space Observatory: Architecture \& Capabilities},} arXiv e-prints,
  arXiv:2601.02556, \dodoi{10.48550/arXiv.2601.02556}

\bibitem[{N. {Sahakyan}(2026){Sahakyan}}]{2026IJMPD..3540009S}
{Sahakyan}, N. 2026, \bibinfo{title}{{AI in the cosmos},} International Journal
  of Modern Physics D, 35, 2540009, \dodoi{10.1142/S0218271825400097}

\bibitem[{N. {Sahakyan} {et~al.}(2025){Sahakyan}, {B{\'e}gu{\'e}}, {Casotto},
  {Dereli-B{\'e}gu{\'e}}, {Vardanyan}, {Khachatryan}, {Giommi}, \&
  {Pe'er}}]{2025ApJ...990..222S}
{Sahakyan}, N., {B{\'e}gu{\'e}}, D., {Casotto}, A., {et~al.} 2025,
  \bibinfo{title}{{Modeling Blazar Broadband Emission with Convolutional Neural
  Networks. III. Proton Synchrotron and Hybrid Models},} \apj, 990, 222,
  \dodoi{10.3847/1538-4357/adf734}

\bibitem[{N. {Sahakyan} {et~al.}(2023){Sahakyan}, {Vardanyan}, \&
  {Khachatryan}}]{2023MNRAS.519.3000S}
{Sahakyan}, N., {Vardanyan}, V., \& {Khachatryan}, M. 2023,
  \bibinfo{title}{{Gradient boosting decision trees classification of blazars
  of uncertain type in the fourth Fermi-LAT catalogue},} \mnras, 519, 3000,
  \dodoi{10.1093/mnras/stac3701}

\bibitem[{N. {Sahakyan} {et~al.}(2024{\natexlab{a}}){Sahakyan},
  {B{\'e}gu{\'e}}, {Casotto}, {Dereli-B{\'e}gu{\'e}}, {Giommi}, {Gasparyan},
  {Vardanyan}, {Khachatryan}, \& {Pe'er}}]{2024ApJ...971...70S}
{Sahakyan}, N., {B{\'e}gu{\'e}}, D., {Casotto}, A., {et~al.}
  2024{\natexlab{a}}, \bibinfo{title}{{Modeling Blazar Broadband Emission with
  Convolutional Neural Networks. II. External Compton Model},} \apj, 971, 70,
  \dodoi{10.3847/1538-4357/ad5351}

\bibitem[{N. {Sahakyan} {et~al.}(2024{\natexlab{b}}){Sahakyan}, {Vardanyan},
  {Giommi}, {B{\'e}gu{\'e}}, {Israyelyan}, {Harutyunyan}, {Manvelyan},
  {Khachatryan}, {Dereli-B{\'e}gu{\'e}}, \& {Gasparyan}}]{2024AJ....168..289S}
{Sahakyan}, N., {Vardanyan}, V., {Giommi}, P., {et~al.} 2024{\natexlab{b}},
  \bibinfo{title}{{Markarian Multiwavelength Data Center (MMDC): A Tool for
  Retrieving and Modeling Multitemporal, Multiwavelength, and Multimessenger
  Data from Blazar Observations},} \aj, 168, 289,
  \dodoi{10.3847/1538-3881/ad8231}

\bibitem[{P. {S{\'a}nchez-S{\'a}ez} {et~al.}(2021){S{\'a}nchez-S{\'a}ez},
  {Lira}, {Mart{\'\i}}, {S{\'a}nchez-Pi}, {Arredondo}, {Bauer}, {Bayo},
  {Cabrera-Vives}, {Donoso-Oliva}, {Est{\'e}vez}, {Eyheramendy}, {F{\"o}rster},
  {Hern{\'a}ndez-Garc{\'\i}a}, {Arancibia}, {P{\'e}rez-Carrasco},
  {Sep{\'u}lveda}, \& {Vergara}}]{2021AJ....162..206S}
{S{\'a}nchez-S{\'a}ez}, P., {Lira}, H., {Mart{\'\i}}, L., {et~al.} 2021,
  \bibinfo{title}{{Searching for Changing-state AGNs in Massive Data Sets. I.
  Applying Deep Learning and Anomaly-detection Techniques to Find AGNs with
  Anomalous Variability Behaviors},} \aj, 162, 206,
  \dodoi{10.3847/1538-3881/ac1426}

\bibitem[{J.~D. {Scargle}(1998){Scargle}}]{1998ApJ...504..405S}
{Scargle}, J.~D. 1998, \bibinfo{title}{{Studies in Astronomical Time Series
  Analysis. V. Bayesian Blocks, a New Method to Analyze Structure in Photon
  Counting Data},} \apj, 504, 405, \dodoi{10.1086/306064}

\bibitem[{B. {Schleicher} {et~al.}(2019){Schleicher}, {Arbet-Engels}, {Baack},
  {Balbo}, {Biland}, {Blank}, {Bretz}, {Bruegge}, {Bulinski}, {Buss}, {Doerr},
  {Dorner}, {Elsaesser}, {Grischagin}, {Hildebrand}, {Linhoff}, {Mannheim},
  {Mueller}, {Neise}, {Neronov}, {Noethe}, {Paravac}, {Rhode}, {Schulz},
  {Sedlaczek}, {Shukla}, {Sliusar}, {Willert}, \&
  {Walter}}]{2019Galax...7...62S}
{Schleicher}, B., {Arbet-Engels}, A., {Baack}, D., {et~al.} 2019,
  \bibinfo{title}{{Fractional Variability{\textemdash}A Tool to Study Blazar
  Variability},} Galaxies, 7, 62, \dodoi{10.3390/galaxies7020062}

\bibitem[{A. {Spurio Mancini} {et~al.}(2022){Spurio Mancini}, {Piras},
  {Alsina}, {Joachimi}, \& {Hobson}}]{2022MNRAS.511.1771S}
{Spurio Mancini}, A., {Piras}, D., {Alsina}, A., {Joachimi}, B., \& {Hobson},
  M.~P. 2022, \bibinfo{title}{{{CosmoPower}: emulating cosmological power
  spectra for accelerated Bayesian inference from next-generation surveys},}
  \mnras, 511, 1771, \dodoi{10.1093/mnras/stac064}

\bibitem[{R. {Teyssier} \& B. {Commer{\c{c}}on}(2019){Teyssier} \&
  {Commer{\c{c}}on}}]{2019FrASS...6...51T}
{Teyssier}, R., \& {Commer{\c{c}}on}, B. 2019, \bibinfo{title}{{Numerical
  Methods for Simulating Star Formation},} Frontiers in Astronomy and Space
  Sciences, 6, 51, \dodoi{10.3389/fspas.2019.00051}

\bibitem[{A. {Tramacere}(2020){Tramacere}}]{2020ascl.soft09001T}
{Tramacere}, A. 2020, {JetSeT: Numerical modeling and SED fitting tool for
  relativistic jets},, Astrophysics Source Code Library, record ascl:2009.001
  \doeprint{2009.001}

\bibitem[{A. {Tramacere} {et~al.}(2009){Tramacere}, {Giommi}, {Perri},
  {Verrecchia}, \& {Tosti}}]{2009A&A...501..879T}
{Tramacere}, A., {Giommi}, P., {Perri}, M., {Verrecchia}, F., \& {Tosti}, G.
  2009, \bibinfo{title}{{Swift observations of the very intense flaring
  activity of Mrk 421 during 2006. I. Phenomenological picture of electron
  acceleration and predictions for MeV/GeV emission},} \aap, 501, 879,
  \dodoi{10.1051/0004-6361/200810865}

\bibitem[{A. {Tramacere} {et~al.}(2011){Tramacere}, {Massaro}, \&
  {Taylor}}]{2011ApJ...739...66T}
{Tramacere}, A., {Massaro}, E., \& {Taylor}, A.~M. 2011,
  \bibinfo{title}{{Stochastic Acceleration and the Evolution of Spectral
  Distributions in Synchro-Self-Compton Sources: A Self-consistent Modeling of
  Blazars' Flares},} \apj, 739, 66, \dodoi{10.1088/0004-637X/739/2/66}

\bibitem[{K.-T. {Tran} {et~al.}(2025){Tran}, {Dao}, {Nguyen}, {Pham},
  {O'Sullivan}, \& {Nguyen}}]{2025arXiv250106322T}
{Tran}, K.-T., {Dao}, D., {Nguyen}, M.-D., {et~al.} 2025,
  \bibinfo{title}{{Multi-Agent Collaboration Mechanisms: A Survey of LLMs},}
  arXiv e-prints, arXiv:2501.06322, \dodoi{10.48550/arXiv.2501.06322}

\bibitem[{A. {Tzavellas} {et~al.}(2024){Tzavellas}, {Vasilopoulos},
  {Petropoulou}, {Mastichiadis}, \& {Stathopoulos}}]{2024A&A...683A.185T}
{Tzavellas}, A., {Vasilopoulos}, G., {Petropoulou}, M., {Mastichiadis}, A., \&
  {Stathopoulos}, S.~I. 2024, \bibinfo{title}{{Application of neural networks
  to synchro-Compton blazar emission models},} \aap, 683, A185,
  \dodoi{10.1051/0004-6361/202348566}

\bibitem[{J. {VanderPlas} {et~al.}(2012){VanderPlas}, {Connolly}, {Ivezic}, \&
  {Gray}}]{2012cidu.conf...47V}
{VanderPlas}, J., {Connolly}, A.~J., {Ivezic}, Z., \& {Gray}, A. 2012,
  \bibinfo{title}{{Introduction to astroML: Machine learning for
  astrophysics},} in 2012 Conference on Intelligent Data Understanding (IEEE),
  47--54, \dodoi{10.1109/CIDU.2012.6382200}

\bibitem[{G. {Vianello} {et~al.}(2015){Vianello}, {Lauer}, {Younk}, {Tibaldo},
  {Burgess}, {Ayala}, {Harding}, {Hui}, {Omodei}, \&
  {Zhou}}]{2015arXiv150708343V}
{Vianello}, G., {Lauer}, R.~J., {Younk}, P., {et~al.} 2015,
  \bibinfo{title}{{The Multi-Mission Maximum Likelihood framework (3ML)},}
  arXiv e-prints, arXiv:1507.08343, \dodoi{10.48550/arXiv.1507.08343}

\bibitem[{V.~A. {Villar} {et~al.}(2021){Villar}, {Cranmer}, {Berger},
  {Contardo}, {Ho}, {Hosseinzadeh}, \& {Lin}}]{2021ApJS..255...24V}
{Villar}, V.~A., {Cranmer}, M., {Berger}, E., {et~al.} 2021, \bibinfo{title}{{A
  Deep-learning Approach for Live Anomaly Detection of Extragalactic
  Transients},} \apjs, 255, 24, \dodoi{10.3847/1538-4365/ac0893}

\bibitem[{L. {Wang} {et~al.}(2023){Wang}, {Yang}, \&
  {Wei}}]{2023arXiv230307678W}
{Wang}, L., {Yang}, N., \& {Wei}, F. 2023, \bibinfo{title}{{Query2doc: Query
  Expansion with Large Language Models},} arXiv e-prints, arXiv:2303.07678,
  \dodoi{10.48550/arXiv.2303.07678}

\bibitem[{Y. {Wang} {et~al.}(2024){Wang}, {Zhang}, {Momtaz}, {Moradi},
  {Rastegarnia}, {Sahakyan}, {Shakeri}, \& {Li}}]{2024arXiv240410019W}
{Wang}, Y., {Zhang}, S.-R., {Momtaz}, A., {et~al.} 2024, {Can AI Understand Our
  Universe? Test of Fine-Tuning GPT by Astrophysical Data},
  \dodoi{10.48550/arXiv.2404.10019}

\bibitem[{H. {Wei} {et~al.}(2025){Wei}, {Sun}, \& {Li}}]{wei2025deepseekocr}
{Wei}, H., {Sun}, Y., \& {Li}, Y. 2025, \bibinfo{title}{{{DeepSeek-OCR}:
  Contexts Optical Compression},} arXiv e-prints, arXiv:2510.18234,
  \dodoi{10.48550/arXiv.2510.18234}

\bibitem[{M. {Wenger} {et~al.}(2000){Wenger}, {Ochsenbein}, {Egret}, {Dubois},
  {Bonnarel}, {Borde}, {Genova}, {Jaschik}, {Lalo{\"e}}, {Lesteven}, \&
  {Monier}}]{2000A&AS..143....9W}
{Wenger}, M., {Ochsenbein}, F., {Egret}, D., {et~al.} 2000,
  \bibinfo{title}{{The SIMBAD astronomical database. The CDS reference database
  for astronomical objects},} \aaps, 143, 9, \dodoi{10.1051/aas:2000332}

\bibitem[{J. {Wu} {et~al.}(2026){Wu}, {O'Brien}, {Li}, {Krafczyk}, {Shah},
  {Wasserman}, {Apley}, {Narayan}, \& {Samia}}]{2026arXiv260304392W}
{Wu}, J., {O'Brien}, J., {Li}, J., {et~al.} 2026, \bibinfo{title}{{SELDON:
  Supernova Explosions Learned by Deep ODE Networks},} arXiv e-prints,
  arXiv:2603.04392, \dodoi{10.48550/arXiv.2603.04392}

\bibitem[{Z. {Zhang} {et~al.}(2026){Zhang}, {Xu}, {Liu}, {Cui}, \&
  {Fan}}]{Zhang2026AstroFlow}
{Zhang}, Z., {Xu}, Y., {Liu}, Y., {Cui}, C., \& {Fan}, D. 2026,
  \bibinfo{title}{{AstroFlow: A customizable workflow management system for
  astronomical data production and case study of EP-WXT},} Astronomy and
  Computing, 55, 101073, \dodoi{10.1016/j.ascom.2026.101073}

\bibitem[{W.~X. {Zhao} {et~al.}(2023){Zhao}, {Zhou}, {Li}, {Tang}, {Wang},
  {Hou}, {Min}, {Zhang}, {Zhang}, {Dong}, {Du}, {Yang}, {Chen}, {Chen},
  {Jiang}, {Ren}, {Li}, {Tang}, {Liu}, {Liu}, {Nie}, \&
  {Wen}}]{2023arXiv230318223Z}
{Zhao}, W.~X., {Zhou}, K., {Li}, J., {et~al.} 2023, \bibinfo{title}{{A Survey
  of Large Language Models},} arXiv e-prints, arXiv:2303.18223,
  \dodoi{10.48550/arXiv.2303.18223}

\end{thebibliography}
\bibliographystyle{aasjournalv7}



\end{document}